%% file: main.tex
\documentclass[3p,times,11pt,sort&compress]{elsarticle}

\usepackage[utf8]{inputenc}
\usepackage{graphicx}
\usepackage{enumerate}
\usepackage{wrapfig}
\usepackage{amsmath,amsthm}
\usepackage{caption,subcaption}

\usepackage{soul}
\usepackage{cancel}
\usepackage{ulem}
\usepackage{algorithm,algpseudocode,varwidth}
\usepackage{amsfonts}
\usepackage{mathtools}
\usepackage[version=4]{mhchem}
\usepackage{mathrsfs}
\usepackage{amssymb}
\usepackage{dsfont}
\usepackage{url}
\usepackage{hyperref}
\usepackage{xcolor}
\usepackage{tcolorbox}
\usepackage{colortbl}
\usepackage{booktabs}
\usepackage{multirow}
\allowdisplaybreaks

\input{shortcuts}

\newcommand\fulltitle{Petrov-Galerkin model reduction for thermochemical \\ nonequilibrium gas mixtures}

\begin{document}

\begin{abstract}
    \input{sections/0_abstract}
\end{abstract}

\begin{frontmatter}

    \title{\fulltitle}
    \author{Ivan Zanardi}
    \ead{zanardi3@illinois.edu}
    \author{Alberto Padovan}
    \ead{padovan3@illinois.edu}
    \author{Daniel J. Bodony}
    \ead{bodony@illinois.edu}
    \author{Marco Panesi\corref{cor1}}
    \ead{mpanesi@illinois.edu}
    
    \cortext[cor1]{Corresponding Author}
    
    \address{University of Illinois at Urbana-Champaign, Department of Aerospace Engineering \\ 104 S Wright St, Urbana, IL 61801, USA}

\end{frontmatter}

\input{sections/1_intro}
\input{sections/2_physics}
\input{sections/3_roms}
\input{sections/4_results}
\input{sections/5_conclusions}

\section*{Acknowledgements}
I.\ Zanardi and M.\ Panesi were supported by the Vannevar Bush Faculty Fellowship OUSD(RE) Grant \#N00014-21-1-295. A.\ Padovan and D.\ Bodony were supported by the National Science Foundation (Characteristic Science Applications for the Leadership Class Computing Facility, Grant \#2139536, on a subaward from the University of Texas, Austin, Texas Advanced Computing Center). The views and conclusions contained herein are those of the authors and should not be interpreted as necessarily representing the official policies or endorsements, either expressed or implied, of the U.S. government.

\appendix
\input{sections/6_appendix}

\bibliography{biblio/global,biblio/hypersonics,biblio/ml_pde,biblio/rom}

\newpage

\renewcommand*{\thetable}{S\arabic{table}}
\renewcommand*{\thefigure}{S\arabic{figure}}
\renewcommand*{\thesection}{S.\arabic{section}}
\renewcommand*{\thesubsection}{\thesection.\arabic{subsection}}
\renewcommand*{\thesubsubsection}{\thesubsection.\arabic{subsubsection}}
\renewcommand*{\theequation}{\thesection.\arabic{equation}}

\setcounter{section}{0}
\setcounter{page}{1}
\setcounter{figure}{0}
\setcounter{equation}{0}


\makeatletter
\let\@title\@empty
\makeatother

\title{Supplementary Material to \\ ``\textit{\fulltitle}''}

\def\ps@pprintTitle{%
     \let\@oddhead\@empty
     \let\@evenhead\@empty
     \def\@oddfoot{\footnotesize\itshape
        Supplementary Data for \ifx\@journal\@empty Elsevier
       \else\@journal\fi\hfill\today}%
     \let\@evenfoot\@oddfoot}
\makeatother

\begin{abstract}
    \input{sections/0_abstract}
\end{abstract}
\begin{keyword}
    \input{sections/keywords}
\end{keyword}

\maketitle

\input{sections/7_supplementary}

\end{document}

%% file: shortcuts.tex
\newcommand{\vb}{\mathbf{b}}
\newcommand{\ve}{\mathbf{e}}
\newcommand{\vf}{\mathbf{f}}
\newcommand{\vg}{\mathbf{g}}
\newcommand{\vh}{\mathbf{h}}
\newcommand{\vm}{\mathbf{m}}
\newcommand{\vq}{\mathbf{q}}

\newcommand{\vy}{\mathbf{y}}
\newcommand{\vz}{\mathbf{z}}

\newcommand{\vmu}{\boldsymbol{\mu}}

\newcommand{\vxi}{\boldsymbol{\xi}}
\newcommand{\vzeta}{\boldsymbol{\zeta}}

\newcommand{\vtheta}{\boldsymbol{\theta}}

\newcommand{\mA}{\mathbf{A}}

\newcommand{\mV}{\mathbf{V}}

\newcommand{\mC}{\mathbf{C}}

\newcommand{\vn}{\mathbf{n}}
\newcommand{\bvn}{\overline{\mathbf{n}}}

\newcommand{\tvn}{\mathbf{n}^\prime}

\newcommand{\vw}{\mathbf{w}}

\newcommand{\tvw}{\mathbf{w}^\prime}

\newcommand{\mPhi}{\boldsymbol{\Phi}}
\newcommand{\mPsi}{\boldsymbol{\Psi}}

\newcommand{\mLam}{\boldsymbol{\Lambda}}
\newcommand{\mTheta}{\boldsymbol{\Theta}}
\newcommand{\mMu}{\mathbf{M}}

\newcommand{\phib}{\boldsymbol{\Phi}}
\newcommand{\psib}{\boldsymbol{\Psi}}

\newcommand{\phifb}{\phib\left(\psib^\intercal\phib\right)^{-1}}

\newcommand{\bbgam}{\overline{\boldsymbol{\gamma}}}

\newcommand{\atom}{\text{\tiny O}}
\newcommand{\mol}{\text{\tiny O$_2$}}

\newcommand{\ordermagn}[1]{$\mathcal{O}\big(#1\big)$}

\newcommand{\eqspace}{\,}

\newtheorem{proposition}{Proposition}
\newtheorem*{docremark}{Remark}

%% file: sections/0_abstract.tex
State-specific thermochemical collisional models are crucial to accurately describe the physics of systems involving nonequilibrium plasmas, but they are also computationally expensive and impractical for large-scale, multi-dimensional simulations.
Historically, computational cost has been mitigated by using empirical and physics-based arguments to reduce the complexity of the governing equations.
However, the resulting models are often inaccurate and they fail to capture the important features of the original physics.
Additionally, the construction of these models is often impractical, as it requires extensive user supervision and time-consuming parameter tuning.
In this paper, we address these issues through an easily implementable and computationally efficient model reduction pipeline based on the Petrov-Galerkin projection of the nonlinear kinetic equations onto a low-dimensional subspace.
Our approach is justified by the observation that kinetic systems in thermal nonequilibrium tend to exhibit low-rank dynamics that rapidly drive the state towards a low-dimensional subspace that can be exploited for reduced-order modeling.
Furthermore, despite the nonlinear nature of the governing equations, we observe that the dynamics of these systems evolve on subspaces that can be accurately identified using the linearized equations about thermochemical equilibrium steady states, and we shall see that this allows us to significantly reduce the cost associated with the construction of the model.
The approach is demonstrated on two distinct thermochemical systems: a rovibrational collisional model for the O$_2$-O system, and a vibrational collisional model for the combined O$_2$-O and O$_2$-O$_2$ systems. 
Our method achieves high accuracy, with relative errors of less than 1\% for macroscopic quantities (i.e., moments) and~10\% for microscopic quantities (i.e., energy levels population), while also delivering excellent compression rates and speedups, outperforming existing state-of-the-art techniques.

%% file: sections/keywords.tex
reduced order modeling \sep thermochemistry \sep nonequilibrium \sep hypersonics

%% file: sections/1_intro.tex
\section{Introduction}
Nonequilibrium flows are encountered in many engineering and science disciplines, such as hypersonic atmospheric entry~\cite{VincentiKruger_Book_1965,Park_Book_1990,Boyd2017NonequilibriumSimulation,Hirschel2009SelectedVehicles} and material processing and manufacturing with low-temperature plasmas~\cite{Harpale_JCP_2015,Harpale_PR_2016}. In these applications, the thermodynamic state of the flow frequently exhibits significant deviations of the internal energy distribution of the gaseous mixture from its Maxwell-Boltzmann equilibrium value. Such deviations can influence key macroscopic properties, including diffusion, viscous stress, heat conduction, internal energy transfer, chemical reactions, and thermal radiation. 
To describe and understand these complex behaviors and their impact, increasingly advanced and expansive mathematical models have been developed~\cite{Cercignani1988TheApplications,Ferziger1972MathematicalGases,Capitelli_Book_2016,Panesi_JCP_2013,Panesi_PR_2014}, encompassing multiple physical phenomena across a wide range of spatio-temporal scales. 
\par
The exact description of thermochemical and radiative processes can be achieved at the microscopic level by considering state-to-state (StS) collisions among atoms, molecules, and electrons, as well as the emission and absorption of photons. This approach is mathematically formalized through the StS master equations~\cite{Nagnibeda_Book_2009,Panesi_JTHT_2011,Panesi_JCP_2013,Panesi_PR_2014,Macdonald_PRF_2016,Capitelli_Book_2016,Macdonald_JFC_2020}. By leveraging quantum chemistry models based on \textit{ab initio} theories~\cite{Wang_JCP_2003,Esposito_CP_2006,Galvao_JPC_2009,Jaffe_AIAA_2009,Venturi_JCPA_2020,Priyadarshini2023EfficientArchitectures}, these equations provide an unprecedented level of accuracy in describing the nonequilibrium state of gases by capturing the microscopic interactions between colliding particles in diverse chemical systems~\cite{Panesi_JCP_2013,Panesi_PR_2014,Munafo_PoP_2013,Kustova_AIP_2019}. While these models provide exceptional accuracy, they are impractical in large-scale multi-dimensional simulations due to the exponential increase in the number of degrees of freedom (i.e., the molecules’ and atoms’ energy levels). As a result, detailed StS calculations are typically employed to develop more efficient reduced-order models to be used in computational fluid dynamics (CFD) simulations.
\par
Reduced order model (ROMs) aim to capture the essential features of the system dynamics while significantly reducing the computational cost associated with the high-fidelity full-order model (FOM). The earliest ROM for thermochemical nonequilibrium, introduced by Landau and Teller~\cite{Landau1936OnDispersion}, was based on several physical assumptions applicable only to simple harmonic oscillator molecules, resulting in a linear relaxation model. 
Building upon this formulation, more sophisticated models have been developed over the years to account for the coupling between vibration and dissociation~\cite{Hammerling1959TheoryNitrogen,Heims1963MomentDissociation,Marrone1963ChemicalLevels,Treanor1962EffectRelaxation}. However, many of these models, known as multi-temperature (MT) models~\cite{Park_Book_1990,Park_JTHT_1989,Park_JTHT_1993,Park_JTHT_1994,Park_JTHT_2001}, rely on crude, semi-empirical assumptions. They are typically constructed based on a rigid separation of internal energy modes (translational, rotational, vibrational, and electronic), without any rigorous derivation from fundamental kinetic equations nor consideration for physical principles and constraints. Given their interpolative nature, these conventional methods are fundamentally inadequate for capturing the behavior of gas particles under strong nonequilibrium conditions, where the distribution of internal energy levels exhibits complex features~\cite{Panesi_JCP_2013,Panesi_PR_2014,Heritier_JCP_2014}. By contrast, state-of-the-art coarse-grained (CG) ROMs~\cite{Magin2012Coarse-grainNitrogen,Panesi_Lani_PofF_2013,Munafo_PR_2014,Munafo_PF_2015,Liu_JCP_2015} provide a more effective solution compared to the traditional MT models. These coarse-grained approaches work by grouping individual energy levels into a smaller number of macroscopic bins. State populations are reconstructed using bin-wise distribution functions that maximize entropy and adhere to constraints based on bin properties (e.g., population, energy, or higher moments). The governing equations for the reduced-order system are derived by summing moments of the StS master equations.
Although these models have demonstrated sufficient accuracy, they have a few limitations. First, while the principle of maximum entropy used to locally reconstruct microscopic scales within each bin perfectly adheres with physical principles, it may not fully capture the complex features of the distribution function. Second, the accuracy of the coarse-grained model is highly dependent on the effectiveness of the clustering. Significant effort, in the form of time-consuming optimization or physics-based analysis of the chemical system, is often required to achieve optimal clustering~\cite{Macdonald_I_JCP_2018,Macdonald_II_JCP_2018,Sahai_JCP_2017,Venturi_JCP_2020,Sharma_PR_2020,Zanardi2023AdaptiveFlows,Jacobsen_JComP_2024}.
\par
This work aims to advance the state of the art of model reduction for thermochemical kinetics by introducing a computationally-efficient pipeline that offers several key improvements over the existing reduced-complexity modeling formulations described above.
First, the proposed method bypasses the need for empirical or physical assumptions, enhancing both accuracy and flexibility across a wide range of nonequilibrium conditions. 
Second, it avoids the need for time-consuming system analysis or optimization to identify the optimal reduced-order representation of the original dynamics. 
The fundamental component of the model reduction pipeline discussed herein is the construction of a Petrov-Galerkin model
obtained by projecting the full-order dynamics (i.e., the StS master equations) onto a low-dimensional linear subspace of the state-space.
Obtaining an accurate Petrov-Galerkin model requires the careful design of a linear projection operator of the form $\mathbb{P} = \phifb \psib^\intercal$, where $\phib$ and $\psib$ are tall rectangular matrices whose spans uniquely define $\mathbb{P}$.
While the most simplistic approach to obtain projection-based model is by \textit{orthogonally} projecting onto the span of Proper Orthogonal Decomposition (POD) modes of high-fidelity data (see, e.g., \cite{Rowley2004model,Barone2009StableFlow} in the context of compressible flows and \cite{Sutherland2009CombustionAnalysis,Parente2013PrincipalSensitivity,Bellemans2015ReductionAnalysis} in the context of thermochemical kinetics), it is well-known that physical systems that exhibit transient growth and high sensitivity to low-energy disturbances require modeling approaches based on \textit{oblique} projections \cite{Rowley2017ModelControl, Otto2022OPTIMIZINGTRAJECTORIES, Otto2023ModelCovariance, Padovan2024Data-drivenDynamics}.
If the full-order system is linear, or if the nonlinear dynamics evolve in the proximity of an equilibrium, appropriate oblique projection operators can be constructed using balanced truncation and its variants~\cite{Moore1981PrincipalReduction,Dullerud_Book_2000,Rowley2005model,Varga2000BalancedSystems,Padovan2024Continuous-timeGramians}, where a Petrov-Galerkin ROM is obtained by balancing the controllability and observability Gramians associated with the underlying linear system.
For nonlinear systems evolving far away from equilibria, Otto et al.~\cite{Otto2023ModelCovariance} recently introduced ``Covariance Balancing Reduction using Adjoint Snapshots'' (CoBRAS), which is a formulation that determines (quasi) optimal oblique projection operators by balancing the state and gradient covariance matrices associated with the full-order solution map.
Given the non-normal nature of the StS master equations, and their high sensitivity to low-energy disturbances, it is essential to seek appropriate oblique projection operators to obtain accurate Petrov-Galerkin models and we therefore choose to adopt the aforementioned CoBRAS formulation.
More specifically, we tailor CoBRAS to the StS master equations, and we assemble the state and gradient covariance matrices using linear trajectories governed by the linearization of the StS equations about thermochemical equilibria.
Using the linearized equations allows us to generate the required forward and adjoint trajectories using the exact solution of linear ordinary differential equations, as opposed to numerically integrating a stiff set of nonlinear equations, and this results in computational savings of several orders of magnitude.
Additionally, despite the fact that these trajectories are a first-order approximation of the nonlinear solution of the StS equations, we shall see that the loss of accuracy is minimal, and we can still capture the relevant subspaces that are necessary to construct an appropriate oblique projector $\mathbb{P}$ for Petrov-Galerkin modeling.
The performance of the proposed models is assessed on two thermochemical systems involving a pure oxygen mixture: a rovibrational collisional (RVC) model and a vibrational collisional (VC) model. 
On both systems, our formulation significantly outperforms state-of-the-art model reduction methods for chemical kinetics, and is capable of providing accurate estimates of the moments of the distribution function and of the evolution of the internal energy level population.
\par
The paper is organized as follows. 
Section \ref{sec:physics} introduces the microscopic master equations that describe gas mixtures undergoing internal excitation, dissociation, or ionization within a 0D isothermal chemical reactor. Section \ref{sec:rom} provides a detailed description of the proposed model reduction approach. Section \ref{sec:conv_rom} presents a brief overview of existing model reduction techniques for nonequilibrium chemical kinetics, and section \ref{sec:num_exp} outlines the numerical results of the ROMs applied to the RVC and VC thermochemical models. 
Finally, section \ref{sec:conclusions} offers concluding remarks and explores potential avenues for future research.

%% file: sections/2_physics.tex
\section{Physical modeling}\label{sec:physics}
We wish to investigate the behavior of a general mixture under sudden heating in an ideal chemical reactor. For this purpose, we make the following assumptions:
\begin{enumerate}[i.]
    \item The 0D reactor is plunged into a thermal bath maintained at constant temperature $T$.
    \item The translational energy mode of the atoms and molecules is assumed to follow a Maxwell-Boltzmann distribution at the temperature $T$ of the thermal bath.
    \item At the beginning of the numerical experiment, the population of the rovibrational energy levels is assumed to follow a Boltzmann distribution at the initial equilibrium temperature $T_0$.
    \item The volume of the chemical reactor is kept constant during the experiment, and the thermodynamic system is closed (no mass exchange with the surrounding environment).
\end{enumerate}
\par
We consider both state-to-state (StS) collisional and radiative processes involving atoms or molecules, denoted by the symbol $A$, which may undergo internal excitation, de-excitation, recombination, ionization, and/or dissociation through interactions with other particles, $B$, which could include electrons, or through photon absorption and emission, also represented by $B$. The internal excitation and de-excitation processes can encompass electronic, rotational, and vibrational transitions. Any ionization or dissociation of $A$ results in the formation of particles $C$ and $D$. It should be noted that $A$, $B$, $C$, and $D$ are generic placeholders and may represent the same type of particles. Superscripts $i$, $j$, $k$, $l$, $p$, and $q$ are used to indicate internal states of the species involved. A generic form of collisional and radiative processes can be expressed as
\begin{enumerate}[i.]
    \item internal excitation and de-excitation:
    \begin{equation}\label{eq:chem.excit}
        \ce{
            $A(i)$ + $B(j)$
            <=>[$k_{ijkl}^{\mathrm{e}}$][$k_{klij}^{\mathrm{e}}$]
            $A(k)$ + $B(l)$
        }\eqspace,
    \end{equation}
    \item ionization, dissociation, and recombination:
    \begin{equation}\label{eq:chem.diss}
        \ce{
            $A(i)$ + $B(j)$
            <=>[$k_{ijpql}^{\mathrm{d}}$][$k_{pqlij}^{\mathrm{r}}$]
            $C(p)$ + $D(q)$ + $B(l)$
        }\eqspace.
    \end{equation}
\end{enumerate}
\par
We also introduce $n_s^i$, $g_s^i$, and $\epsilon_s^i$, representing the population (or number density), degeneracy, and energy level of state $i\in\mathcal{I}_s$ for species $s\in\mathcal{S}$, respectively. Here, $\mathcal{I}_s$ represents the set of energy levels for species $s$, and $\mathcal{S}$ denotes the set of chemical species in the mixture. All energy levels $\epsilon_s^i$ are referenced to a common zero-point energy. Therefore, for each species in the gas mixture, the dissociation or ionization energy is accounted for within $\epsilon_s^i$. Each energy level is treated as an individual species, and their production rates can be computed using the zeroth-order reaction rate theory, leading to the formulation of the master equations. For instance, the governing equation for the population density of state $i$ for species $A$ can be expressed as
\begin{equation}\label{eq:master_eq}
\frac{d n_A^i}{d t}  = \sum_{j,k,l} \left(-k_{ijkl}^{\mathrm{e}} n_A^i n_B^j+k_{klij}^{\mathrm{e}} n_A^k n_B^l\right)
+ \sum_{j,l,p,q} \left(-k_{ijpql}^{\mathrm{d}} n_A^i n_B^j+k_{pqlij}^{\mathrm{r}} n_C^p n_D^q n_B^l\right)\eqspace.
\end{equation}
Here, $k_{ijkl}^{\mathrm{e}}$ and $k_{klij}^{\mathrm{e}}$ represent the Maxwellian-distribution-based state-to-state excitation and de-excitation rate coefficients for process \eqref{eq:chem.excit}, while $k_{ijpql}^{\mathrm{d}}$ and $k_{pqlij}^{\mathrm{r}}$ represent the ionization/dissociation and recombination rate coefficients for process \eqref{eq:chem.diss}. These microscopic state-to-state rate coefficients obey the principles of detailed balance (or microreversibility) and depend on the translational temperature $T$. The forward (exothermic) rate coefficients were calculated using the quasiclassical trajectory (QCT) method~\cite{Jaffe_AIAA_2015}. Equivalent master equations can be written for the internal states of species $B$, $C$, and $D$ using a similar form to equation \eqref{eq:master_eq}. Further details on notation and specific equations will be provided in section \ref{sec:num_exp} for each thermochemical system investigated.

\subsection{State-space representation}
For the scope of this work, it is convenient to express the governing equations~\eqref{eq:master_eq} in compact state-space form as follows
\begin{equation} \label{eq:fom}
    \frac{d}{dt}\vn(t) = \vh(\vn(t);T)\eqspace,
    \quad \vn(0) = \vn_0\left(\vzeta\right)\eqspace,
\end{equation}
where the state vector $\vn(t)\in\mathbb{R}^N$ represents the number densities associated with each state $i\in\mathcal{I}_s$ for all species $s\in\mathcal{S}$, so that $N=\sum_{s\in\mathcal{S}}\left|\mathcal{I}_s\right|$ and $\left|\mathcal{I}_s\right|$ denotes the cardinality of the set $\mathcal{I}_s$. 
The initial condition $\vn(0)\in\mathbb{R}^N$ is defined by the function $\vn_0:\mathbb{R}^p\rightarrow\mathbb{R}^N$, which takes a set of parameters $\vzeta\in\mathbb{R}^p$ as input. These parameters include the density $\rho$, initial equilibrium temperature $T_0$, and the initial species mass fractions, $w_{s_0}$, for each $s\in\mathcal{S}$.
For isothermal chemical reactors where $dT/dt = d\rho/dt = 0$, the dynamics in \eqref{eq:fom} admit thermochemical equilibrium solutions $\overline\vn(\rho,T)$ that are functions of density and temperature.
To emphasize the importance of these two parameters and to adopt a more standard dynamical-systems notation, we pass the parametric density dependence of the solution $\vn(t)$ from the initial condition to the dynamics.
Thus, using the linear change of coordinates
\begin{equation}\label{eq:n_to_w}
    \vn(t)=\rho\,\text{diag}(\vm)^{-1}\vw(t)\eqspace,
\end{equation}
where $\vw(t)\in\mathbb{R}^N$ is the vector of mass fractions and $\vm\in\mathbb{R}^N$ is the (constant) vector of masses, equation~\eqref{eq:fom} becomes
\begin{equation} \label{eq:fom.param}
    \frac{d}{dt}\vw(t) = \vf(\vw(t);\vtheta)\eqspace,
    \quad \vw(0) = \vw_0\left(\vmu\right)\eqspace,
\end{equation}
where $\vtheta=\left(T,\rho\right)$ and $\vmu$ contains the initial equilibrium temperature $T_0$ and the initial species mass fractions.
It is also useful to define a vector of outputs $\vy$
\begin{equation} \label{eq:fom.output}
    \vy(t) = \mC \vw(t)\eqspace, 
\end{equation}
which, in general, contains physical quantities that we are interested in. 
In the present work, we let $\vy(t)\in\mathbb{R}^{(m+1)\left|\mathcal{S}\right|}$ represent the moments of orders $0$ through $m$ of the distribution function for each species in the mixture. 
More specifically, the $j$-th moment associated with species $s$ is defined through the output matrix~$\mC$ as follows
\begin{equation}
\label{eq:output_y}
    \vy_{s,j} = \left(\mC \vw\right)_{s,j} = \frac{1}{j!}\sum_{i\in\mathcal{I}_s} \frac{1}{M_s}\left(\epsilon_s^i\right)^j w_s^i \eqspace ,
\end{equation}
where $M_s$ is the species molar mass and the species energy levels $\epsilon_s^i$ have units of eV. 

%% file: sections/3_roms.tex
\section{Projection-based model reduction}\label{sec:rom}
The numerical solution of the nonlinear governing equation \eqref{eq:fom.param} is often computationally prohibitive due to the large number of degrees of freedom (i.e., the molecules' and atoms' energy levels), and to the presence of extremely fast time scales that impose severe numerical restrictions on temporal integration schemes. 
These issues can be mitigated by computing reduced-order models that can be integrated in time at a fraction of the computational cost of the full-order model \eqref{eq:fom.param}.
In this work, these models are computed using a Petrov-Galerkin procedure, where the dynamics are (obliquely) projected onto a low-dimensional subspace.

\subsection{Petrov-Galerkin model reduction}\label{sec:rom:pg}
In (linear) Petrov-Galerkin model reduction, the state $\vw(t) \in\mathbb{R}^N$ is constrained to a $r$-dimensional subspace of $\mathbb{R}^N$ defined by the range of the rank-$r$ linear projection operator $\mathbb{P}$.
That is, $\hat\vw(t) = \mathbb{P}\vw(t) \in \mathbb{R}^N$.
Injecting this ansatz into \eqref{eq:fom.param}, it is easy to see that the dynamics of $\hat\vw$ are governed by
\begin{equation}
\label{eq:pg_rom}
    \begin{aligned}
        \frac{d}{dt}\hat\vw(t) &= \mathbb{P}\vf(\mathbb{P}\hat\vw(t);\vtheta)\eqspace,
        \quad \hat\vw(0) = \mathbb{P}\vw_0(\vmu)\eqspace, \\
        \hat\vy(t) &= \mC \mathbb{P}\hat\vw(t)\eqspace,
    \end{aligned}
\end{equation}
Equation \eqref{eq:pg_rom} is known as a Petrov-Galerkin model of \eqref{eq:fom.param}.
While the state $\hat\vw(t)$ is $N$-dimensional (i.e., the same dimension as the original state $\vw(t)$), an equivalent reduced-order representation of the dynamics in \eqref{eq:pg_rom} can be written in terms of an $r$-dimensional vector of coefficients $\hat\vz(t)$.
In particular, since any rank-$r$ linear projector $\mathbb{P}$ can be written as $\mathbb{P} = \phifb\psib^\intercal$, where $\phib$ and $\psib$ are rank-$r$ matrices of size $N\times r$ matrices, defining $\hat\vz(t) = \psib^\intercal \hat{\vw}(t)$ and left-multiplying the first equation in \eqref{eq:pg_rom} by $\psib^\intercal$ we obtain
\begin{equation}
\label{eq:rom}
    \begin{aligned}
        \frac{d}{dt}\hat\vz(t) &= \psib^\intercal\vf(\phifb \hat\vz(t);\vtheta)\eqspace,
        \quad \hat\vz(0) = \psib^\intercal\vw_0(\vmu)\eqspace,\\
        \hat\vy(t) &= \mC \phifb\hat\vz(t)\eqspace,
    \end{aligned}
\end{equation}
which is the Petrov-Galerkin reduced-order model that we sought.
The accuracy of this model depends exclusively on the choice of trial and test subspaces spanned by $\phib$ and $\psib$, respectively, so it is essential to choose these subspaces appropriately.
In this work, we define these subspaces using the recently introduced CoBRAS formulation \cite{Otto2023ModelCovariance}, which we describe in the next subsection.

\subsection{Covariance Balancing Reduction using Adjoint Snapshots (CoBRAS)}\label{sec:rom:cobras}
Let us consider the following map
\begin{equation}
    \label{eq:F}
    F: \mathbb{R}^N \to \mathbb{R}^q: \vw(t_0;\vtheta,\vmu)\mapsto \left(\vy(t_0;\vtheta,\vmu),\vy(t_1;\vtheta,\vmu),\ldots,\vy(t_{L-1};\vtheta,\vmu)\right)\eqspace,
\end{equation}
which sends a state $\vw(t_0;\vtheta,\vmu)$ along a trajectory of \eqref{eq:fom.param} to a sequence of $L$ measurements $\vy(t_k;\vtheta,\vmu)$ collected at times $t_k \geq t_0$. (Notice that $t_0$ is not necessarily equal to $0$.)
A similar map may be defined for the Petrov-Galerkin dynamics \eqref{eq:pg_rom}
\begin{equation}
    \label{eq:Fhat}
    \hat{F}: \mathbb{R}^N \to \mathbb{R}^q: \mathbb{P}\vw(t_0;\vtheta,\vmu)\mapsto \left(\hat\vy(t_0;\vtheta,\vmu),\hat\vy(t_1;\vtheta,\vmu),\ldots,\hat\vy(t_{L-1};\vtheta,\vmu)\right)\eqspace.
\end{equation}
Within CoBRAS, we measure the accuracy of the Petrov-Galerkin model in the mean square sense, and we therefore seek the projection operator $\mathbb{P}$ that minimizes the mean square error $\mathbb{E}\left[F(\vw) - \hat{F}\left(\mathbb{P}\vw\right)\right]$.
Direct optimization of this cost function using, e.g., gradient descent, would be computationally expensive so we proceed by leveraging a useful result in \cite{Zahm2020Gradient-BasedFunctions}.
In particular, under the assumption that the states $\vw(t_0;\vtheta,\vmu)$ are normally distributed with covariance $W_s$, Proposition 2.5 in \cite{Zahm2020Gradient-BasedFunctions} shows that the mean square approximation error is upper-bounded by 
\begin{equation}
\label{eq:bound}
    \text{trace}\left(W_g \left(I - \mathbb{P}\right)W_s\left(I - \mathbb{P}\right)^\intercal\right)\eqspace,
\end{equation}
where $W_g = \mathbb{E}\left[\nabla F \nabla F^\intercal\right]$ is the gradient covariance and $\nabla F = D_{\vw} F(\vw)^\intercal \in \mathbb{R}^{N\times L\, \mathrm{dim}(\vy)}$.
Minimizing equation~\eqref{eq:bound} is a much simpler task, and Theorem 2.3 in~\cite{Otto2023ModelCovariance} provides a closed-form expression for the optimal~$\mathbb{P}$.
Specifically, given the following factorizations $W_s = X X^\intercal$ and $W_g = Y Y^\intercal$ of the state and gradient covariances, the optimal rank-$r$ projector may be expressed as $\mathbb{P} = \phib \psib^\intercal$ with 
\begin{equation}
    \label{eq:phi_psi}
    \phib = X V_r \Sigma_r^{-1/2}\eqspace,
    \quad \psib = Y U_r \Sigma_r^{-1/2}\eqspace,
\end{equation}
where $U_r$, $V_r$ and $\Sigma_r$ are the leading $r$ left and right singular vectors and singular values of the product $Y^\intercal X$.
Notice that $\psib^\intercal \phib$ is equal to the identity matrix by construction, so that $\mathbb{P}$ is indeed a projection.
In the next subsection, we describe an efficient numerical procedure tailored specifically for the dynamical system in~\eqref{eq:fom.param} to compute (an approximation of) the state and gradient covariance matrices $W_s$ and $W_g$.


\subsection{Computing the covariance matrices}\label{sec:rom:covmat}

In this section, we present a computationally-efficient approach to estimate the covariance factors $X$ and $Y$ that are needed to compute the CoBRAS projection \eqref{eq:phi_psi}.
The state $\vw$ along trajectories of \eqref{eq:fom.param} is a function of the temporal variable $t_0$ and of the physical parameters $\vtheta$ and $\vmu$.
Thus, if we draw $t_0$, $\vtheta$ and $\vmu$ from continuous mutually-independent distributions with probability density functions $f_{\mathrm{T}_0}$, $f_{\mTheta}$, and $f_{\mMu}$, the state covariance is defined as
\begin{equation}
\label{eq:state_cov}
    W_s \coloneqq \mathbb{E}\left[\vw\vw^\intercal\right] = \int \vw(t_0;\vtheta,\vmu)\,\vw(t_0;\vtheta,\vmu)^\intercal\, f_{\mathrm{T}_0}(t_0)\, f_{\mTheta}(\vtheta)\, f_{\mMu}(\vmu)\, dt_0 \, d\vtheta \, d\vmu \approx X X^\intercal \eqspace,
\end{equation}
where $X$ is a numerical-quadrature factor that is defined as in equations (20) and (21) of \cite{Rowley2005model}.
The computational burden associated with the numerical approximation of \eqref{eq:state_cov} lies in computing $\vw(t_0;\vtheta,\vmu)$ along trajectories of \eqref{eq:fom.param}, which requires integrating the numerically-stiff governing equations \eqref{eq:fom.param} from time $0$ to time $t_0$.
Similar considerations hold for the estimation of the gradient covariance matrix, which may be written as
\begin{equation}
\label{eq:grad_cov}
    W_g \coloneqq \mathbb{E}\left[\nabla F\nabla F^\intercal\right] = \int \nabla F(\vw(t_0;\vtheta,\vmu)) \, \nabla F(\vw(t_0;\vtheta,\vmu))^\intercal \, f_{\mathrm{T}_0}(t_0)\, f_{\mTheta}(\vtheta)\, f_{\mMu}(\vmu)\, dt_0 \, d\vtheta \, d\vmu \approx Y Y^\intercal \eqspace,
\end{equation}
where $Y$ is a numerical-quadrature factor defined similarly to $X$ in \eqref{eq:state_cov}, and a closed-form expression for~$\nabla F$ is provided in the proposition below.

\begin{proposition}
\label{prop:adjoint}    
    Let $\vg^{(k)}_j \coloneqq \left(D_{\vw} y_j(t_k;\vtheta,\vmu)\right)^\intercal$ denote the $j$-th column of the $k$-th block of $\nabla F \in \mathbb{R}^{N\times L\, \mathrm{dim}(\vy)}$, with $j \in \{1,2,\ldots, \mathrm{dim}(\vy)\}$ and $k\in \{0,2,\ldots, L-1\}$.
    Then $\vg^{(k)}_j = \vxi(t_0)$, where $\vxi(t)$ satisfies the (backward-in-time) adjoint equation
    \begin{equation}
        \label{eq:adjoint}
        -\frac{d}{dt}\vxi(t) = D\vf(\vw(t;\vtheta,\vmu))^\intercal\vxi(t)\eqspace,
        \quad \vxi(t_k) = \mC^\intercal \ve_j\eqspace,
        \quad t\in [t_0,t_k]\eqspace,
    \end{equation}
    and $\ve_j$ is the $j$-th unit vector of the standard basis of $\mathbb{R}^{\mathrm{dim}(\vy)}$.
\end{proposition}

\begin{proof}
    The proof is given in \ref{app:proof_of_prop}.
\end{proof}
The adjoint equation \eqref{eq:adjoint} shares the same numerical stiffness as the forward nonlinear dynamics \eqref{eq:fom.param}, and this can make its numerical integration expensive.
In order to reduce the overall computational cost required to estimate the state and gradient covariance matrices, we propose to work with the \textit{linearized} dynamics of \eqref{eq:fom.param}.
This will let us leverage the exact solution of linear ordinary differential equations, thereby bypassing the need for numerical integration of the high-dimensional stiff governing equations.
Additionally, we shall see momentarily that the (approximate) gradient $\nabla F$ will become a function of $\vtheta$ only, so that the (approximate) definition of $W_g$ in \eqref{eq:grad_cov} will no longer require integrating over $t_0$ and $\vmu$.
\par
As previously mentioned, for a given value of $\vtheta$, the dynamics in \eqref{eq:fom.param} exhibit a linearly stable thermochemical equilibrium solution, which we shall denote $\overline{\vw}(\vtheta)$.
Then, the linear dynamics of about $\overline\vw(\vtheta)$ are governed by
\begin{equation}\label{eq:fom.param.lin}
    \frac{d}{dt}\tvw_{\vtheta}(t) = \underbrace{D \vf(\overline{\vw}(\vtheta))}_{\coloneqq \mA_{\vtheta}} \tvw_{\vtheta}(t)\eqspace,
    \quad \tvw_{\vmu,\vtheta}(0) = \vw_0(\vmu) - \overline{\vw}(\vtheta)\eqspace,
\end{equation}
where $\mA_{\vtheta}$ is the state Jacobian evaluated about the equilibrium solution.
Using the variation of constants formula, we can then approximate $\vw(t_0;\vtheta,\vmu)$ in \eqref{eq:fom.param} as follows
\begin{equation}
    \label{eq:vw_exp}
    \vw(t_0;\vtheta,\vmu) \approx \overline{\vw}(\vtheta) + e^{\mA_{\vtheta}t_0} \tvw_{\vmu,\vtheta}(0)\eqspace.
\end{equation}
Similarly, we can approximate $\vg^{(k)}_j$ in Proposition \ref{prop:adjoint} using the time-invariant adjoint equation
\begin{equation}
    \label{eq:adjoint_lin}
    -\frac{d}{dt}\vxi(t) = \mA_{\vtheta}^\intercal\vxi(t)\eqspace,
    \quad \vxi(t_k) = \mC^\intercal \ve_j\eqspace,
    \quad t\in [t_0,t_k]\eqspace,
\end{equation}
where $\mA_{\vtheta}$ is precisely the zeroth-order approximation of $D\vf(\vw(t;\vtheta,\vmu))$.
Once again, using the variation of constants formula, we have
\begin{equation}
\label{eq:vg_exp}
    \vg^{(k)}_j = e^{\mA_{\vtheta}^\intercal (t_k - t_0)}\mC^\intercal \ve_j\eqspace.
\end{equation}
Here, we see that $\vg^{(k)}_j$ is a parametric function of $\vtheta$ only.
More specifically, we remark that $\vg^{(k)}_j$ is not a function of $t_0$, but rather of $t_k - t_0$, and this is consistent with the time-invariant nature of equation \eqref{eq:adjoint_lin}.
It follows that the (approximate) gradient covariance matrix in \eqref{eq:grad_cov} can be written as
\begin{equation}
\label{eq:grad_cov_approx}
    W_g \approx \int \nabla F(\vtheta) \, \nabla F(\vtheta)^\intercal \, f_{\mTheta}(\vtheta)\, d\vtheta \eqspace.
\end{equation}
In equations \eqref{eq:vw_exp} and \eqref{eq:vg_exp} the matrix exponential can be computed efficiently using the eigendecomposition of $\mA_{\vtheta}$.
The overall procedure to obtain the projection factors $\mPhi$ and $\mPsi$ is illustrated in Algorithm \ref{alg:proj}.
\begin{docremark}
    In this paper, we consider systems that exhibit one single equilibrium for a given value of $\vtheta$. 
    In systems where multiple equilibria coexist, the same approximations illustrated in equations \eqref{eq:vw_exp} and \eqref{eq:vg_exp} hold around any one of the coexisting steady states and the user may choose freely which steady state to use for the linearization.
\end{docremark}
\begin{algorithm}[!htb]
\caption{Computing the factor $\mPhi$ and $\mPsi$ for the linear projection operator $\mathbb{P} = \mPhi \mPsi^\intercal$}\label{alg:proj}
\begin{algorithmic}[1]
\Require \begin{varwidth}[t]{0.94\linewidth}
Set of quadrature points for temperatures and densities ($\mathcal{P}$), initial condition parameters ($\mathcal{M}$), and initial times ($\mathcal{\mathcal{T}}_0$), along with their associated quadrature weights, to approximate the integrals in equations~\eqref{eq:state_cov} and~\eqref{eq:grad_cov}.
\end{varwidth}
\vskip 5pt
\State Initialize dynamic set $\mathcal{W}_0$.
\For{$\vmu \in \mathcal{M}$}
    \State Compute the parametrized initial condition $\vw_0(\vmu)$.
    \State Add $\vw_0$ to the set $\mathcal{W}_0$.
\EndFor
\State Initialize dynamic arrays $X$ and $Y$.
\For{$\vtheta \in \mathcal{P}$}
    \State Compute the equilibrium steady-state $\overline{\vw}(\vtheta)$.
    \State Construct matrix $\mA_{\vtheta}$ and compute its eigendecomposition $\mA_{\vtheta} = \mV \mLam \mV^{-1}$.
    \State \emph{Computing state covariance}
    \For{$\vw_0 \in \mathcal{W}_0$}
        \For{$t_0 \in \mathcal{\mathcal{T}}_0$}
            \State Compute $\vw(t_0; \vtheta, \vmu)$ using equation \eqref{eq:vw_exp} and $e^{\mA_{\vtheta}t_0} = \mV e^{\mLam t_0}\mV^{-1}$.
            \State Append $\sqrt{\delta_s} \vw(t_0; \vtheta, \vmu)$ to $X$, where
            $$\delta_s = f_{\mathrm{T}_0}(t_0) f_{\mTheta}(\vtheta) f_{\mMu}(\vmu) w_{t_0} w_{\vtheta} w_{\vmu}\eqspace,$$
            \State and $w_{t_0}$, $w_{\vtheta}$, and $w_{\vmu}$ are the quadrature weights.
        \EndFor
    \EndFor
    \State \emph{Computing gradient covariance}
    \For{$j \in \{1,2,\ldots, \mathrm{dim}(\vy)\}$}
        \For{$k \in \{0,2,\ldots, L-1\}$}
            \State Compute $\vg_j(t_k;\vtheta)$ using equation \eqref{eq:vg_exp} and $e^{\mA^\intercal_{\vtheta}(t_k-t_0)} = \mV^{-\intercal} e^{\mLam (t_k-t_0)}\mV^\intercal$.
            \State Append $\sqrt{\delta_g} \vg_j(t_k; \vtheta)$ to $Y$, where
            $\delta_g = f_{\mTheta}(\vtheta) w_{\vtheta}\eqspace$.
        \EndFor
    \EndFor
\EndFor
\State Perform SVD $Y^\intercal X=U\Sigma V^\intercal$.
\State Compute matrices $\mPhi$ and $\mPsi$ using equation \eqref{eq:phi_psi}.
\end{algorithmic}
\end{algorithm}

\section{Existing model reduction formulations}\label{sec:conv_rom}
The conventionally used model reduction techniques primarily focus on predicting macroscopic quantities, such as total species mass and internal energy, which are essential in flow measurements and simulations. These quantities can be modeled using additional conservation equations, beyond the standard hydrodynamic equations, to account for variations in chemical composition and the nonequilibrium relaxation of energy modes. In the most general formulation, this additional set of governing equations can be derived by taking successive moments of the master equations \eqref{eq:master_eq}, as follows
\begin{equation}\label{eq:rom_conv_gov_eq}
    \sum_{i\in\mathcal{I}_g\subseteq\mathcal{I}_s}\left(\epsilon_s^i\right)^m\frac{dn_s^i}{dt} = \Omega^g_{s,m}
    \eqspace,
    \quad\forall \;g\in\left\{1,\dots,N_g\right\} \eqspace.
\end{equation}
In this context, $g$ represents a pseudo-species, which refers to a particular species internal degree of freedom treated as a state variable. The parameter $m$ indicates the moment order, while $\Omega^g_{s,m}$ denotes the reactive source term of order $m$ for the pseudo-species $g$. Based on the assumptions used to define the subset $\mathcal{I}_g\subseteq\mathcal{I}_s$, two distinct simplified models can be derived.

\subsection{Coarse-grained models}
If the subset $\mathcal{I}_g$ represent a group of states, the approach is named coarse-grained (CG) modeling~\cite{Panesi_Lani_PofF_2013,Liu_JCP_2015,Munafo_EPJ_2012,Sahai_JCP_2017,Venturi_JCP_2020,Sharma_PR_2020,Macdonald_I_JCP_2018,Macdonald_II_JCP_2018}. The construction of such a model involves a two-step procedure: (i) grouping energy states into $N_g$ macroscopic bins (or groups) according to a specified strategy, and (ii) defining a bin-wise distribution function to represent the population within each group, subject to a series of moment constraints. 
Various clustering techniques have been proposed for step (i)~\cite{Magin2012Coarse-grainNitrogen,Sahai_JCP_2017,Venturi_JCP_2020,Jacobsen_JComP_2024} and in this work, we adopt the advanced spectral clustering technique proposed by Sahai et al.~\cite{Sahai_JCP_2017}. 
For step (ii), Liu et al.~\cite{Liu_JCP_2015} applied the maximum entropy principle to derive a log-linear representation of the bin-wise distribution function, yielding a thermalized local Boltzmann distribution for each bin. This distribution is expressed as follows
\begin{equation}
    \ln \frac{g_s^i}{n_s^i}=\alpha_s^g+\beta_s^g \epsilon_s^i \eqspace,
    \quad i \in \mathcal{I}_g \eqspace,
\end{equation}
where the bin-specific coefficients $\alpha_s^g$ and $\beta_s^g$ are expressed as a function of the macroscopic group constraints (i.e., number density and energy). The populations and internal energies of the different bins serve as the unknowns in the reduced-order system, leading to a total of $2\times N_g$ additional conservation equations. For further details, please refer to \cite{Liu_JCP_2015}.
\par
This model reduction technique effectively maintains positivity and guarantees positive entropy production by design. 
However, these advantages come with practical challenges due to the use of a nonlinear function of exponential nature. Specifically, coarse-grained rates must be interpolated across group energies (or temperatures), and this can increase computational costs and introduce potential errors and instabilities, as it necessitates prior knowledge of the relevant temperature ranges.

\subsection{Multi-temperature models}
In the specific case of binning one group per internal energy mode, which is a particular case of the CG approach, multi-temperature (MT) models are employed~\cite{Park_Book_1990, Park_JTHT_1993, Park_JTHT_2001}. 
This framework adds an additional conservation equation for the species total number density, along with individual conservation equations for each energy mode (vibrational, rotational, and electronic). While this method is still commonly used in the CFD community, many of these models are based on simplistic, semi-empirical assumptions. They often rely on a rigid separation of internal energy modes without rigorous derivation from fundamental kinetic equations or consideration of physical principles. In this study, we adopt a more rigorous approach, utilizing the recent quasi-steady-state (QSS) method~\cite{Panesi_JCP_2013}, which computes its kinetic database directly from state-to-state calculations.

%% file: sections/4_results.tex
\section{Numerical experiments}\label{sec:num_exp}
In this section, we evaluate the performance of the proposed CoBRAS model on two distinct thermochemical systems. 
Both of these involve a pure oxygen mixture, denoted as $\mathcal{S} = \{\text{O}, \text{O}_2\}$, with both species in their electronic ground state.
\begin{enumerate}[i.]
    \item The first system is a rovibrational collisional (RVC) model, which considers only molecule-atom collisions and has been extensively studied in prior works~\cite{Panesi_JCP_2013,Liu_JCP_2015,Sahai_JCP_2017,Venturi_JCP_2020}. This model serves as a benchmark to compare CoBRAS with traditional approaches, including coarse-graining (CG), multi-temperature models (MT), and the POD-Galerkin approach of Sutherland and Parente~\cite{Sutherland2009CombustionAnalysis, Parente2013PrincipalSensitivity}. 
    \item The second system is a vibrational collisional (VC) model, which includes both molecule-molecule and molecule-atom collisions. The resulting equations are quadratic in the O$_2$ energy level population. 
    Despite the nonlinearity, we demonstrate that the linearized dynamics around thermochemical equilibrium remain representative of the subspaces on which the nonlinear dynamics evolve, and this justifies the use of the procedure described in section \ref{sec:rom:covmat} to approximate the covariance matrices.
    Here, we compare CoBRAS against POD-Galerkin only since we have no access to validated solvers that implement the CG approach on this system.
\end{enumerate}
\par
In all (Petrov-)Galerkin models implemented in this section, we seek a reduced-order representation of the O$_2$ energy levels, while the O dynamics are accounted for exactly (see sections \ref{suppl:rvc.rom} and \ref{suppl:vc.rom} in the Supplementary Material).
All test cases were implemented in Python, utilizing PyTorch for GPU acceleration and employing the SciPy \texttt{integrate.solve\_ivp} function with the LSODA method for time integration. All the CG and MT solutions were computed using the \textsc{plato} (PLAsma in Thermodynamic nOnequilibrium) library~\cite{Munafo_AIAA_2023}.

\subsection{O$_2$-O rovibrational collisional model}\label{sec:num_exp:rvc_model}
In our first numerical experiment, we investigate the rovibrational excitation and dissociation of an O$_2$ molecule colliding with an O atom. The O$_2$ molecule has 6\,115 rovibrational energy levels, resulting in a total of $N=6\,116$ degrees of freedom for the system.
The kinetic database for the O$_2$-O system comprises two types of processes:
\begin{enumerate}[i.]
    \item collisional dissociation,
    \begin{equation}\label{eq:coll.atom.diss}
        \ce{
            O_2$(i)$ + O
            <=>[${}^{\mathrm{a}}k_{i}^{\mathrm{d}}$][${}^{\mathrm{a}}k_{i}^{\mathrm{r}}$]
            O + O + O
        }\eqspace,
    \end{equation}
    \item collisional excitation, including both inelastic (nonreactive) and exchange processes,
    \begin{equation}\label{eq:coll.atom.excit}
        \ce{
            O_2$(i)$ + O
            <=>[${}^{\mathrm{a}}k_{ij}^{\mathrm{e}}$][${}^{\mathrm{a}}k_{ji}^{\mathrm{e}}$]
            O_2$(j)$ + O
        }\eqspace.
    \end{equation}
\end{enumerate}
The RVC model here described is governed by the following set of equations
\begin{align}
    \frac{dn_i}{d t} =
        & \sum_{j \in \mathcal{I}} \left(-{}^{\mathrm{a}}k_{i j}^{\mathrm{e}} n_in_\atom
        + {}^{\mathrm{a}}k_{j i}^{\mathrm{e}} n_j n_\atom \right) \nonumber\\
        & -{}^{\mathrm{a}}k_i^{\mathrm{d}}n_in_\atom + {}^{\mathrm{a}}k_i^{\mathrm{r}}n_\atom^3 \eqspace, \label{eq:rvc.fom.ni} \\
    \frac{dn_\atom}{dt} = & - 2\sum\limits_{i\in\mathcal{I}}\frac{dn_i}{d t} \eqspace. \label{eq:rvc.fom.no}
\end{align}
For interested readers, the linearized FOM equations necessary to implement the method described in Sections \ref{sec:rom:pg} to \ref{sec:rom:covmat}, as well as the derivation of the ROM, are available in Supplementary Material sections \ref{suppl:rvc.linfom} and \ref{suppl:rvc.rom}.
\par
The bounds and distributions used to sample training and testing trajectories for the RVC model are provided in table \ref{table:rvc.mu_space}. Here, $w_{\atom_0}$ denotes the initial mass fraction of O, with $w_{\mol}=1-w_{\atom}$ always applicable. For training, we use Arrhenius-fitted rates derived from quasi-classical trajectory (QCT) calculations, evaluated at 10 different temperatures uniformly sampled within the range of $[5,15]\times 10^3$ K. This range includes the boundary temperatures, with a step size of $\Delta T=1\,111.11$ K. For testing, we select 10 randomly chosen temperatures within the same range using LHS. A total of $M=1\,000$ testing trajectories are sampled, with 100 trajectories for each selected temperature.
\begin{table}[!htb]
    \centering
    \begin{tabular}{c|cccc}
        \toprule
        & $T$ [K] & $\rho$ [kg/m$^{3}$] & $w_{\atom_0}$ & $T_0$ [K] \\
        \midrule
        Minimum & 5\,000 & $10^{-4}$ & 0.01 & 500 \\
        Maximum & 15\,000 & 1 & 1 & 5\,000 \\
        Distribution & Uniform & Log-uniform & Uniform & Log-uniform \\
        \bottomrule
    \end{tabular}
    \caption{\textit{Sampled parameter space for the RVC model}. Sampling bounds and distributions for each parameter used to generate the training and testing trajectories.}
    \label{table:rvc.mu_space}
\end{table}

\subsubsection{Linearized FOM performances}\label{sec:num_exp:rvc:lin_fom}
In this section, we assess the performance of the linearized FOM employed in constructing the Petrov-Galerkin ROM. 
In particular, we wish to demonstrate that the procedure illustrated in section \ref{sec:rom:covmat} leads to significant computational savings while still capturing the relevant subspaces for Petrov-Galerkin modeling.
The computational cost is tabulated in table \ref{table:rvc.lin_fom.costs}, where we compare the cost required to generate one (or more) trajectories of the nonlinear FOM in equation \eqref{eq:fom} and the cost of evaluating \eqref{eq:vw_exp} using the matrix exponential.
In particular, the third column of the table shows that once the eigendecomposition of $\mA_{\vtheta}$ is available, evaluating \eqref{eq:vw_exp} is four orders of magnitude faster than time-stepping \eqref{eq:fom}.
For a more realistic comparison, however, we should also account for the cost associated with computing the eigendecomposition, which scales with the cube of the matrix dimension.
This operation is indeed costly, but fortunately, it can be performed once (per equilibrium) and then held in memory (or stored) for future use.
The last column in table \ref{table:rvc.lin_fom.costs} shows that, despite the costly eigendecomposition, we still achieve a factor of 400 speed-up when generating 200 trajectories.
\begin{table}[!htb]
    \centering
    \begin{tabular}{c|c|c|c|c}
        \toprule
        Model & Operation & Device & Time [s] - 1 Run & Time [s] - 200 Runs \\
        \midrule
        FOM & Simulation & CPU - 8 Threads & $1.25 \times 10^{2}$ & $2.50 \times 10^{4}$ \\
        \arrayrulecolor{gray}\midrule
        \multirow{2}*{Linearized FOM}
        & Eigendecomposition & CPU - 8 Threads & $6.17 \times 10^{1}$ & $6.17 \times 10^{1}$ \\
        & Simulation & CPU - 8 Threads & $2.19 \times 10^{-2}$ & $4.38$ \\
        \bottomrule
    \end{tabular}
    \caption{\textit{Comparison of computational cost: FOM vs. linearized FOM for the RVC Model}. This table illustrates the performance differences between the FOM and its linearized counterpart across various numerical operations.}
    \label{table:rvc.lin_fom.costs}
\end{table}
\par
\begin{figure}[htb!]
\centering
\includegraphics[width=0.35\textwidth]{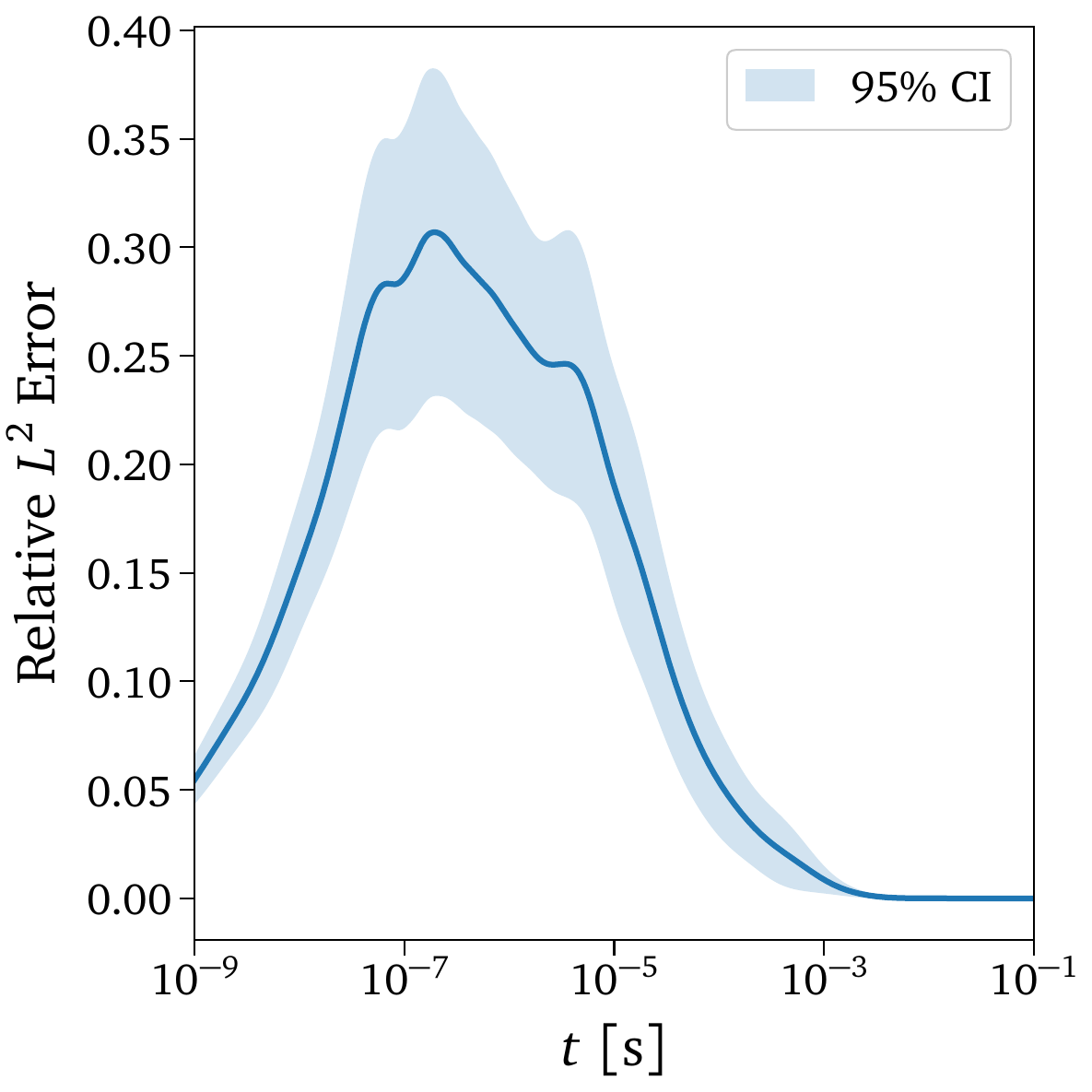}
\caption{\textit{Relative $L^2$ error of the linearized FOM for the RVC model}. This figure presents the time evolution of the mean relative $L^2$ error, along with a 95\% confidence interval derived from 1\,000 testing trajectories, using the linearized FOM.}
\label{fig:rvc_linfom_err}
\end{figure}
In order to assess the effectiveness of the linearized dynamics in identifying the appropriate subspaces for model reduction, we simply compute the mean relative $L^2$ error
\begin{equation}\label{eq:l2_err}
    \text{e}(t)=\frac{1}{M}\sum_{k=1}^M
    \frac{\| \, \vw^{(k)}(t)-\vw^{(k)}_{\text{lin}}(t) \, \| _2}{\|\,\vw^{(k)}(t)\,\|_2+\varepsilon} \eqspace,
\end{equation}
between $\vw_{\text{lin}}$ given by \eqref{eq:vw_exp} and the solution $\vw$ of \eqref{eq:fom} over $M=1\,000$ trajectories.
Here, we use $\varepsilon=10^{-7}$ to avoid division by zero.
The relative $L^2$ error is particularly relevant here because the methodology described in sections \ref{sec:rom:pg} to \ref{sec:rom:covmat} produce optimal test and trial bases in the $L^2$ sense. With a 95\% confidence interval, the maximum relative $L^2$ error is approximately 30\%. This finding suggests that the linearized equations are sufficiently accurate to identify appropriate subspaces for Petrov-Galerkin projection of the full-order nonlinear equations.

\subsubsection{Models for macroscopic quantities}
\begin{figure}[htb!]
\centering
\begin{subfigure}[htb!]{0.35\textwidth}
    \caption*{\hspace{7mm}\small CoBRAS} \vskip 5pt
    \includegraphics[width=\textwidth]{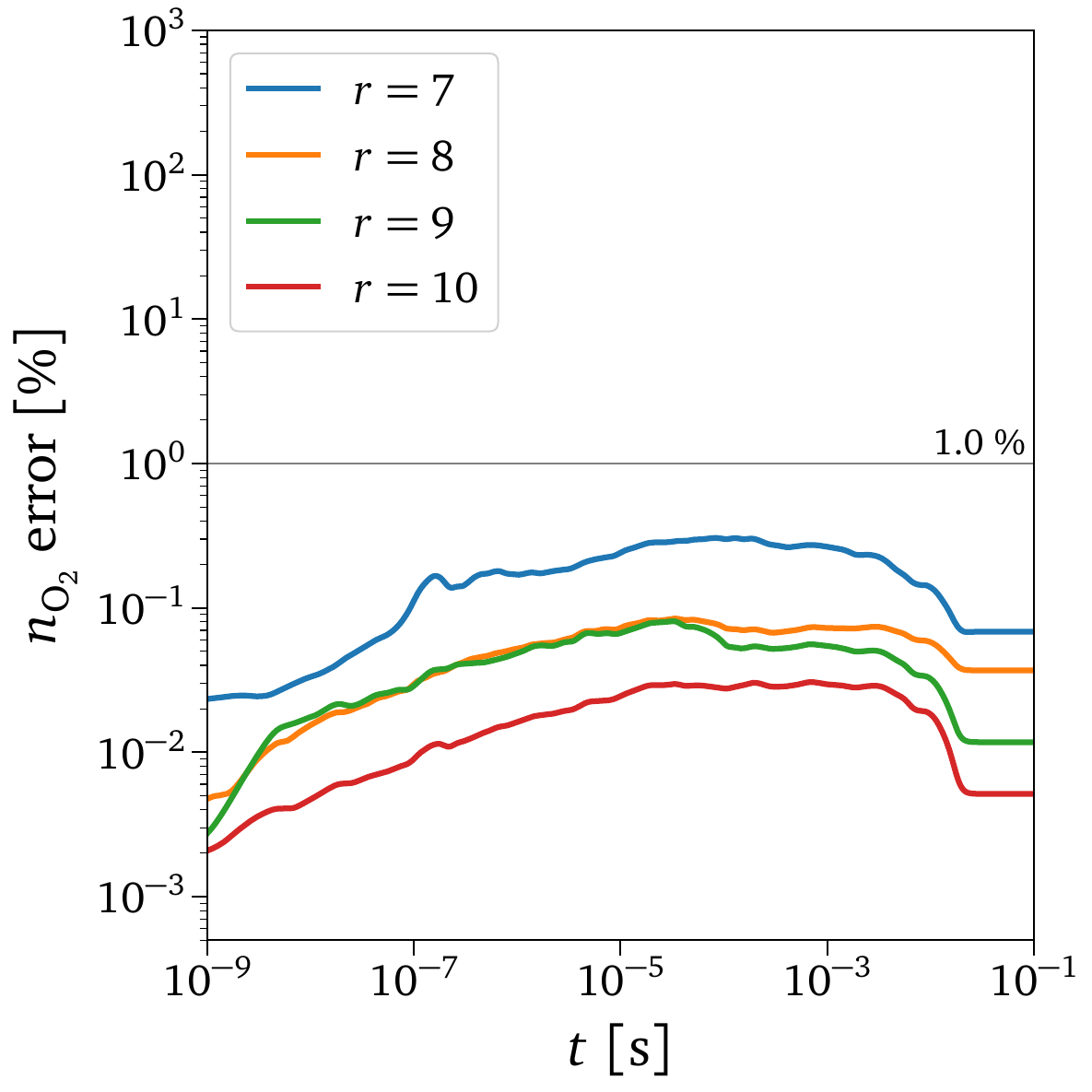}
\end{subfigure}
\quad
\begin{subfigure}[htb!]{0.35\textwidth}
    \caption*{\hspace{7mm}\small POD} \vskip 5pt
    \includegraphics[width=\textwidth]{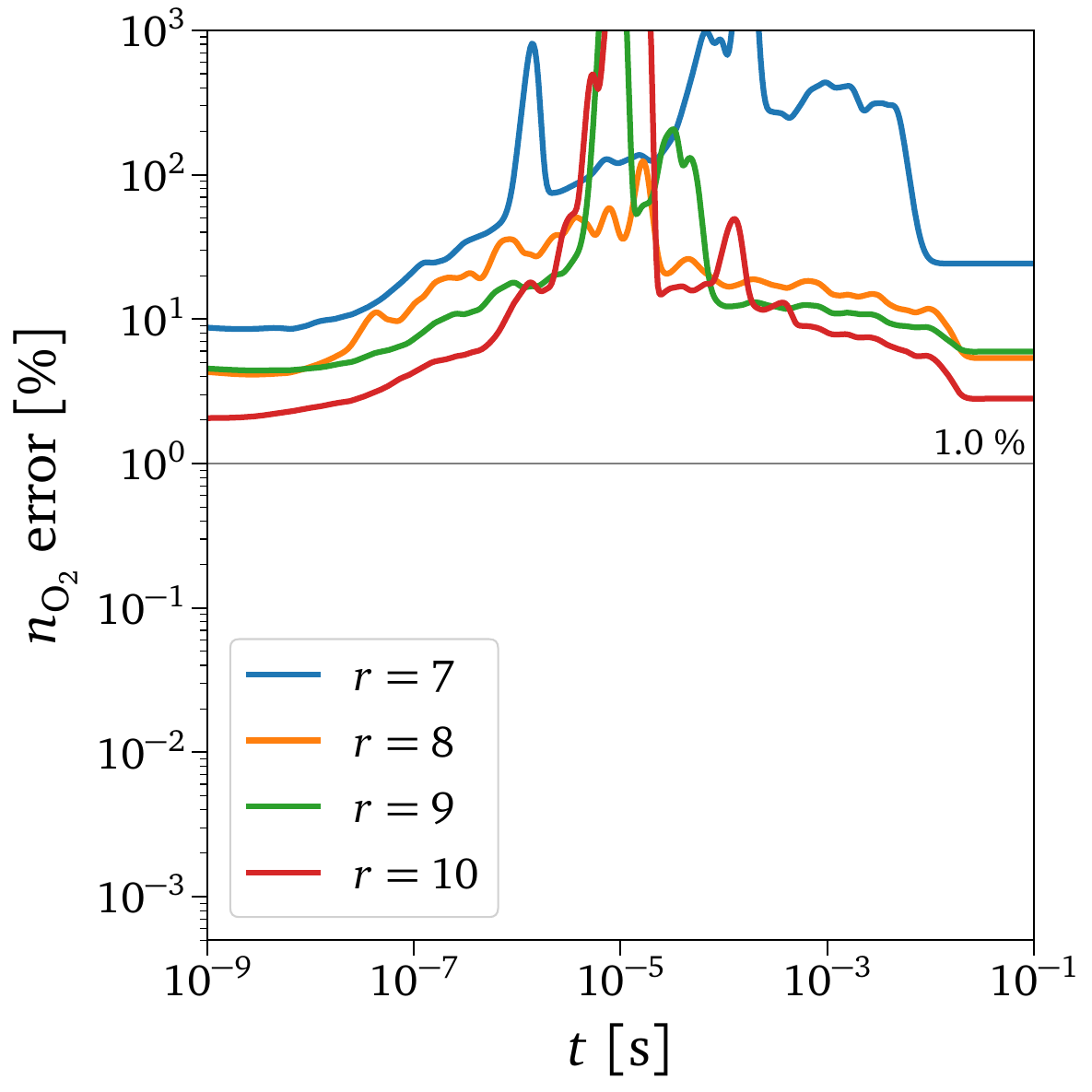}
\end{subfigure}
\\[5pt]
\begin{subfigure}[htb!]{0.35\textwidth}
    \includegraphics[width=\textwidth]{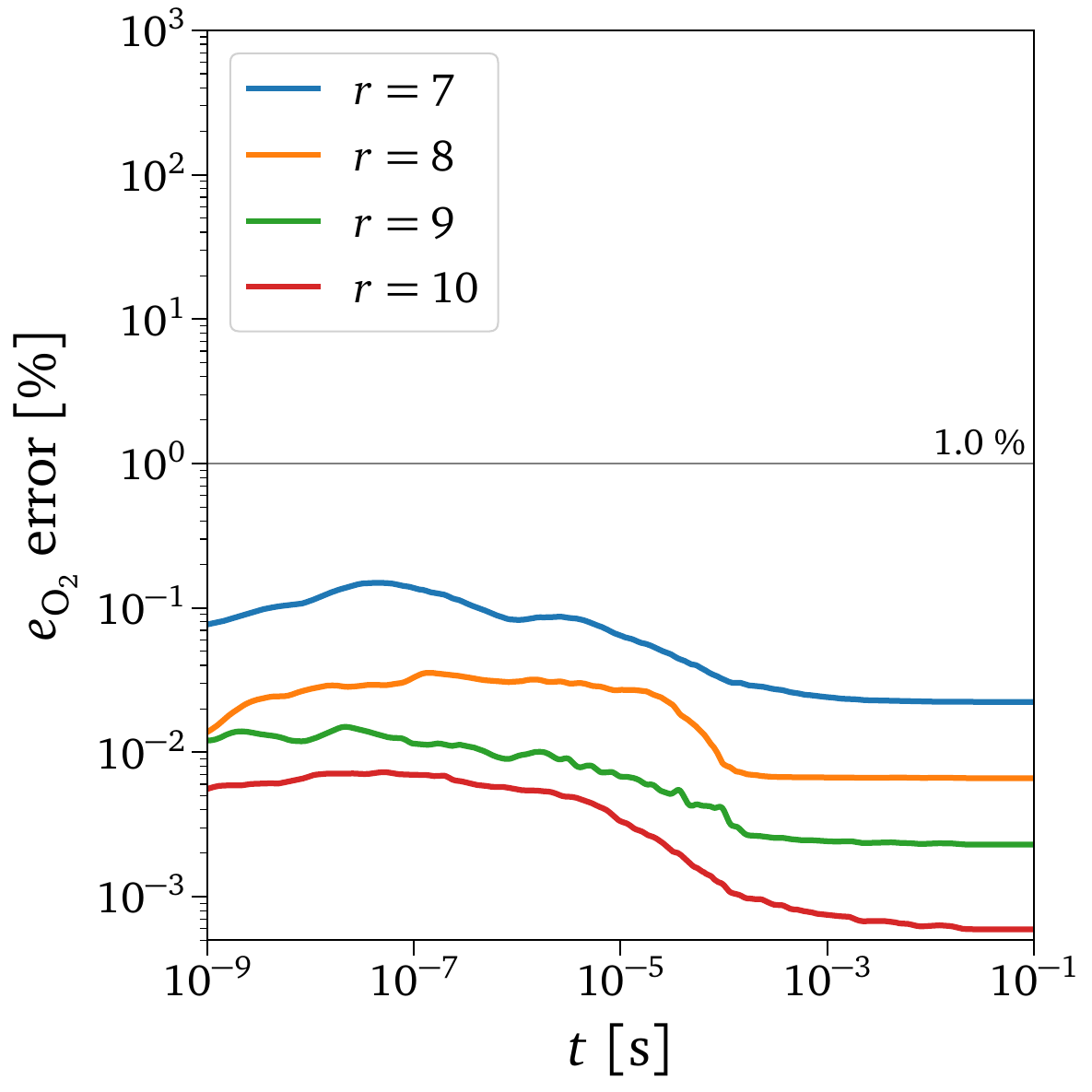}
\end{subfigure}
\quad
\begin{subfigure}[htb!]{0.35\textwidth}
    \includegraphics[width=\textwidth]{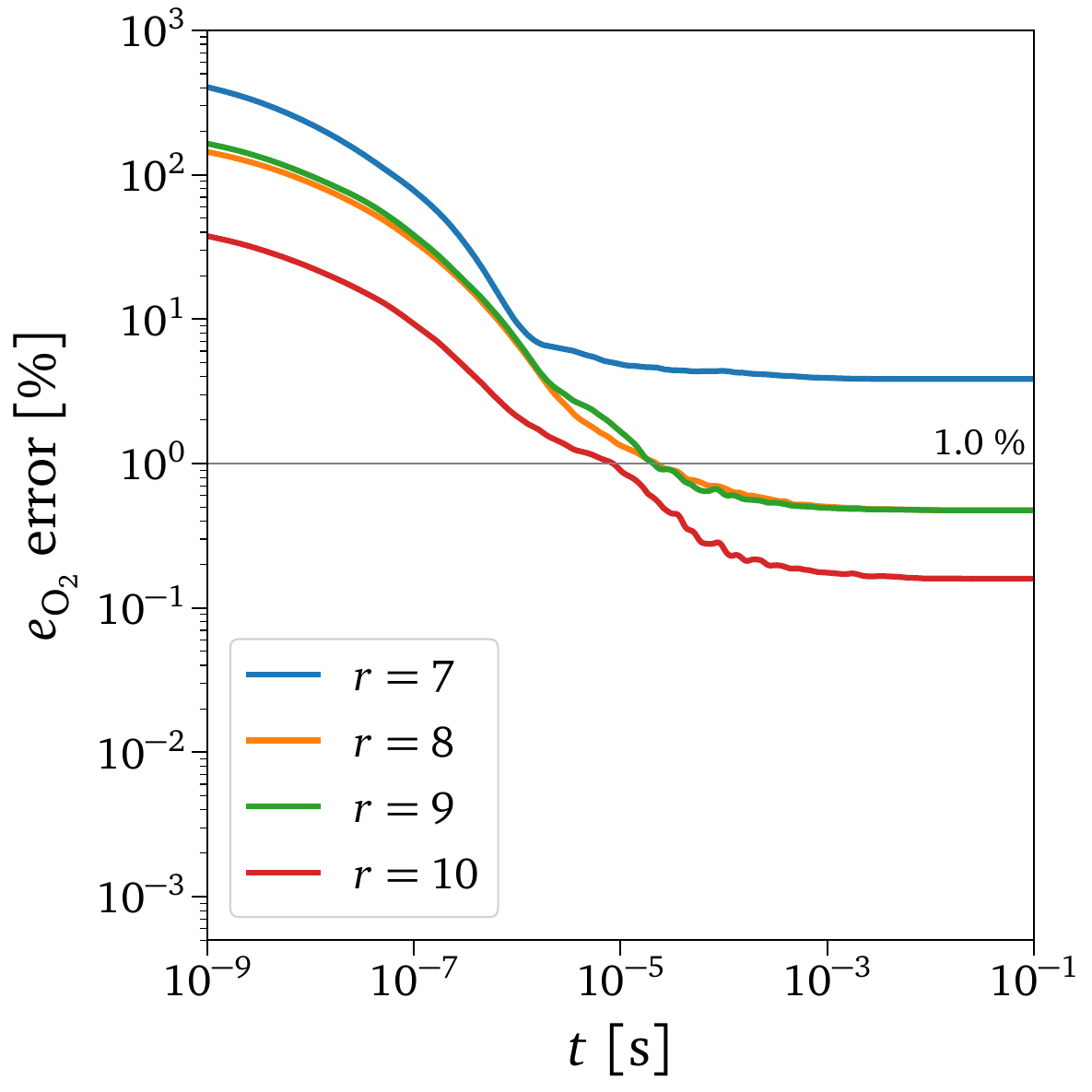}
\end{subfigure}
\caption{\textit{Mean moments error for RVC model using CoBRAS and POD ROMs}. Time evolution of the mean relative errors for the zeroth (top row) and first-order (bottom row) moments of the O$_2$ energy level population across 1\,000 testing trajectories. Results are shown for different dimensions $r$ of the CoBRAS (left panels) and POD (right panels) ROM systems.}
\label{fig:rvc_mean_moms_err}
\end{figure}

In this section, we seek models that accurately predict the time evolution of the system's macroscopic quantities such as mass and energy, and we therefore choose $\vy \in\mathbb{R}^2$ in \eqref{eq:fom.output} to contain the zeroth and first moments (i.e., the mass and energy, respectively) of the distribution.
Figure \ref{fig:rvc_mean_moms_err} illustrates the time evolution of the mean relative error between the FOM output $y_j(t)$ and the predicted output $\hat{y}_j(t)$ from CoBRAS (left panels) and POD (right panels) for the $j = 0$ (top row) and $j=1$ (bottom row) moments of O$_2$ energy level population. 
This analysis is based on $M = 1\,000$ different testing trajectories across various reduced system dimensions, $r$. The figure is essential for understanding how the model performs over time and across different reduced dimensions. 
With CoBRAS, the error remains consistently low, under 1\%, even for a small reduced dimension of $r = 7$, representing a compression of approximately 870 times. 
By contrast, the POD method yields predictions with errors consistently exceeding 1\%. 
Similar qualitative performance is observed for the time evolution of the energy (bottom row), where we see that CoBRAS maintains low errors across all times, while POD exhibits errors that are consistently above 10\%.
\par
\begin{figure}[htb!]
\centering
\begin{subfigure}[htb!]{0.35\textwidth}
    \includegraphics[width=\textwidth]{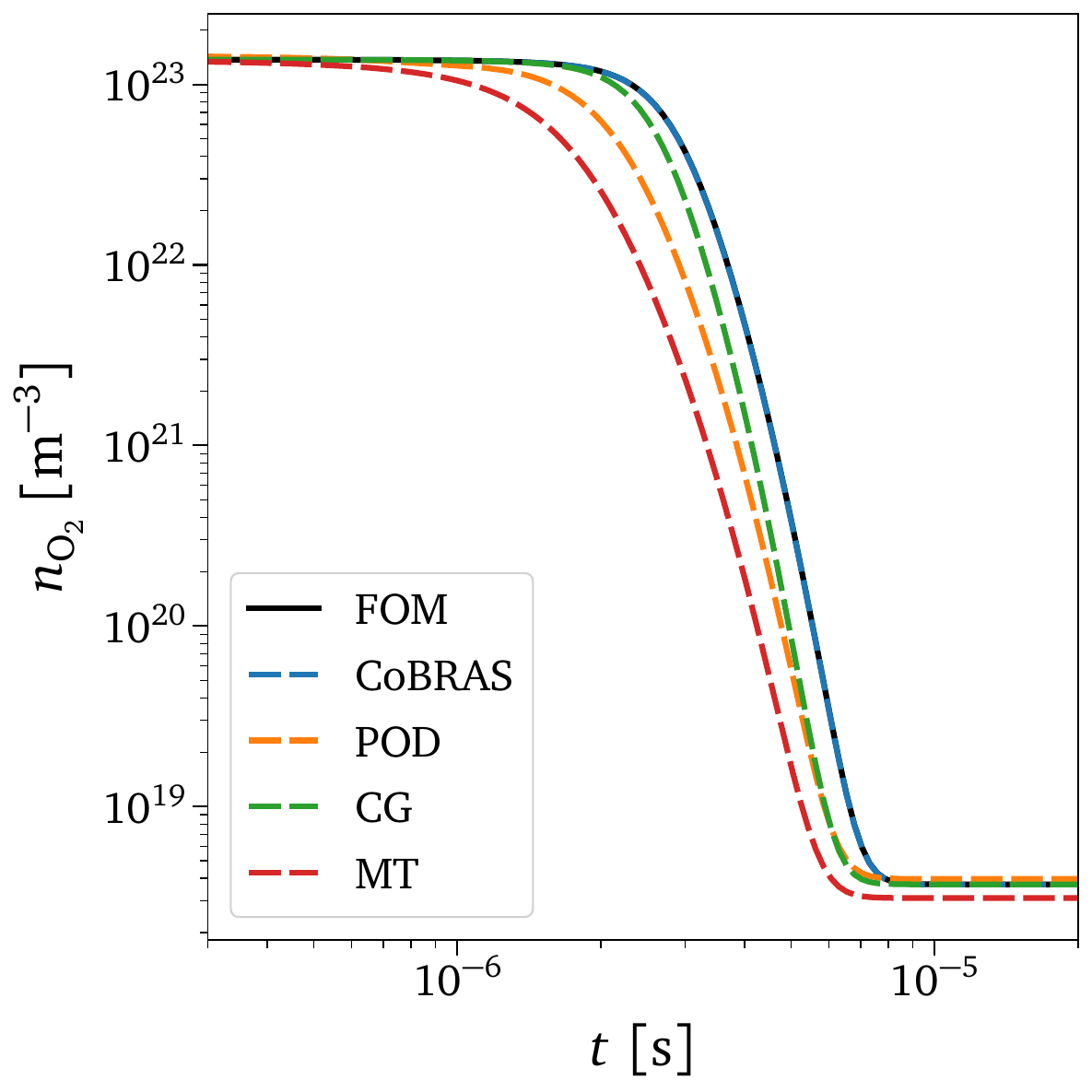}
\end{subfigure}
\quad
\begin{subfigure}[htb!]{0.35\textwidth}
    \includegraphics[width=\textwidth]{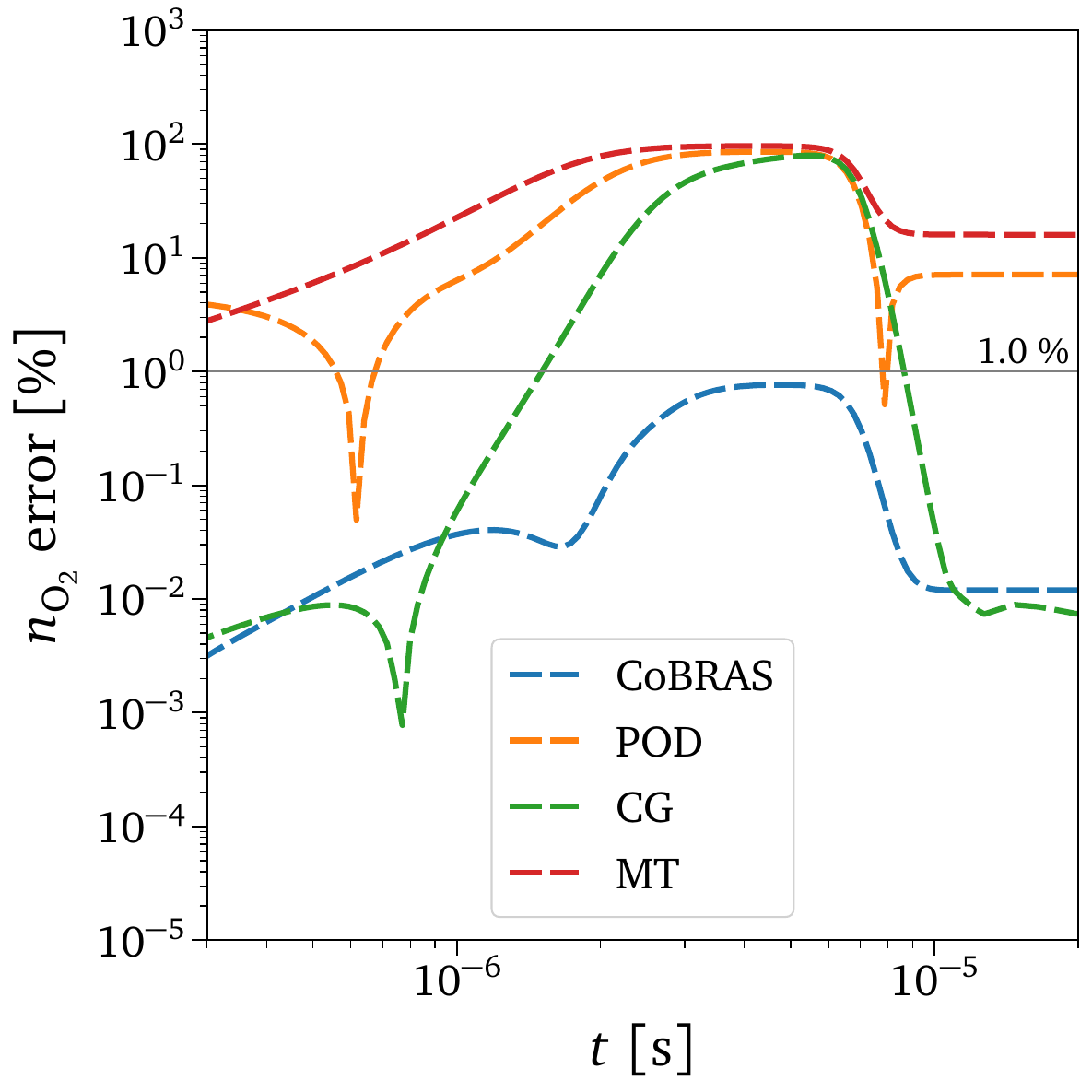}
\end{subfigure}
\\[5pt]
\begin{subfigure}[htb!]{0.35\textwidth}
    \includegraphics[width=\textwidth]{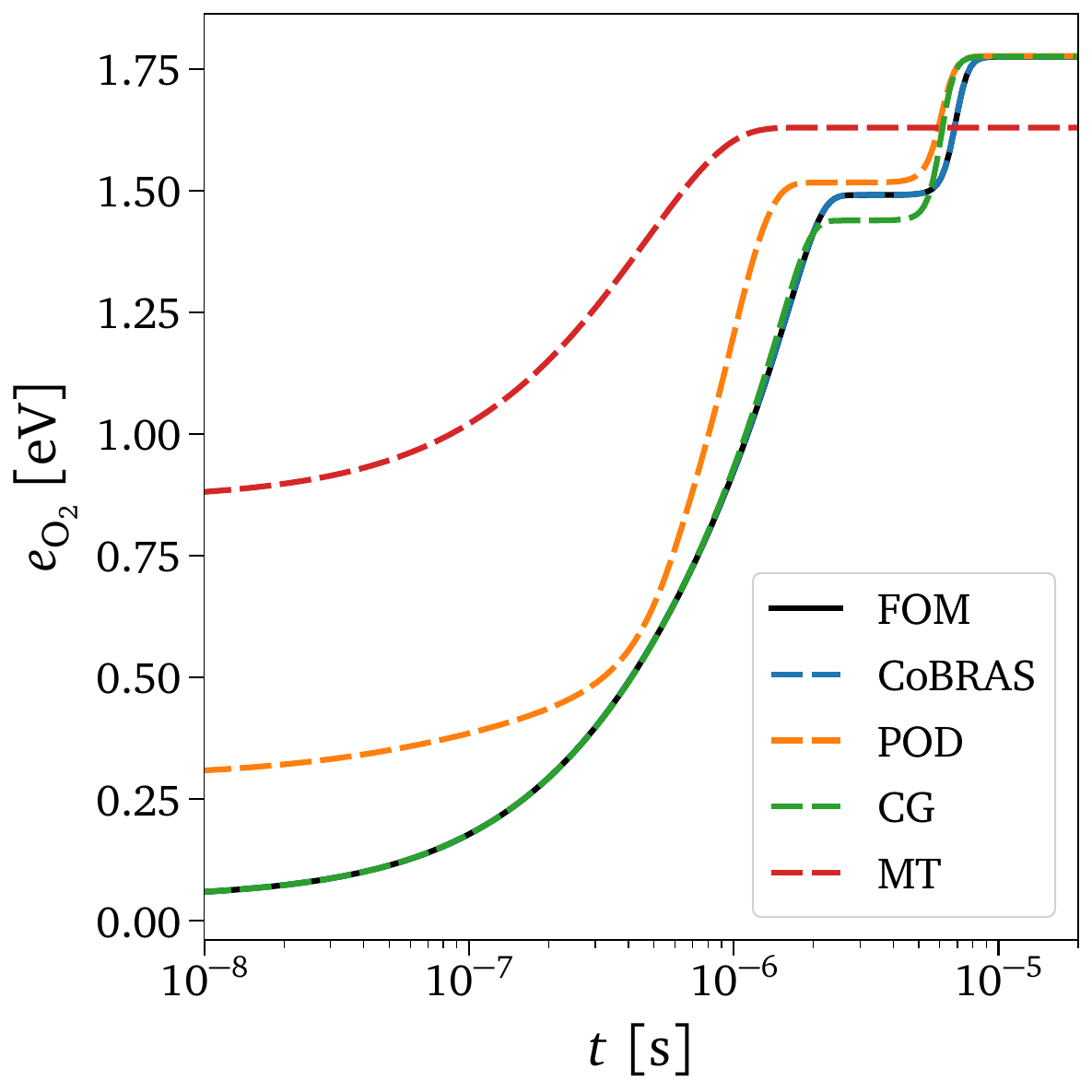}
\end{subfigure}
\quad
\begin{subfigure}[htb!]{0.35\textwidth}
    \includegraphics[width=\textwidth]{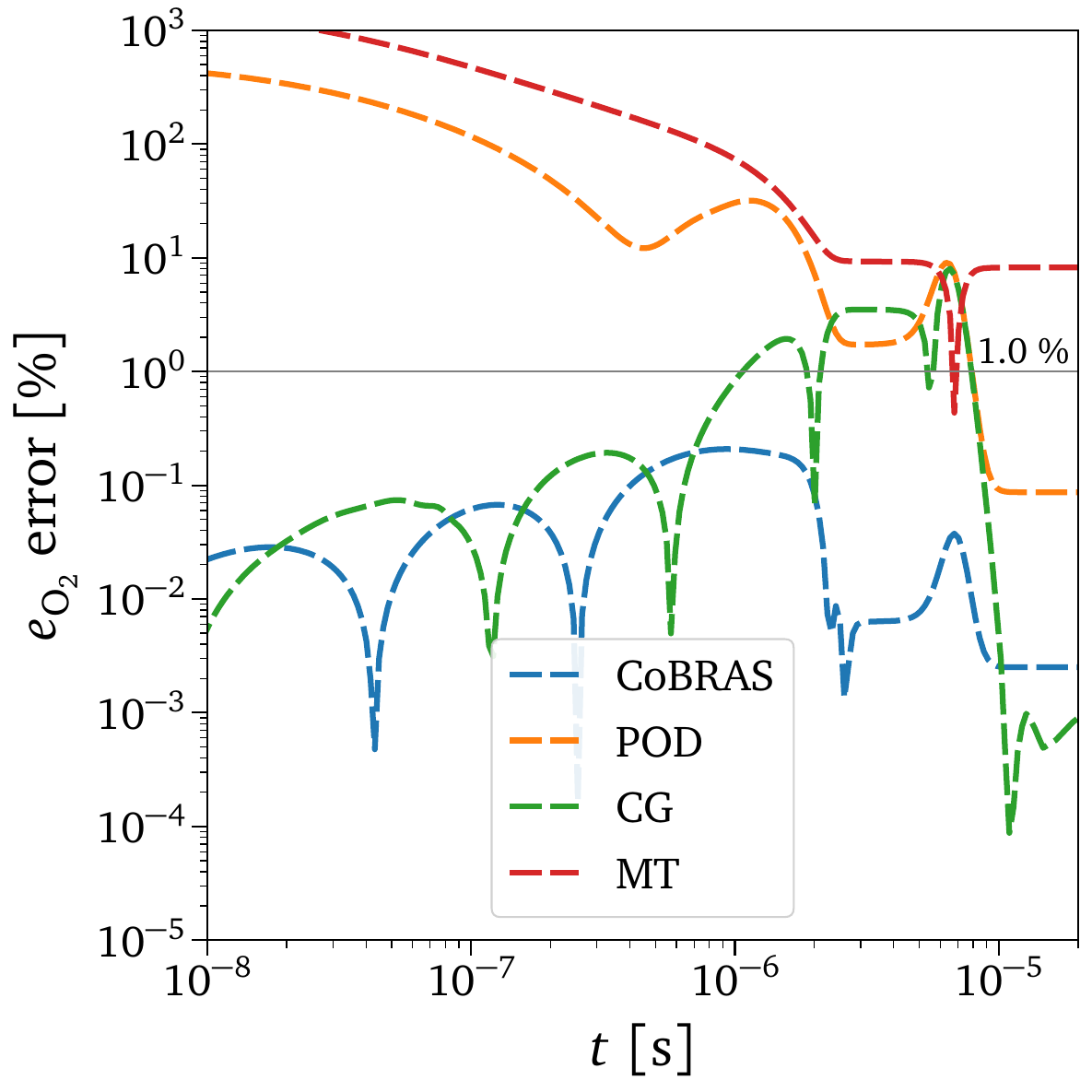}
\end{subfigure}
\caption{\textit{FOM vs. ROMs for RVC model: moments evolution}. Time evolution of the zeroth (top row) and first-order (bottom row) moments of the O$_2$ energy level population in an initially cold gas with $T=10\,000$ K, as calculated using the FOM and various ROMs. The left panels display the computed moment evolution, while the right panels show the relative errors of the ROMs compared to the FOM. The CoBRAS and POD ROMs use a reduced model dimension of $r = 8$, the CG ROM uses $r = 24$, and the MT ROM uses $r = 2$.}
\label{fig:rvc_mom_cold}
\end{figure}
Figure \ref{fig:rvc_mom_cold} presents a comparative analysis of various reduced-order modeling approaches for a test case with initially cold gas. 
The test condition includes a system temperature of $T=10\,000$ K, an initial pressure of $p_0 = 1\,000$ Pa, initial molar concentration of atomic oxygen $x_{\atom_0} = 0.05$, and initial equilibrium temperature $T_0 = 500$ K. 
In the left panels, we show the time evolution of the zeroth and first moments, while the right panels illustrate the relative error.
Additional comparisons at various testing temperatures for the same initial conditions are provided in figure \ref{fig:rvc_mom_cold_temps} of the Supplementary Material. These figures are crucial for evaluating the performance of CoBRAS technique against the POD method, the traditional ROMs, and the reference FOM solution. The dimensions of the reduced-order models are as follows: $r = 8$ for CoBRAS and POD ROMs, $r = 24$ for CG (using 12 bins), and $r = 2$ for MT. The left panels demonstrate that for both moments, the CoBRAS-based ROM closely matches the FOM, with a maximum error consistently below 1\%. Additionally, CoBRAS outperforms the POD, CG, and MT approaches, each of which fails to accurately predict the system dynamics. 
These performance differences become even more pronounced in the error evolution depicted in the right panels. For the zeroth-order moment, the point-wise error for POD, CG, and MT approaches 100\%, whereas for the first-order moment, it is approximately 10\% for CG and over 100\% for POD and MT. The high error in MT’s energy evolution is due to the separation of energy modes, where the rotational mode is in equilibrium with the translational mode at temperature $T$, while the vibrational energy is modeled with a fictitious temperature $T_v$. This separation of energy modes is a fundamental assumption distinguishing MT from the other ROMs, rooted in physical considerations not discussed here.

\subsubsection{Models for microscopic quantities}
\label{sec:micro_1}
\begin{figure}[htb!]
\centering
\begin{subfigure}[htb!]{0.35\textwidth}
    \caption*{\hspace{7mm}\small CoBRAS} \vskip 5pt
    \includegraphics[width=\textwidth]{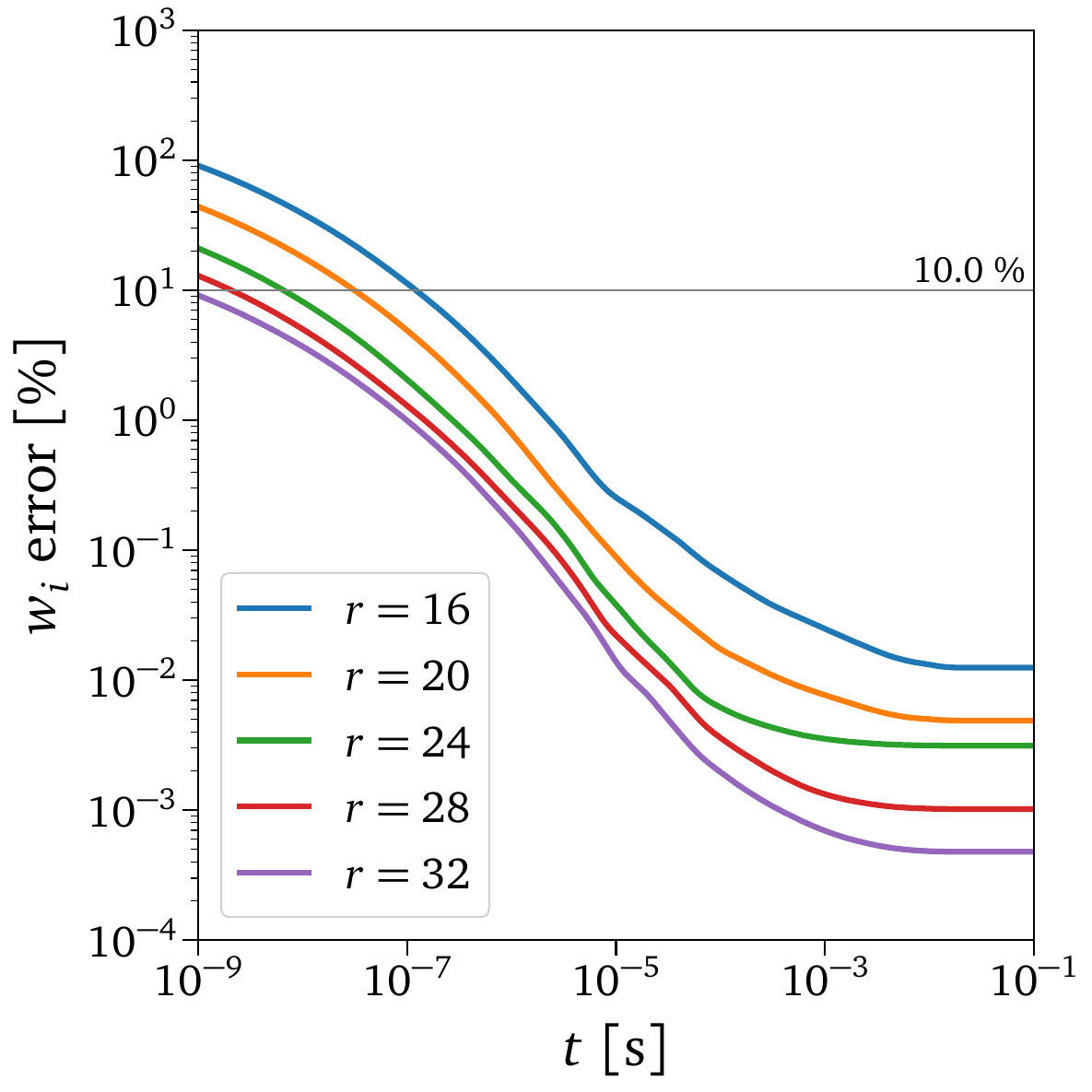}
\end{subfigure}
\quad
\begin{subfigure}[htb!]{0.35\textwidth}
    \caption*{\hspace{7mm}\small POD} \vskip 5pt
    \includegraphics[width=\textwidth]{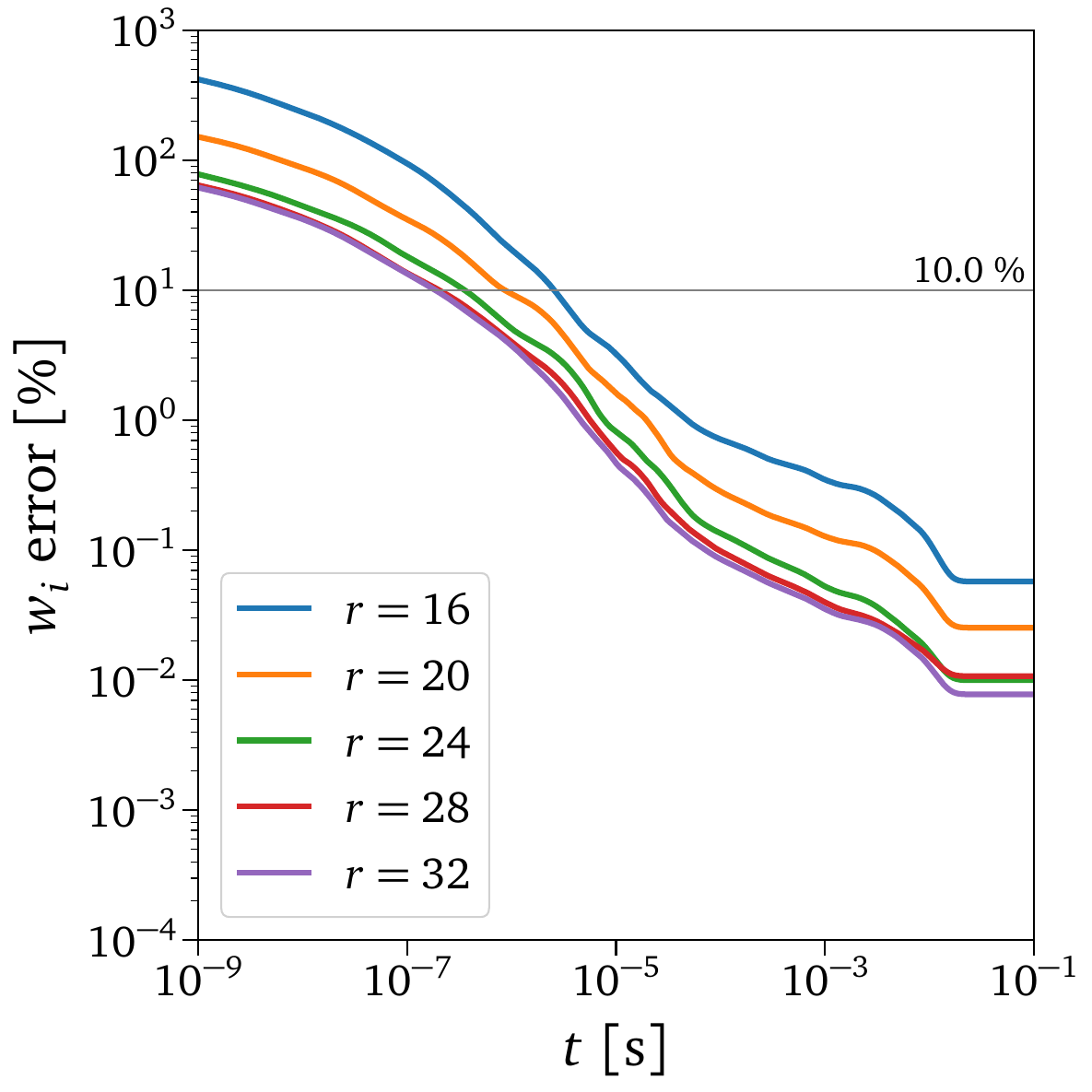}
\end{subfigure}
\caption{\textit{Mean relative error in the distribution function for RVC model using CoBRAS and POD ROMs}. Time evolution of the mean relative error in the O$_2$ energy level population across 1\,000 testing trajectories. Results are shown for different dimensions $r$ of the CoBRAS (left panel) and POD (right panel) ROM systems.}
\label{fig:rvc_mean_dist_err}
\end{figure}

Here, we seek models that can provide an accurate description of the \textit{whole} distribution along trajectories.
In principle, this can be done by choosing the output $\vy$ to be equal to the state vector $\vw$, but the evaluation of the gradient covariance would be rather expensive (given that $\nabla F$ scales with $\mathrm{dim}(\vy)$). 
Instead, we choose $\vy\in\mathbb{R}^{10}$ to contain the first ten moments of the distribution, since user experience suggests that a model that is capable of tracking these moments is likely capable of producing accurate estimates of the distribution itself.
Figure \ref{fig:rvc_mean_dist_err} presents the time evolution of the mean relative error between the FOM solution $\vw(t)$ and the predicted solutions $\hat\vw(t)$ obtained from CoBRAS (left panel) and POD (right panel) for the O$_2$ energy level population. 
The error is calculated over $M=1\,000$ trajectories and for different reduced-system dimension $r$. 
The figure highlights the model's accuracy over time and across different reduced dimensions, demonstrating its effectiveness in predicting microscopic quantities.
The error analysis is consistent with the considerations made for figure \ref{fig:rvc_mean_moms_err}, indicating that the overall error remains below 10\% when using CoBRAS with $r \geq 28$, which corresponds to a compression factor of nearly 220. In contrast, the POD technique consistently yields results with higher errors compared to those produced by CoBRAS. 
It is worth remarking that neither POD nor CoBRAS guarantee positivity of the predicted state $\hat\vw$, and we have observed that non-physical negative values may appear at very early times $t < 10^{-8}$ s.
This is particularly true at low initial temperatures $T_0 \le 500$, where the distribution spans approximately 20 orders of magnitude $\Big(10^3 - 10^{23}\Big)$, and the tail of the predicted distribution may take negative values.
\par
\begin{figure}[htb!]
\centering
\begin{subfigure}[htb!]{0.35\textwidth}
    \includegraphics[width=\textwidth]{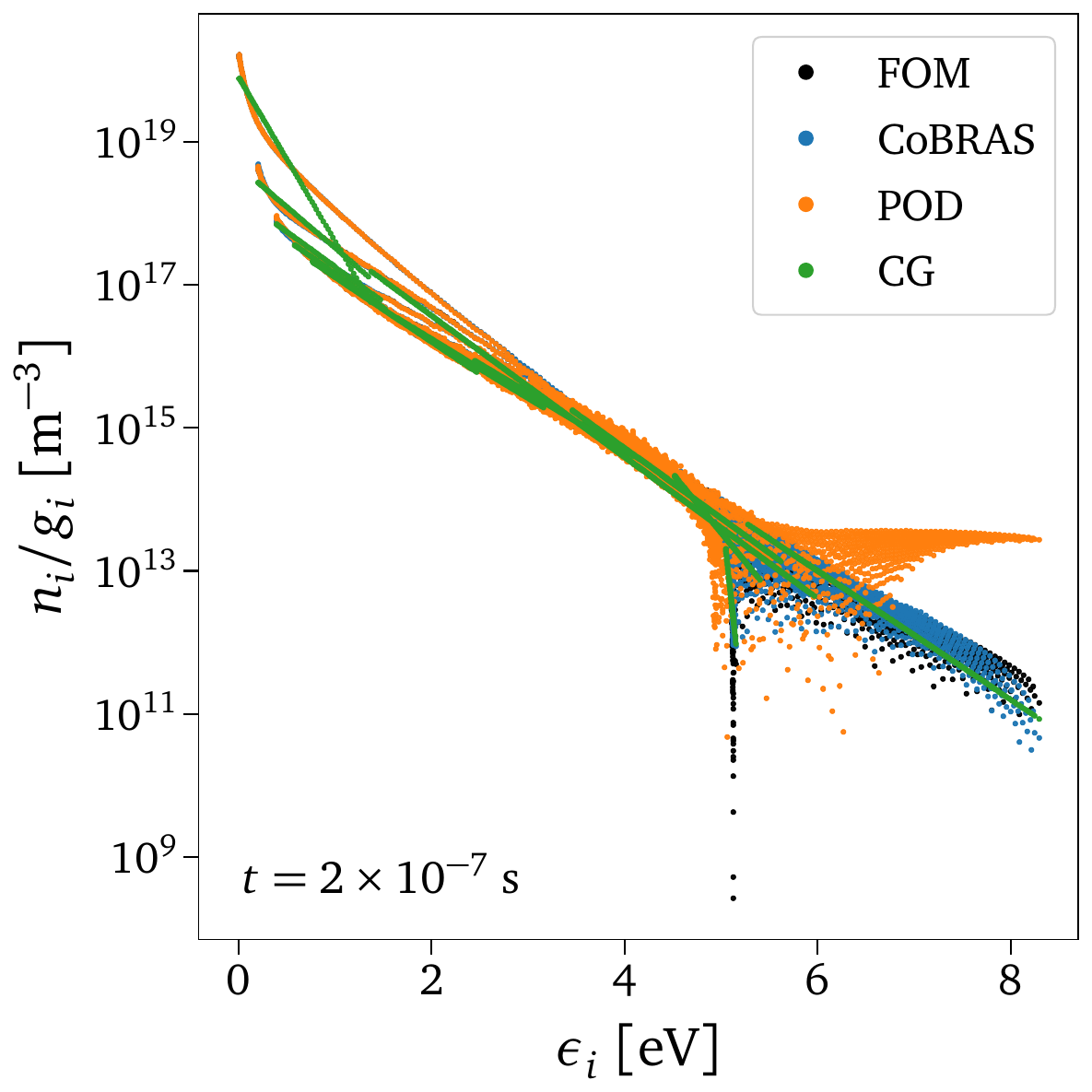}
\end{subfigure}
\quad
\begin{subfigure}[htb!]{0.35\textwidth}
    \includegraphics[width=\textwidth]{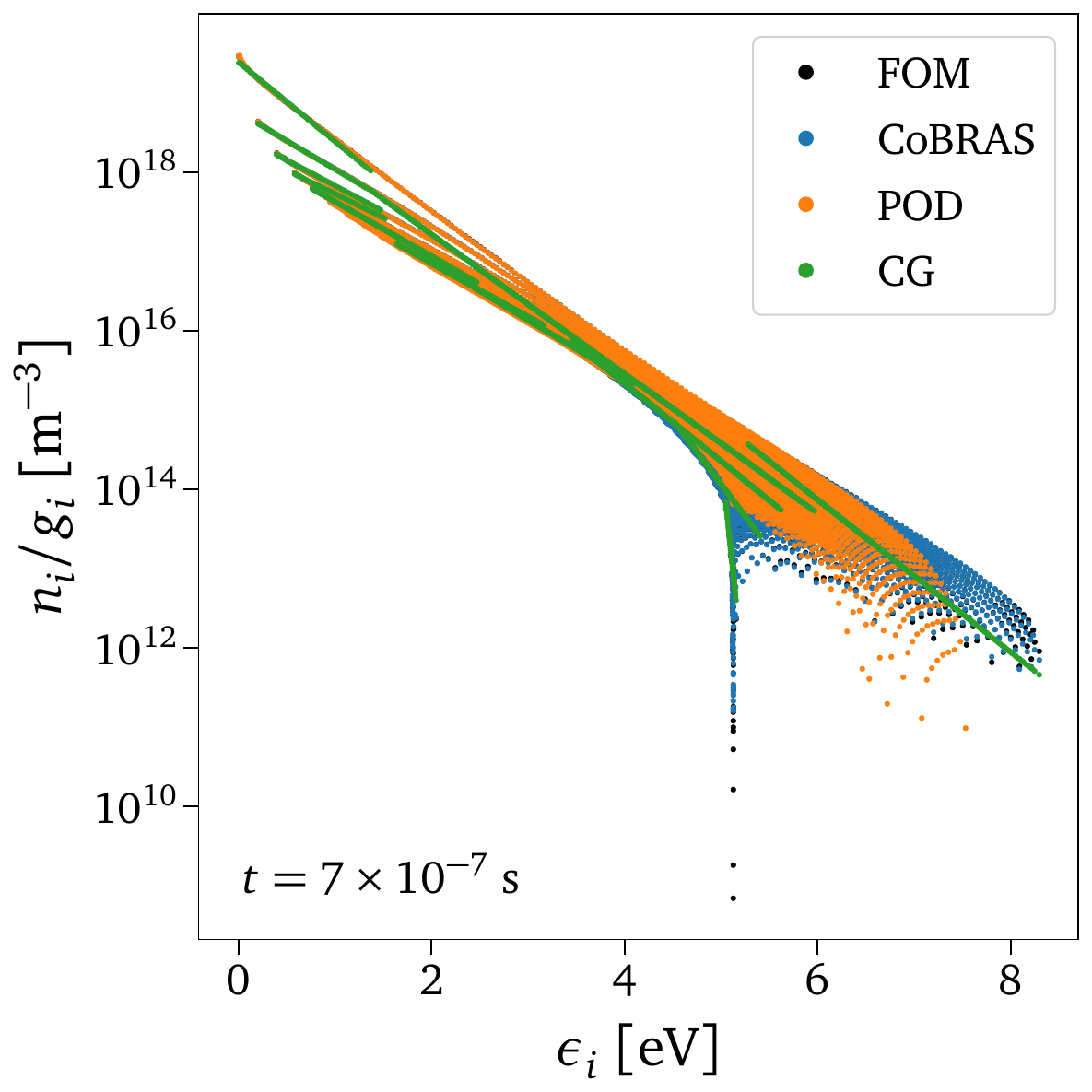}
\end{subfigure}
\\[5pt]
\begin{subfigure}[htb!]{0.35\textwidth}
    \includegraphics[width=\textwidth]{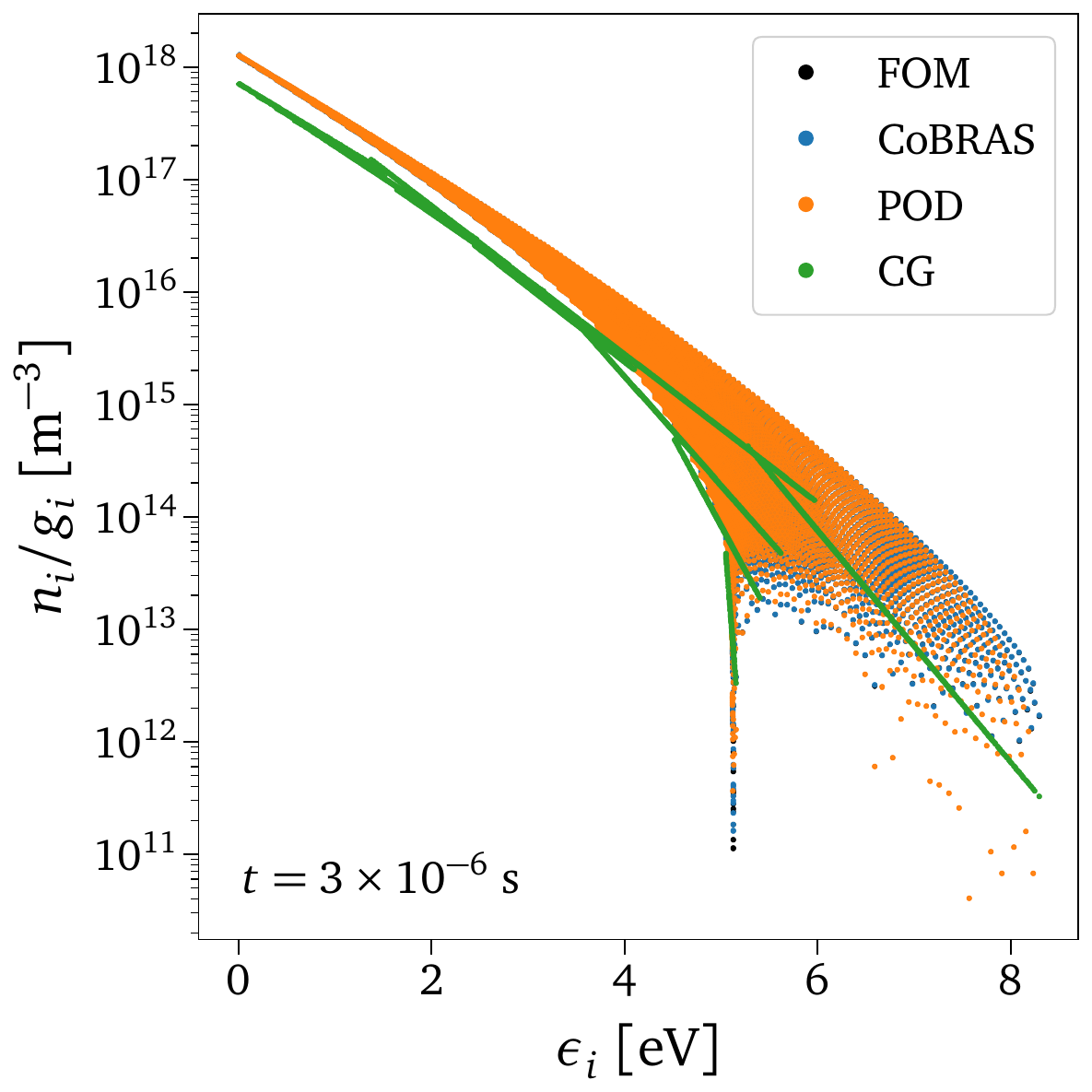}
\end{subfigure}
\quad
\begin{subfigure}[htb!]{0.35\textwidth}
    \includegraphics[width=\textwidth]{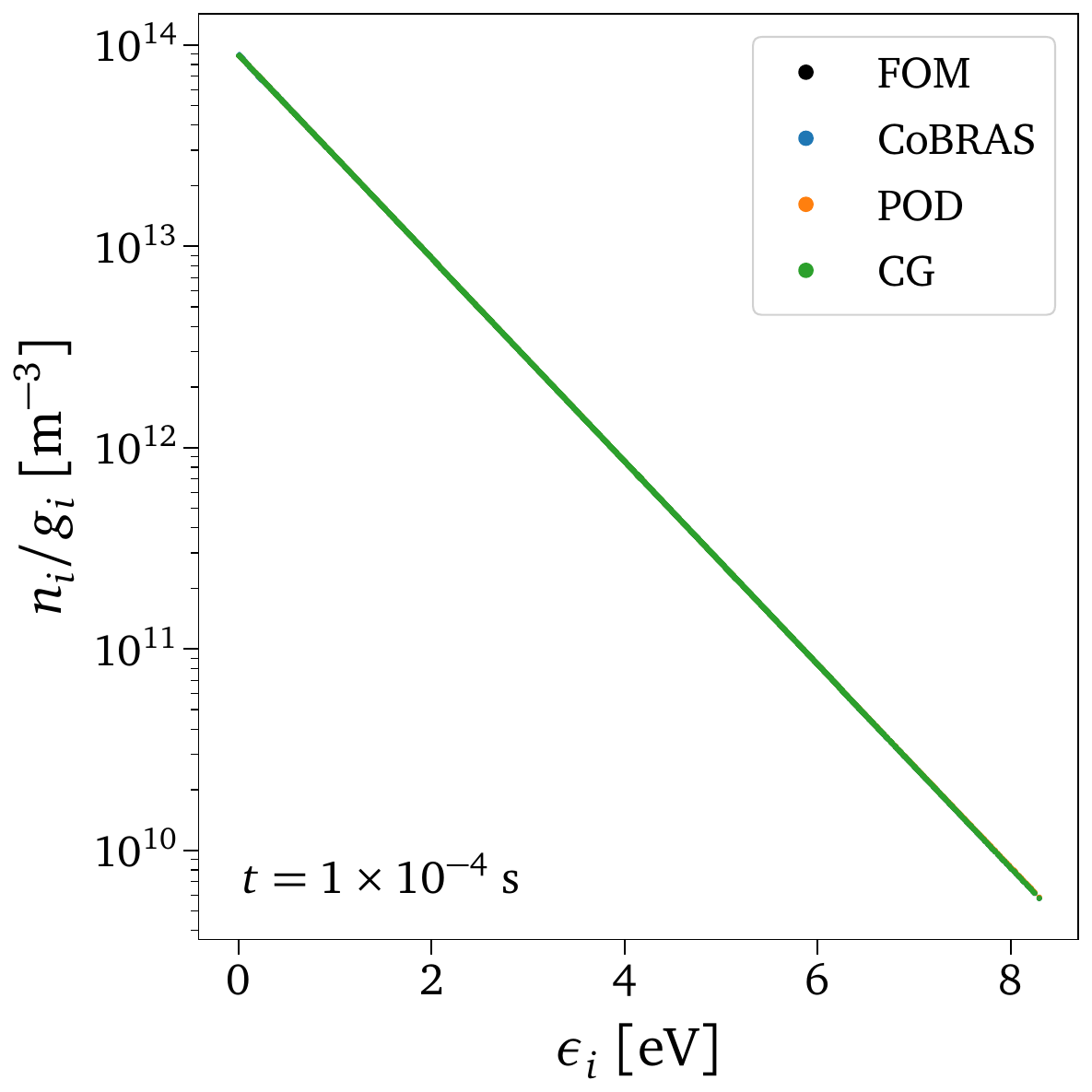}
\end{subfigure}
\caption{\textit{FOM vs. ROMs for RVC model: distribution snapshots}. Time snapshots of the O$_2$ energy level population in an initially cold gas with $T=10\,000$ K, calculated using FOM, CoBRAS, POD, and CG reduced order models. All ROMs have a dimension of $r=24$.}
\label{fig:rvc_dist_cold}
\end{figure}
Figure \ref{fig:rvc_dist_cold} presents four snapshots of the O$_2$ internal state population taken from the FOM and from CoBRAS, POD, and the CG reduced-order models, each operating at a reduced dimensionality of $r=24$ (with 12 bins for the CG approach). 
This figure is central for evaluating the effectiveness of the CoBRAS-based ROM in capturing microscopic scales of the physical system compared to POD and the conventional CG technique. 
This test case corresponds to the scenario illustrated in figure \ref{fig:rvc_mom_cold}. 
Further comparisons at various testing temperatures for the same initial conditions are provided in figure \ref{fig:rvc_dist_cold_temps} of the Supplementary Material. 
In qualitative terms, CoBRAS displays consistently higher fidelity to the FOM across all snapshots, effectively capturing the microscopic details of the distribution function with greater accuracy than the other ROMs. 
Additionally, although the POD model demonstrates improved performance compared to the CG model, it still does not reach the accuracy achieved by the CoBRAS model. 
The discrepancy between CoBRAS and CG is particularly pronounced at $t=3 \times 10^{-6}$ s, where the quasi-steady-state (QSS) distribution is present. 
Here CoBRAS matches the reference solution very accurately, whereas the CG shows a vertical shift indicative of faster dynamics, as previously noted in figure \ref{fig:rvc_mom_cold}. 
Notably, all ROMs successfully reconstruct the equilibrium distribution at $t=10^{-4}$ s.

\subsubsection{Computational performance}
Table \ref{table:rvc.rom.costs} presents estimates of the total number of FLOPs required for solving the FOM and the CoBRAS-based ROM of different dimensions, along with the stiffness of the corresponding systems. The results indicate that using the ROM can reduce FLOPs by up to $10^6$ for RHS evaluations, which mainly involve a matrix-vector product with \ordermagn{d^2} complexity (see equations \eqref{eq:rvc.fom.vec.ni} and \eqref{eq:rvc.fom.vec.no} in the Supplementary Material), and by $10^9$ for solving the linear system in the BDF scheme with \ordermagn{d^3} complexity. 
(Here, $d$ denotes the system's dimension.)
Additionally, the ROM generates a less stiff system of equations, enabling the use of larger time steps during integration.
\begin{table}[!htb]
    \centering
    \begin{tabular}{c|c|c|c|c}
        \toprule
        \multirow{2}*{Model} & Dimension & RHS Evaluation & BDF Scheme & Fastest Timescale \\
         & $d$ & \ordermagn{d^2} & \ordermagn{d^3} & $1/\lvert\lambda_{\min}\rvert$ \\
        \midrule
        FOM & 6116 & $3.74\times 10^7$ & $2.28\times 10^{11}$ & $1.36\times 10^{-10}$ \\
        \arrayrulecolor{gray}\midrule
        \multirow{2}*{CoBRAS} & 25 & $6.25\times 10^2$ & $1.56\times 10^4$ & $5.35\times 10^{-10}$ \\
        & 9 & $8.10\times 10^1$ & $7.29\times 10^2$ & $1.29\times 10^{-9}$ \\
        \bottomrule
    \end{tabular}
    \caption{\textit{Comparison of computational cost: FOM vs. CoBRAS ROM for the RVC Model}. Total FLOPs for essential numerical operations and system stiffness estimates are provided for the FOM and various dimensions of the CoBRAS model.}
    \label{table:rvc.rom.costs}
\end{table}

\subsection{O$_2$-O$_2$ and O$_2$-O vibrational collisional model}\label{sec:num_exp:vc_model}
In this experiment, we focus on the vibrational excitation and dissociation of O$_2$ molecules during collisions with either O atoms or other O$_2$ molecules. The O$_2$ molecule has 45 vibrational energy levels, resulting in a total of $N=46$ degrees of freedom for the system. We opted not to use the RVC model, as it would require managing an impractical kinetic database with billions of possible transitions. Instead, we employed the VC model, which is based on the RVC model but relies on the simplifying assumption of equilibrium (Maxwell-Boltzmann) distribution of the rotational levels,
\begin{equation}
    \frac{n_i}{n_v}=\frac{q_i}{Q_v}\eqspace, \quad v \in \mathcal{V}\eqspace, \quad i \in \mathcal{I}_v\eqspace,
\end{equation}
with 
\begin{equation}
    q_i = g_i \exp \left(-\frac{\epsilon_i}{k_B T_v}\right)
\end{equation}
and $Q_v=\sum_{i \in \mathcal{I}_v} q_i$ being the partition function of the vibrational level $v$. $\mathcal{V}$ denotes the set of all vibrational levels, while $\mathcal{I}_v$ refers to the set of rotational levels with the same vibrational level $v$. In the VC model, it is  further assumed that rotation and translation are in equilibrium, such that $T_v = T$ for all $v \in \mathcal{V}$. Each vibrational level can be characterized by an energy $\epsilon_v$ and a fictitious degeneracy $g_v$, which are determined as follows
\begin{align}
    \epsilon_v & =\frac{1}{Q_v}\sum_{i \in \mathcal{I}_v}q_i\epsilon_i \eqspace , \\
    g_v & = Q_v \exp \left(\frac{\epsilon_v}{k_B T}\right) \eqspace .
\end{align}
These values are then used in the computation of backward rates by enforcing detailed balance.
\par
The kinetic database includes O$_2$-O collisions as described in section \ref{sec:num_exp:rvc_model}, properly reduced for the equilibrium assumptions of rotational levels, as well as O$_2$-O$_2$ collisions, which involve three types of processes:
\begin{enumerate}[i.]
    \item collisional excitation, including both inelastic (nonreactive) and exchange processes,
    \begin{equation}\label{eq:coll.mol.excit}
        \ce{
            O_2$(v)$ + O_2$(w)$
            <=>[${}^{\mathrm{m}}k_{vwpq}^{\mathrm{e}}$][${}^{\mathrm{m}}k_{pqvw}^{\mathrm{e}}$]
            O_2$(p)$ + O_2$(q)$
        }\eqspace,
    \end{equation}
    \item combined collisional dissociation and excitation,
    \begin{equation}\label{eq:coll.mol.diss_excit}
        \ce{
            O_2$(v)$ + O_2$(w)$
            <=>[${}^{\mathrm{m}}k_{vwp}^{\mathrm{ed}}$][${}^{\mathrm{m}}k_{pvw}^{\mathrm{er}}$]
            O_2$(p)$ + O + O
        }\eqspace,
    \end{equation}
    \item collisional dissociation,
    \begin{equation}\label{eq:coll.mol.diss}
        \ce{
            O_2$(v)$ + O_2$(w)$
            <=>[${}^{\mathrm{m}}k_{vw}^{\mathrm{d}}$][${}^{\mathrm{m}}k_{vw}^{\mathrm{r}}$]
            O + O + O + O
        }\eqspace,
    \end{equation}
\end{enumerate}
where $w$, $p$ and $q$ represent different vibrational levels. The VC model here described is governed by the following set of equations:
\begin{align}
    \frac{dn_v}{d t} =
        & \sum_{w,p,q \in \mathcal{V}} \left(-{}^{\mathrm{m}}k_{vwpq}^{\mathrm{e}} n_v n_w
        + {}^{\mathrm{m}}k_{pqvw}^{\mathrm{e}} n_p n_q \right) \nonumber\\
        & + \sum_{w,p \in \mathcal{V}} \left(-{}^{\mathrm{m}}k_{vwp}^{\mathrm{ed}} n_v n_w
        + {}^{\mathrm{m}}k_{pvw}^{\mathrm{er}} n_p n_\atom^2 \right) \nonumber\\
        & + \sum_{w \in \mathcal{V}} \left(- {}^{\mathrm{m}}k_{vw}^{\mathrm{d}} n_v n_w + {}^{\mathrm{m}}k_{vw}^{\mathrm{r}} n_\atom^4 \right) \nonumber\\
        & + \sum_{w \in \mathcal{V}} \left(-{}^{\mathrm{a}}k_{vw}^{\mathrm{e}} n_v n_\atom
        + {}^{\mathrm{a}}k_{wv}^{\mathrm{e}} n_w n_\atom \right) \nonumber\\
        & -{}^{\mathrm{a}}k_v^{\mathrm{d}}n_vn_\atom + {}^{\mathrm{a}}k_v^{\mathrm{r}}n_\atom^3 \eqspace, \label{eq:vc.fom.ni} \\
    \frac{dn_\atom}{dt} = & - 2\sum\limits_{v\in\mathcal{V}}\frac{dn_v}{d t}\label{eq:vc.fom.no}
    \eqspace.
\end{align}
For interested readers, the linearized FOM equations necessary to implement the method described in Sections \ref{sec:rom:pg} to \ref{sec:rom:covmat}, as well as the derivation of the ROM, are available in Supplementary Material sections \ref{suppl:vc.linfom} and \ref{suppl:vc.rom}.
\par
The bounds and distributions used for sampling the training and the $M=1\,000$ testing trajectories for the VC model are provided in table \ref{table:vc.mu_space}. For this thermochemical system, we consider a single temperature of $T = 10\,000$ K.
\begin{table}[!htb]
    \centering
    \begin{tabular}{c|cccc}
        \toprule
        & $T$ [K] & $\rho$ [kg/m$^{3}$] & $w_{\atom_0}$ & $T_0$ [K] \\
        \midrule
        Minimum & 10\,000 & $10^{-4}$ & 0 & 500 \\
        Maximum & 10\,000 & 1 & 1 & 10\,000 \\
        Distribution & - & Log-uniform & Uniform & Log-uniform \\
        \bottomrule
    \end{tabular}
    \caption{\textit{Sampled parameter space for the VC model}. Sampling bounds and distributions for each parameter used to generate the training and testing trajectories.}
    \label{table:vc.mu_space}
\end{table}

\subsubsection{Linearized FOM performances}\label{sec:num_exp:vc:lin_fom}

\begin{figure}[htb!]
\centering
\includegraphics[width=0.35\textwidth]{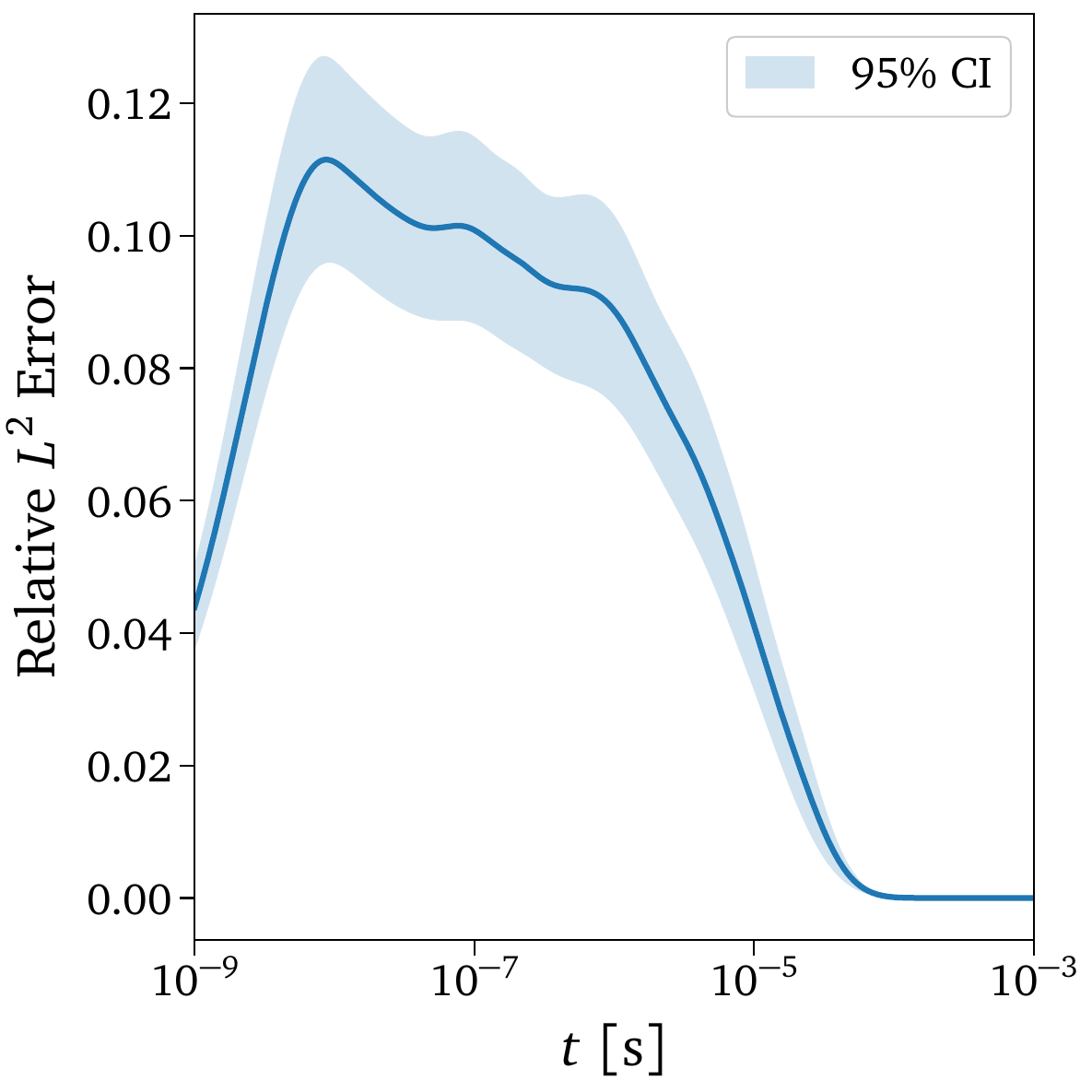}
\caption{\textit{Relative $L^2$ error of the linearized FOM for the VC model}. This figure presents the time evolution of the mean relative $L^2$ error, along with a 95\% confidence interval derived from 1\,000 testing trajectories, using the linearized FOM.}
\label{fig:vc_linfom_err}
\end{figure}

In line with the analysis in section \ref{sec:num_exp:rvc:lin_fom}, we assess the efficienty and accuracy of the linearized FOM to estimate the state and gradient covariance matrices associated with the VC model.
As in the previous section, we observe that using the linearized FOM leads to a significant reduction in computational cost (see table \ref{table:vc.lin_fom.costs}) while also providing a very good approximation of the subspaces on which the nonlinear dynamics evolve (see figure \ref{fig:vc_linfom_err}).
\begin{table}[!htb]
    \centering
    \begin{tabular}{c|c|c|c|c}
        \toprule
        Model & Operation & Device & Time [s] - 1 Run & Time [s] - 200 Runs \\
        \midrule
        FOM & Simulation & CPU - 4 Threads & $6.14 \times 10^{-2}$ & $1.23 \times 10^{1}$ \\
        \arrayrulecolor{gray}\midrule
        \multirow{2}*{Linearized FOM}
        & Eigendecomposition & CPU - 4 Threads & $7.36 \times 10^{-4}$ & $7.36 \times 10^{-4}$ \\
        & Simulation & CPU - 4 Threads & $5.16 \times 10^{-5}$ & $1.03 \times 10^{-2}$ \\
        \bottomrule
    \end{tabular}
    \caption{\textit{Comparison of computational cost: FOM vs. linearized FOM for the VC Model}. This table illustrates the performance differences between the FOM and its linearized counterpart across various numerical operations.}
    \label{table:vc.lin_fom.costs}
\end{table}
\par

\subsubsection{Models for macroscopic quantities}
\begin{figure}[htb!]
\centering
\begin{subfigure}[htb!]{0.35\textwidth}
    \caption*{\hspace{7mm}\small CoBRAS} \vskip 5pt
    \includegraphics[width=\textwidth]{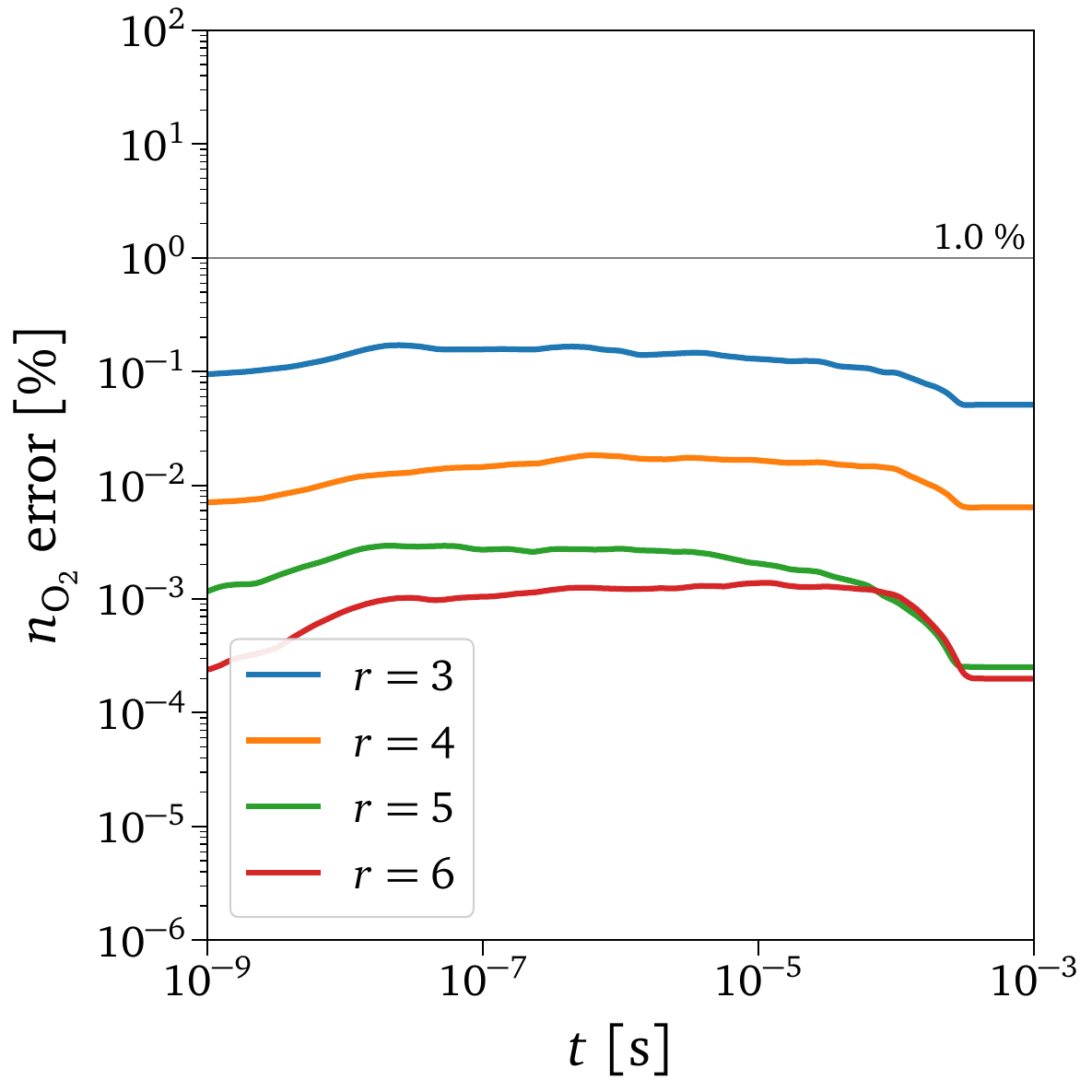}
\end{subfigure}
\quad
\begin{subfigure}[htb!]{0.35\textwidth}
    \caption*{\hspace{7mm}\small POD} \vskip 5pt
    \includegraphics[width=\textwidth]{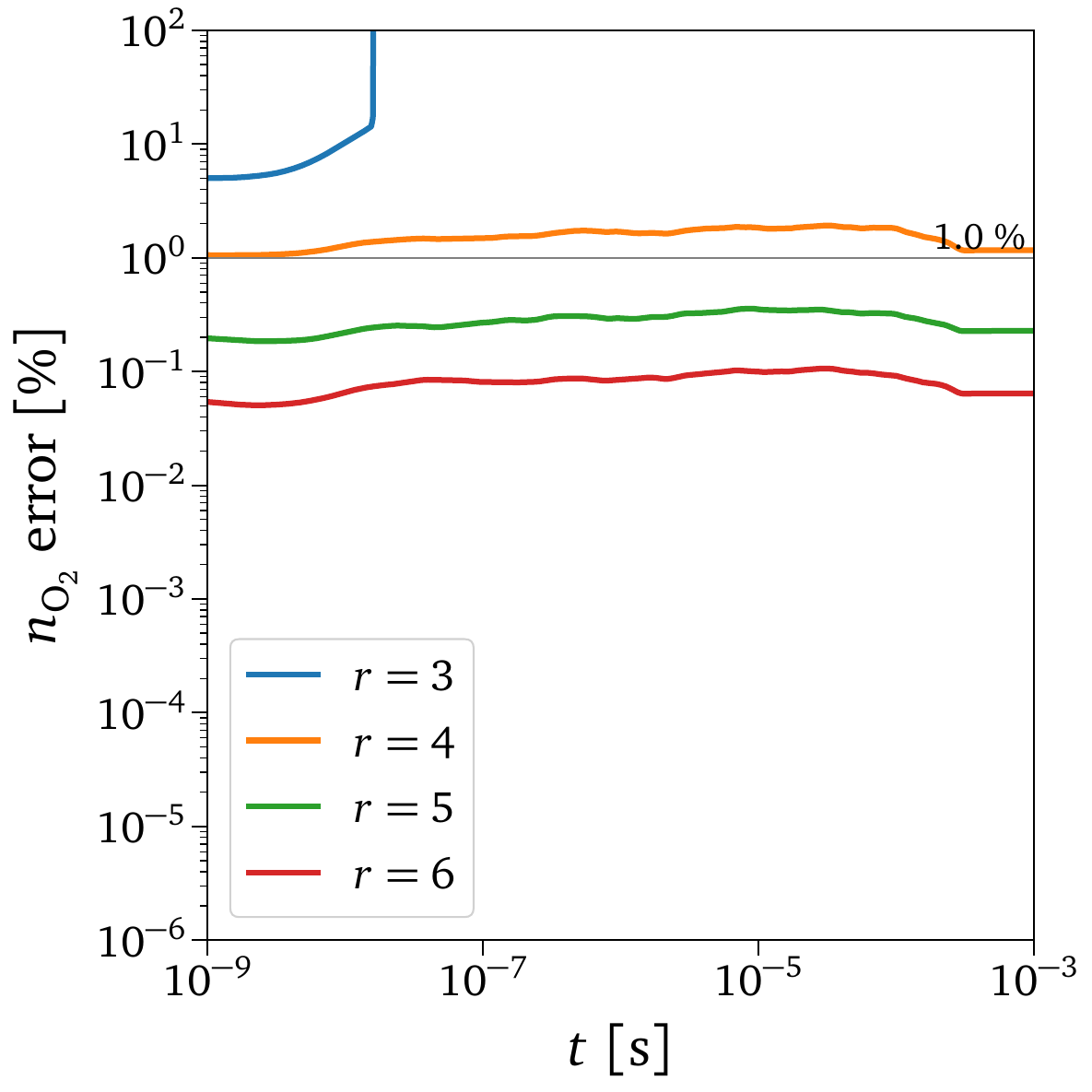}
\end{subfigure}
\\[5pt]
\begin{subfigure}[htb!]{0.35\textwidth}
    \includegraphics[width=\textwidth]{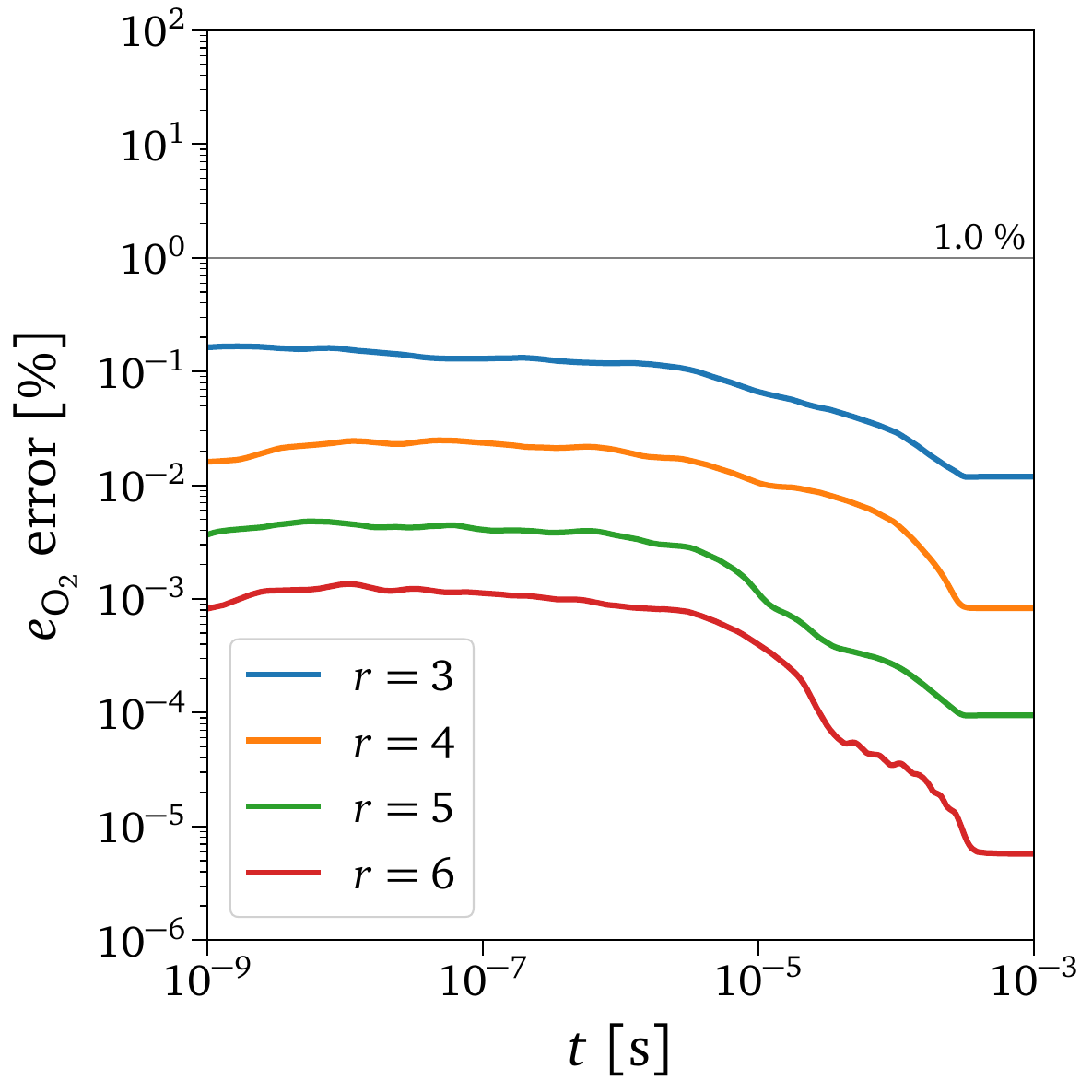}
\end{subfigure}
\quad
\begin{subfigure}[htb!]{0.35\textwidth}
    \includegraphics[width=\textwidth]{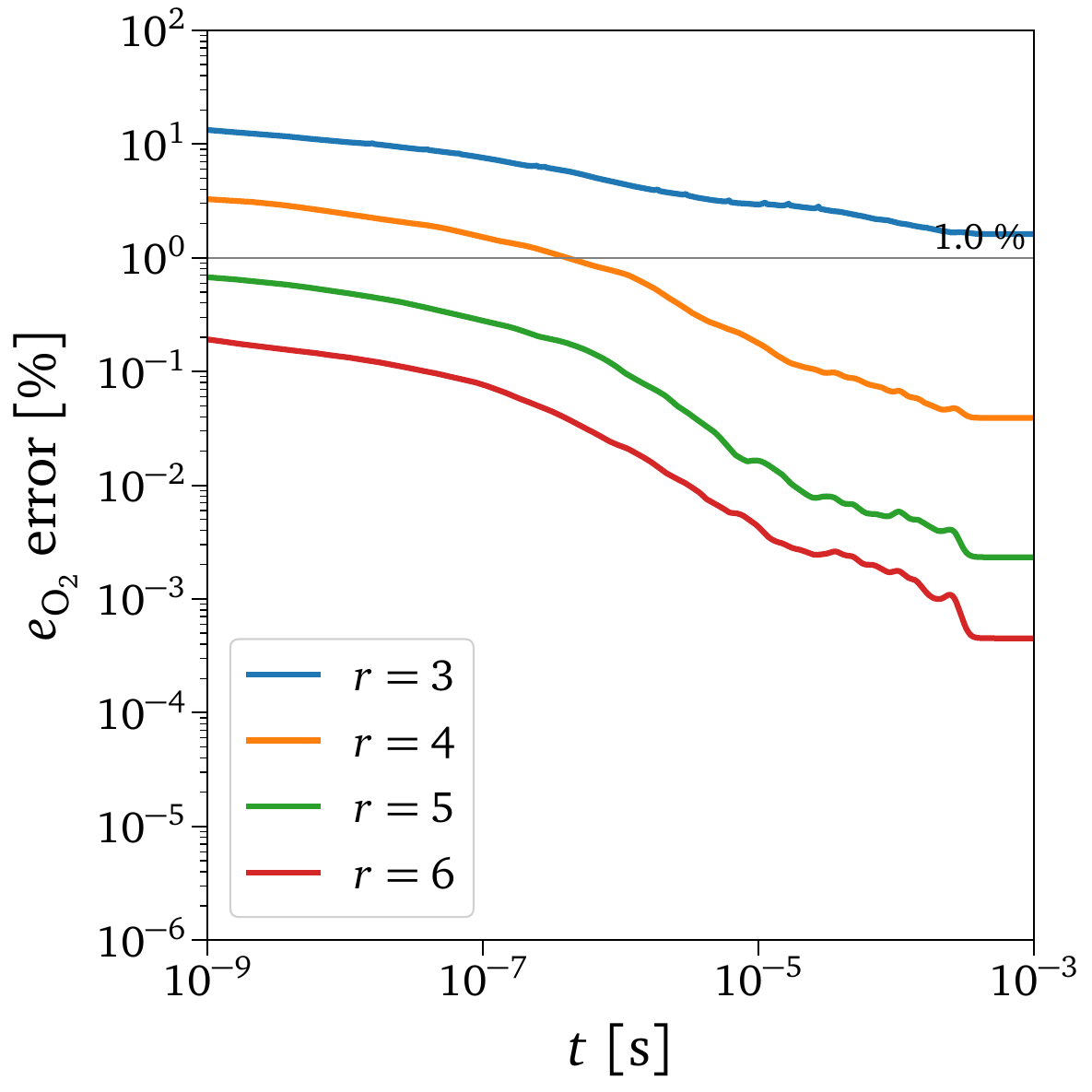}
\end{subfigure}
\caption{\textit{Mean moments error for VC model using CoBRAS and POD ROMs}. Time evolution of the mean relative errors for the zeroth (top row) and first-order (bottom row) moments of the O$_2$ energy level population across 1\,000 testing trajectories. Results are shown for different dimensions $r$ of the CoBRAS (left panels) and POD (right panels) ROM systems.}
\label{fig:vc_mean_moms_err}
\end{figure}

In this section, we seek models that accurately predict the time evolution of the system's macroscopic quantities such as mass and energy, and we therefore choose $\vy \in\mathbb{R}^2$ in \eqref{eq:fom.output} to contain the zeroth and first moments (i.e., the mass and energy, respectively) of the distribution.
Figure \ref{fig:vc_mean_moms_err} presents the time evolution of the mean relative error between the FOM output $y_j(t)$ and the predicted output $\hat{y}_j(t)$ from the CoBRAS (left panels) and POD (right panels) ROMs for the $j=0$ (top row) and $j=1$ (bottom row) moments of the O$_2$ energy level population in the VC model. 
This analysis is conducted over $M = 1\,000$ test trajectories for various reduced system dimensions, $r$. 
The figure provides insight into the model's accuracy over time and across different reduced dimensions when evaluating macroscopic quantities. 
For CoBRAS, the overall error remains consistently low, staying below 1\% for both moments even with a reduced system size of $r = 3$, achieving a compression rate above 11. 
By contrast, the POD ROM is significantly more inaccurate and it may even be unstable for very low dimensions (see the $r = 3$ curve in the top right panel).
Similar qualitative behavior is observed in the bottom row of figure \ref{fig:vc_mean_moms_err}, which is the analog of the top row for the first-order moment (i.e., the energy of the O$_2$ population).
\par
\begin{figure}[htb!]
\centering
\begin{subfigure}[htb!]{0.35\textwidth}
    \includegraphics[width=\textwidth]{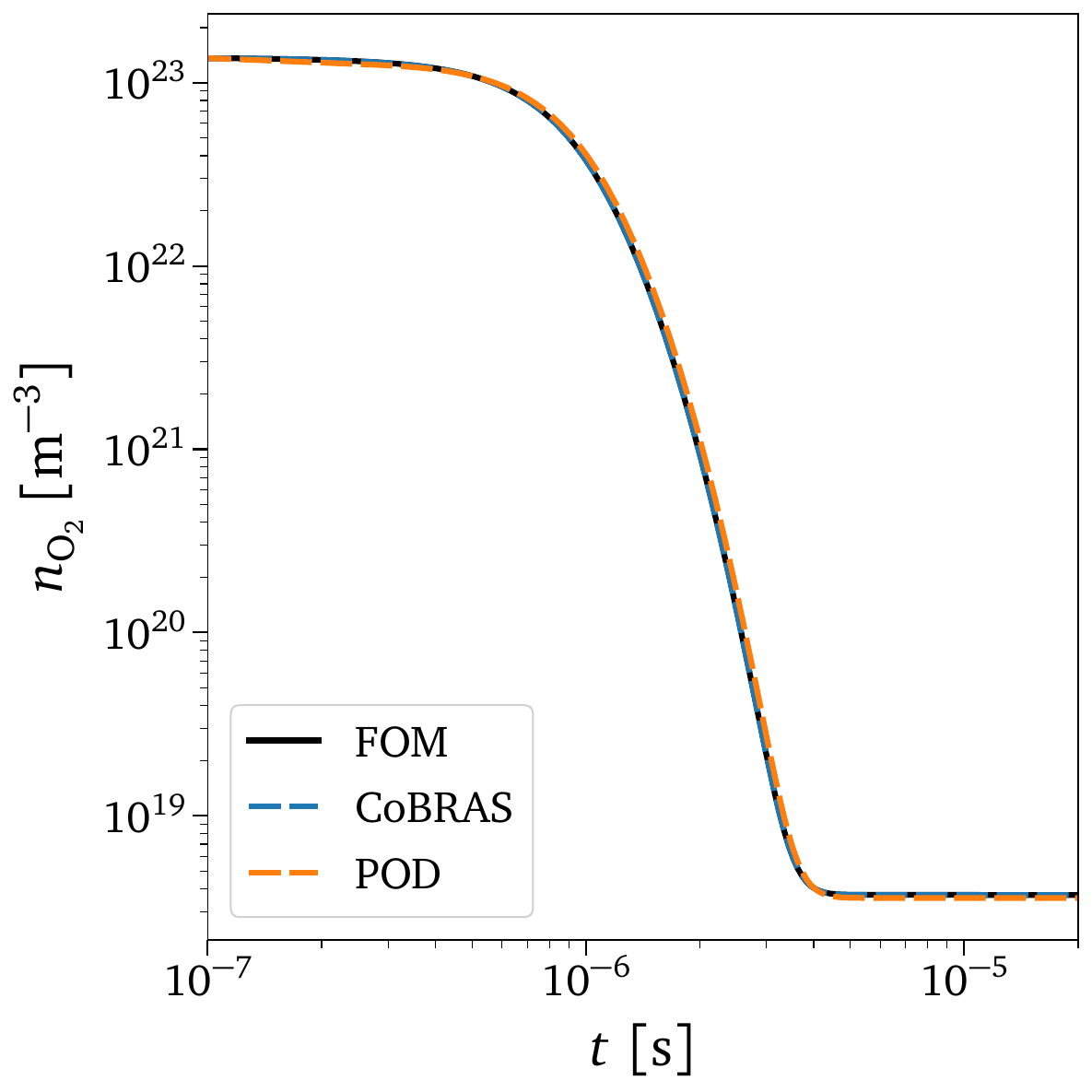}
\end{subfigure}
\quad
\begin{subfigure}[htb!]{0.35\textwidth}
    \includegraphics[width=\textwidth]{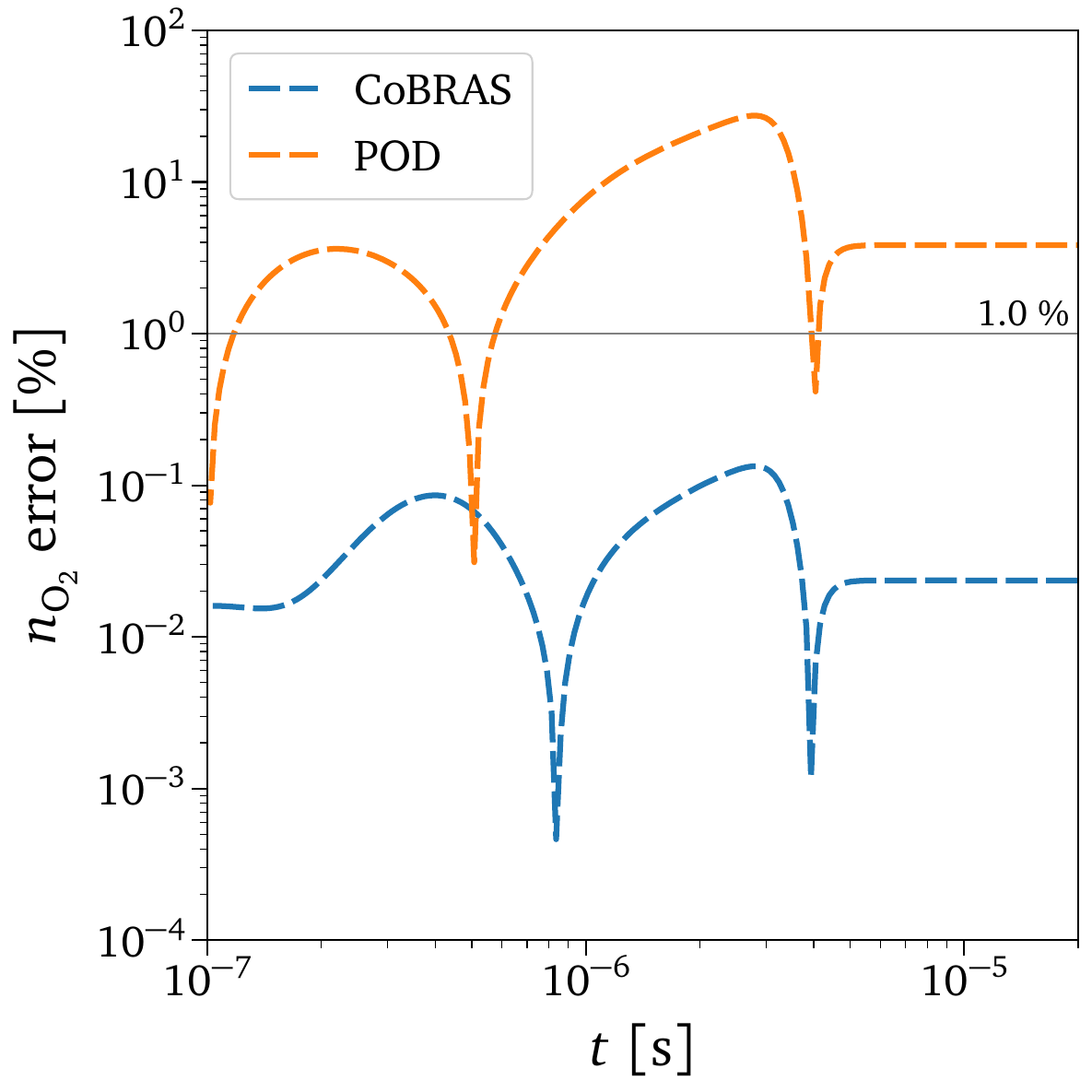}
\end{subfigure}
\\[5pt]
\begin{subfigure}[htb!]{0.35\textwidth}
    \includegraphics[width=\textwidth]{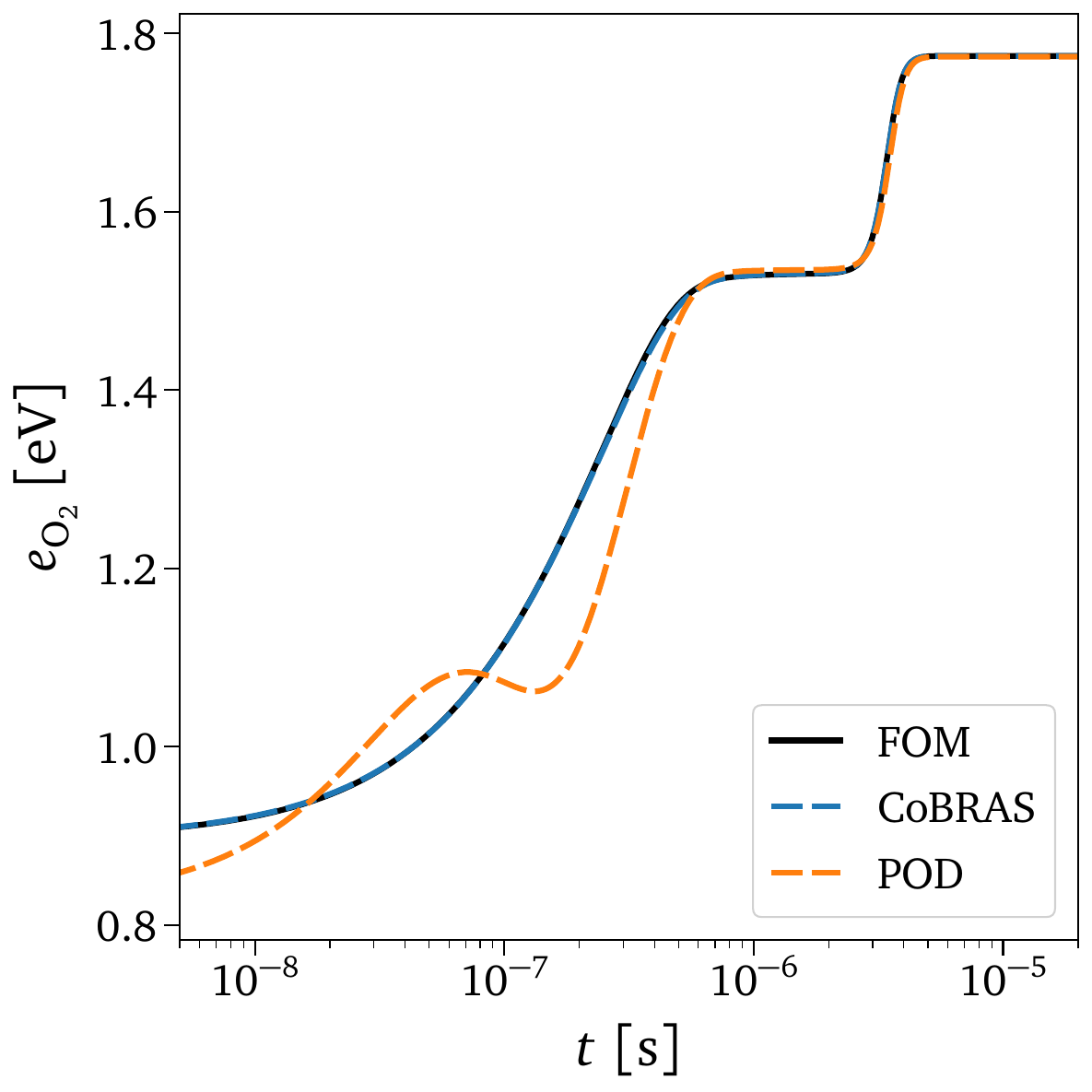}
\end{subfigure}
\quad
\begin{subfigure}[htb!]{0.35\textwidth}
    \includegraphics[width=\textwidth]{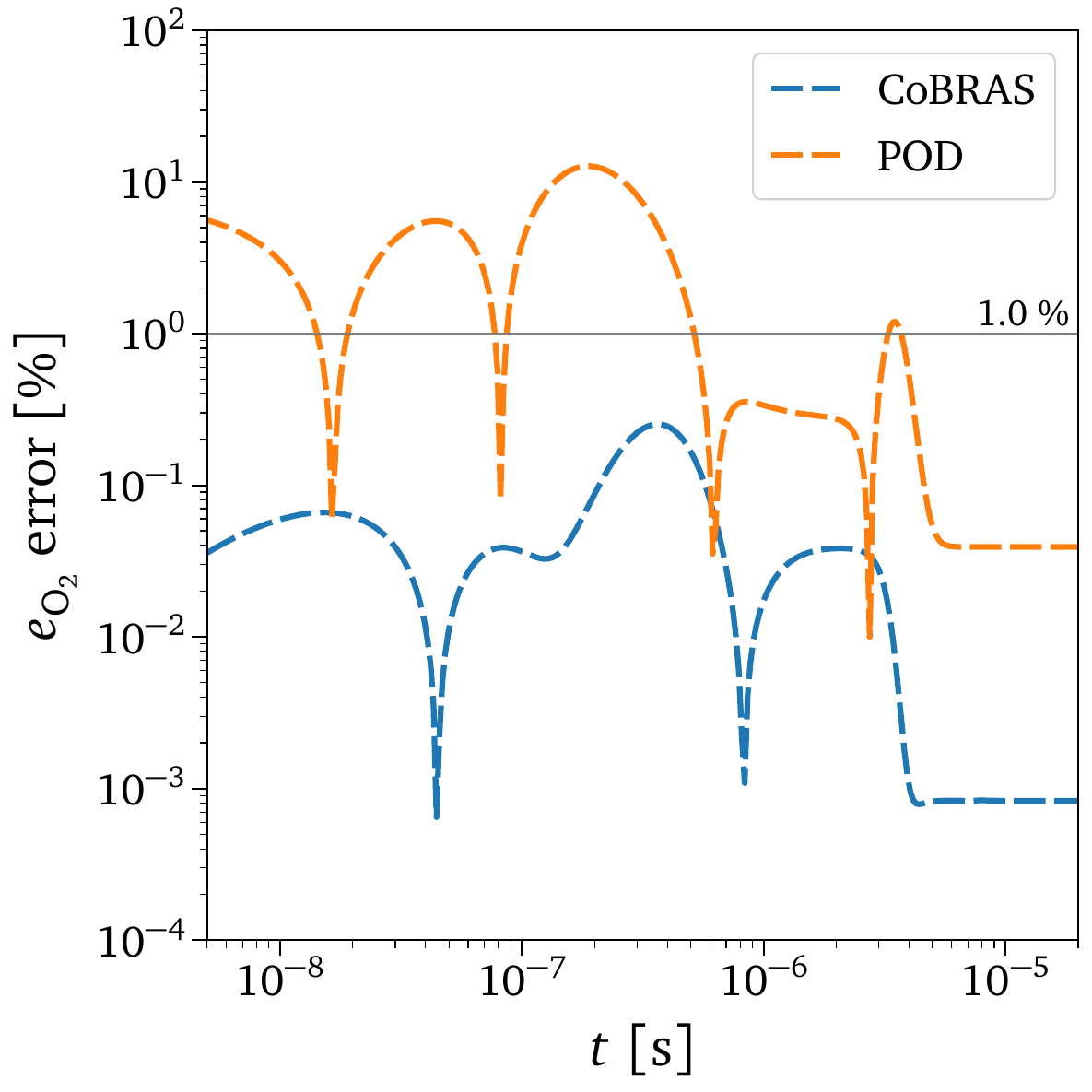}
\end{subfigure}
\caption{\textit{FOM vs. ROMs for VC model: moments evolution}. Time evolution of the zeroth (top row) and first-order (bottom row) moments of the O$_2$ energy level population in an initially cold gas with $T=10\,000$ K, as calculated using the FOM and various ROMs. The left panels display the computed moment evolution, while the right panels show the relative errors of the ROMs compared to the FOM. The CoBRAS and POD ROMs utilize a reduced model dimension of $r=4$.}
\label{fig:vc_mom_cold}
\end{figure}
Figures \ref{fig:vc_mom_cold} provides a comparative analysis of different reduced-order modeling techniques for a test case involving an initially cold gas, focusing on the zeroth and first-order moments of the O$_2$ internal energy distribution function. 
In the left panels, we show the time evolution of these moments, while the right panels display the relative error.
The initial conditions match those in figure \ref{fig:rvc_mom_cold}, with the ROM dimension set to $r=4$. These figures are essential for assessing the performance of the CoBRAS technique in comparison to the POD method and the reference FOM solution for macroscopic quantities. As shown in the left panels, CoBRAS closely matches the FOM results, maintaining a maximum error below 1\% over the entire time frame for both moments. This demonstrates the superior accuracy of CoBRAS relative to POD, for which the error exceeds 10\%.

\subsubsection{Models for microscopic quantities}
\begin{figure}[htb!]
\centering
\begin{subfigure}[htb!]{0.35\textwidth}
    \caption*{\hspace{7mm}\small CoBRAS} \vskip 5pt
    \includegraphics[width=\textwidth]{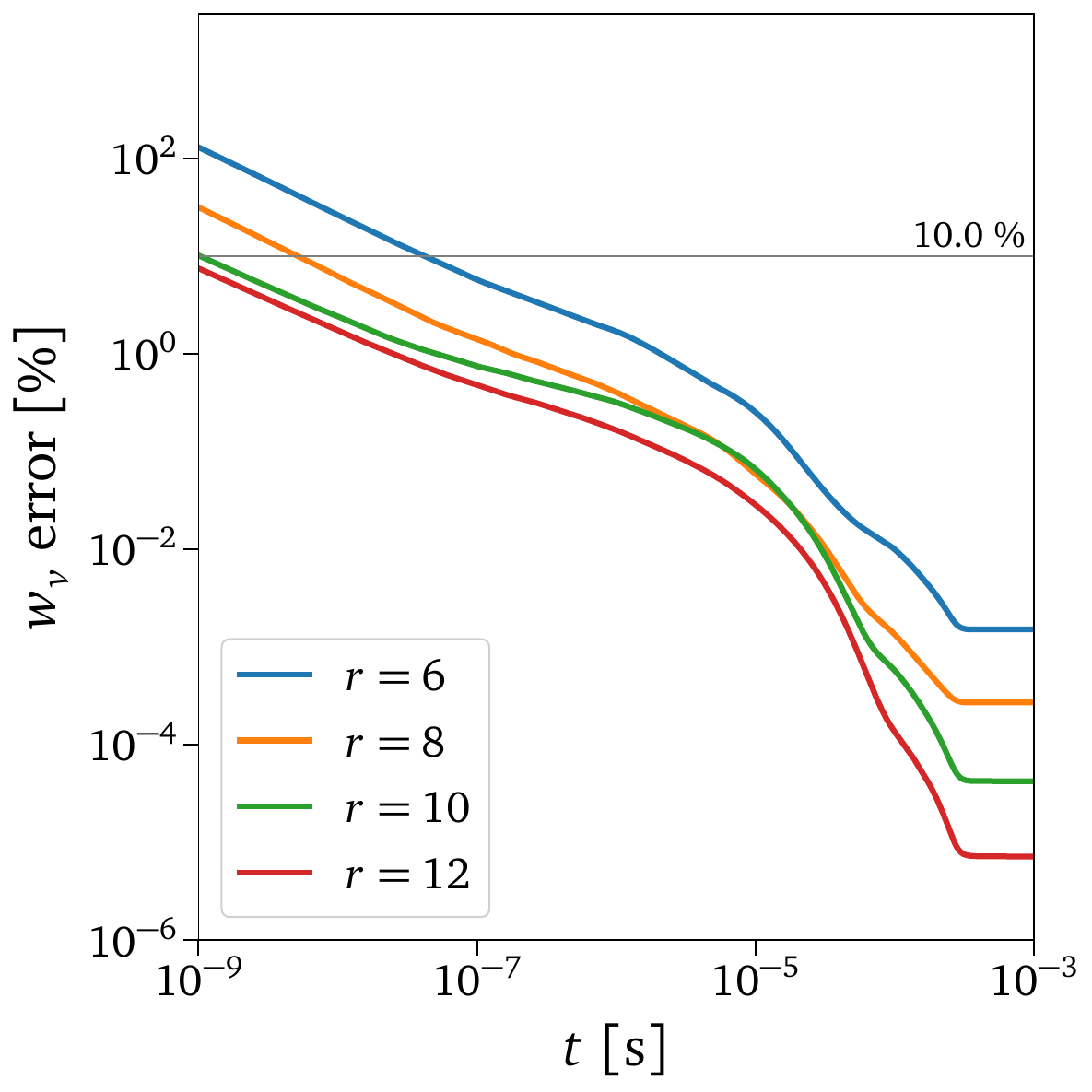}
\end{subfigure}
\quad
\begin{subfigure}[htb!]{0.35\textwidth}
    \caption*{\hspace{7mm}\small POD} \vskip 5pt
    \includegraphics[width=\textwidth]{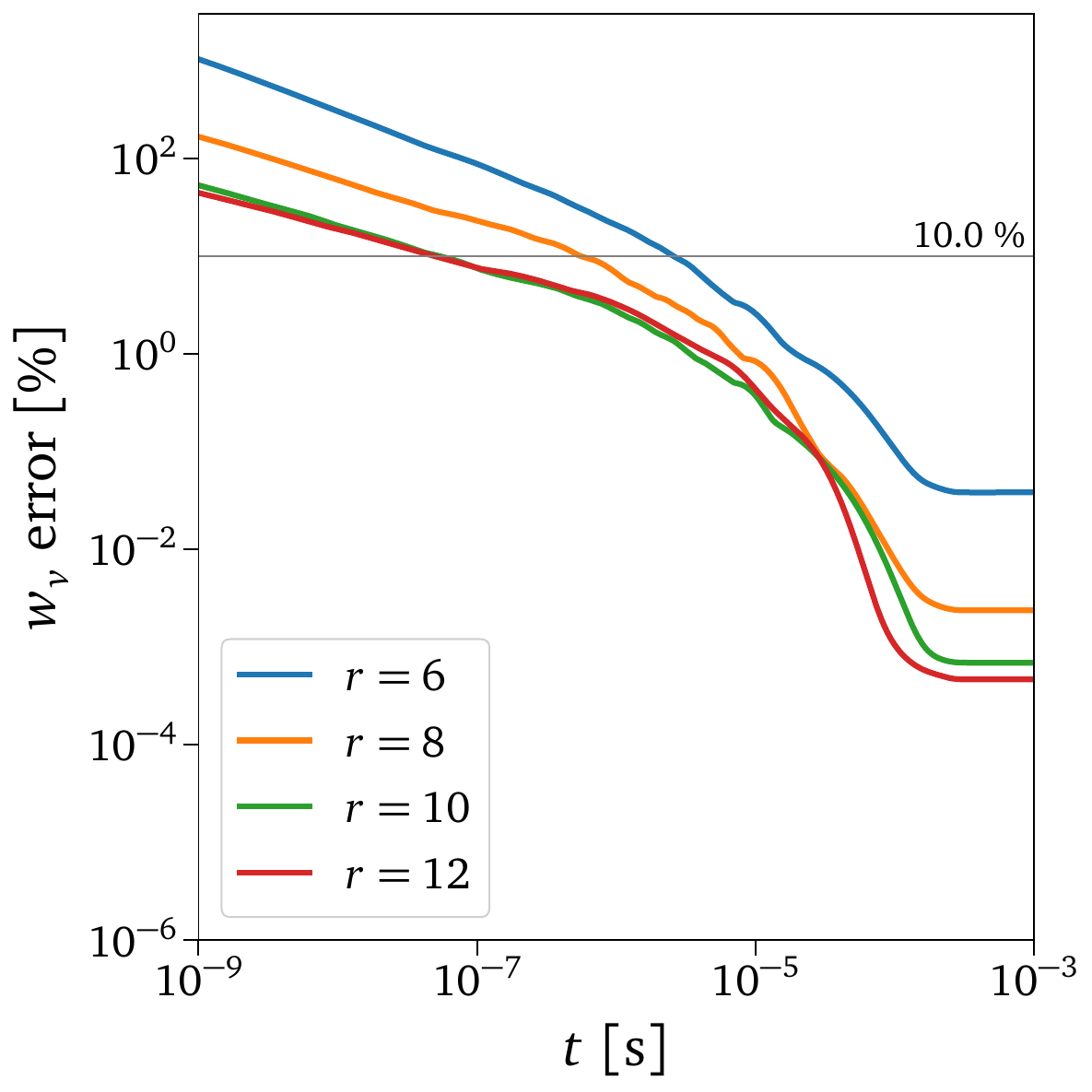}
\end{subfigure}
\caption{\textit{Mean relative error in the distribution function for VC model using CoBRAS and POD ROMs}. Time evolution of the mean relative error in the O$_2$ energy level population across 1\,000 testing trajectories. Results are shown for different dimensions $r$ of the CoBRAS (left panel) and POD (right panel) ROM systems.}
\label{fig:vc_mean_dist_err}
\end{figure}

Here, we seek models that can provide an accurate description of the \textit{whole} distribution along trajectories.
We follow the same rationale as in section \ref{sec:micro_1} and we choose $\vy\in\mathbb{R}^{10}$.
Figure \ref{fig:vc_mean_dist_err} illustrates the time evolution of the mean relative error between the FOM solution $\vw(t)$ and the predicted solutions $\hat\vw(t)$ obtained from the CoBRAS (left panel) and POD (right panel) ROMs for the O$_2$ energy level population. 
The comparison uses the same set of $M=1\,000$ trajectories as in the previous section, evaluating the ROMs for various reduced system dimensions $r$. 
This figure emphasizes the accuracy of the models over time and across different reduced dimensions in capturing microscopic quantities. 
As for the RVC system that we considered in the previous sections, the overall error remains below 10\% for most of the dynamics when employing CoBRAS with $r \geq 10$, corresponding to a compression factor of nearly 4.5. 
By contrast, POD-Galerkin consistently produces results with higher errors. 
As already noted in the RVC system, neither POD-Galerkin nor CoBRAS guarantee the positivity of the predicted state $\hat\vw$ and this can lead to non-physical negative values at very early times.
\par
\begin{figure}[htb!]
\centering
\begin{subfigure}[htb!]{0.35\textwidth}
    \includegraphics[width=\textwidth]{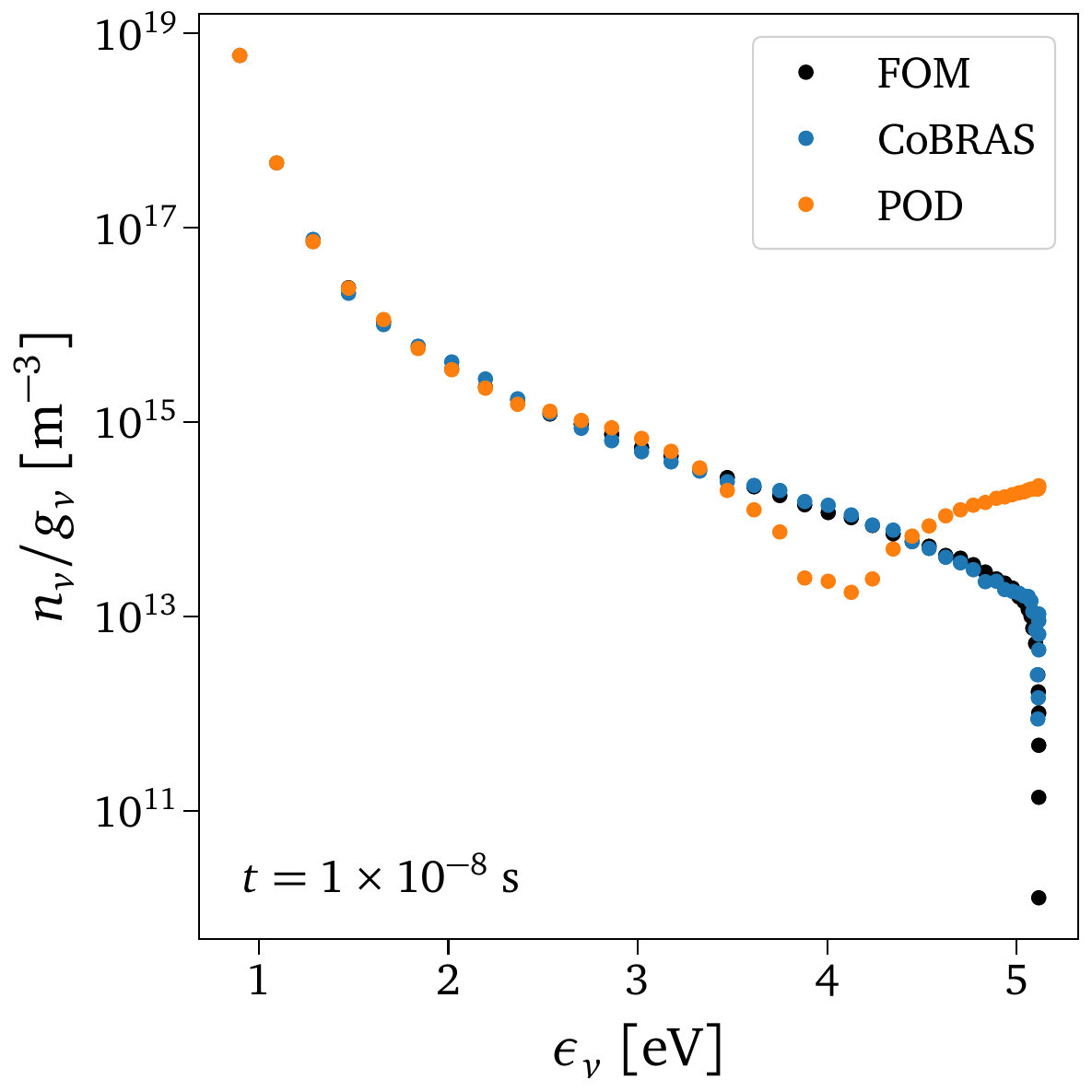}
\end{subfigure}
\quad
\begin{subfigure}[htb!]{0.35\textwidth}
    \includegraphics[width=\textwidth]{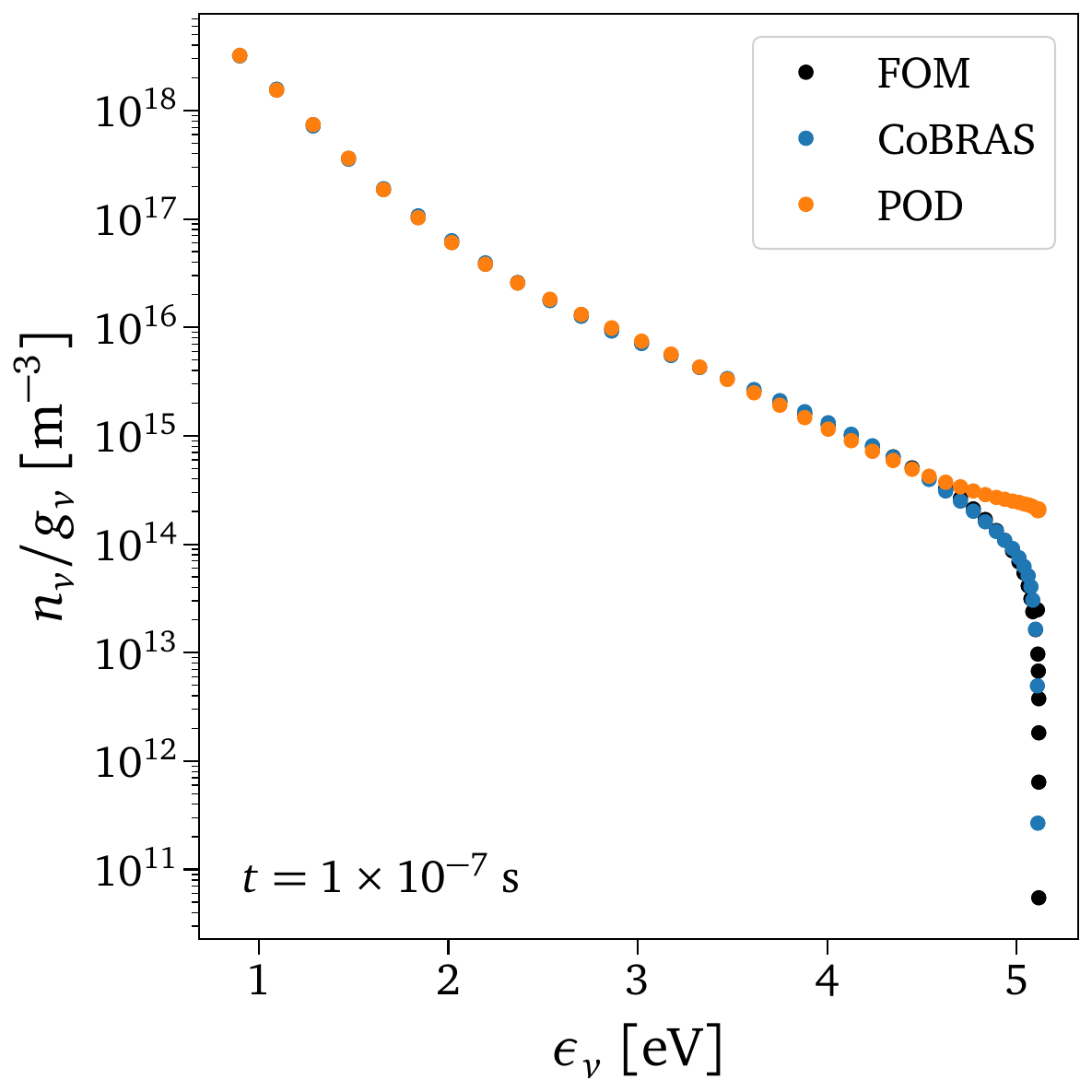}
\end{subfigure}
\\[5pt]
\begin{subfigure}[htb!]{0.35\textwidth}
    \includegraphics[width=\textwidth]{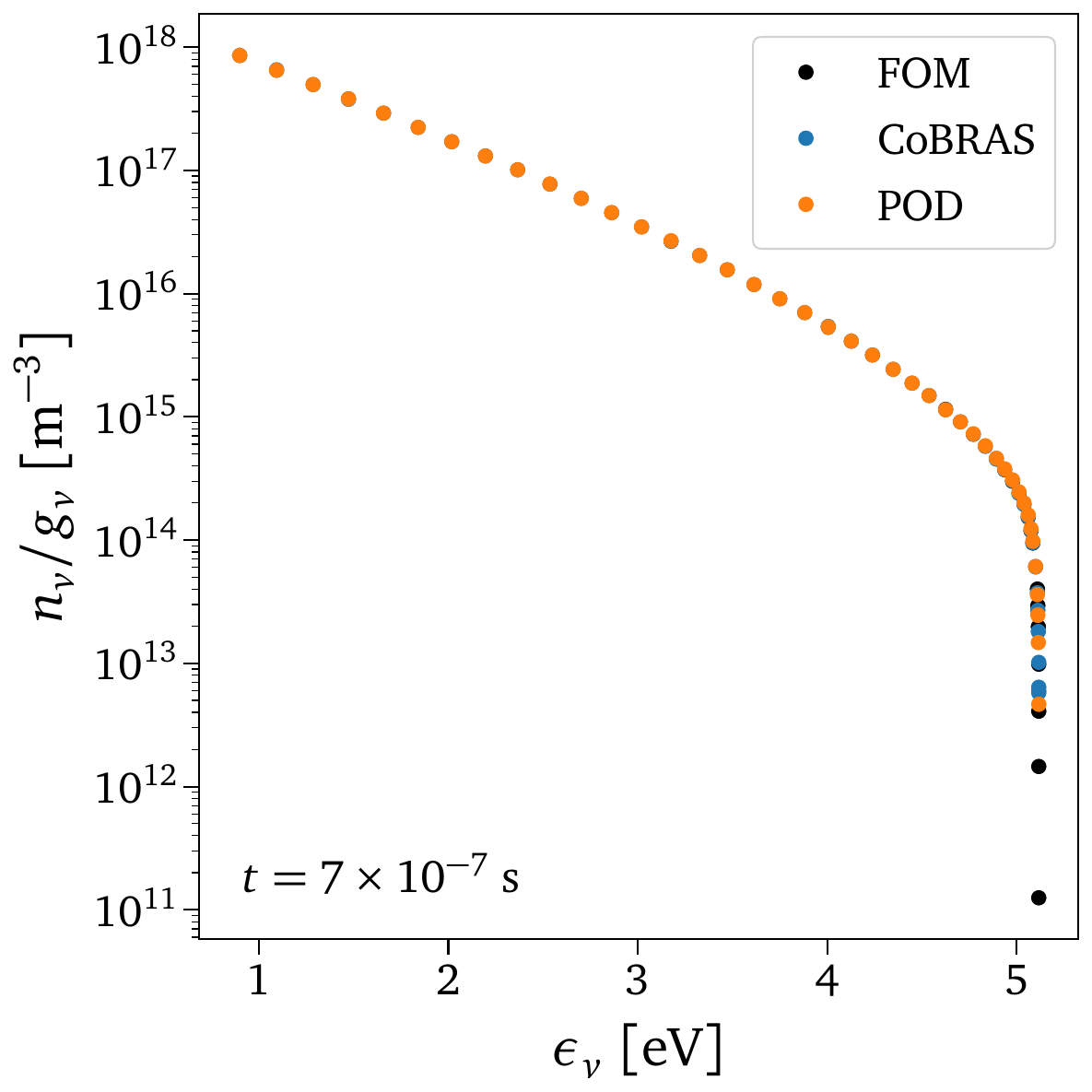}
\end{subfigure}
\quad
\begin{subfigure}[htb!]{0.35\textwidth}
    \includegraphics[width=\textwidth]{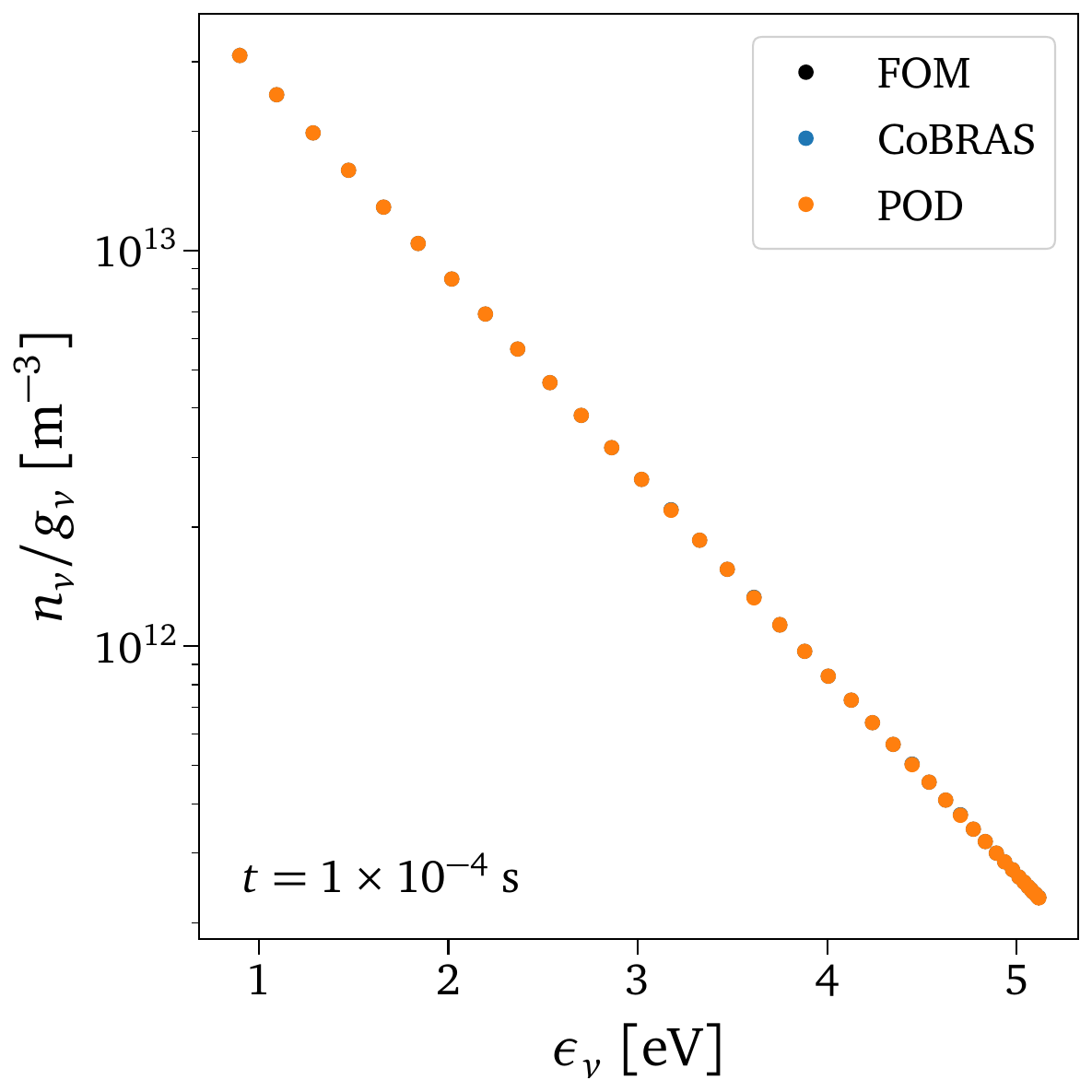}
\end{subfigure}
\caption{\textit{FOM vs. ROMs for VC model: distribution snapshots}. Time snapshots of the O$_2$ energy level population in an initially cold gas with $T=10\,000$ K, calculated using FOM, CoBRAS, and CG reduced order models. All ROMs have a dimension of $r=8$.}
\label{fig:vc_dist_cold}
\end{figure}
Figure \ref{fig:vc_dist_cold} displays four snapshots of the distribution function calculated using the FOM, along with CoBRAS and POD ROMs, with the reduced model dimension set to $r=8$. This case corresponds to the one presented in figure \ref{fig:vc_mom_cold}. For all snapshots, CoBRAS closely matches the FOM results, whereas POD exhibits significant discrepancies in the tails (i.e., high energy levels) of the distribution function. Notably, all ROMs successfully reconstruct the equilibrium distribution at $t=10^{-4}$ s.

\subsubsection{Computational performance}
Table \ref{table:vc.rom.costs} provides estimates of the total FLOPs required to solve both the FOM and the CoBRAS-based ROM across various dimensions, along with the stiffness of the corresponding systems. The results demonstrate that the ROM can achieve a reduction of over $10^2$ FLOPs for both RHS evaluations, primarily involving a tensor-matrix product with a complexity of \ordermagn{d^3} (see equations \eqref{eq:vc.fom.vec.ni} and \eqref{eq:vc.fom.vec.no} in the Supplementary Material), and for solving the linear system in the BDF scheme. Furthermore, the ROM produces a less stiff system of equations, allowing for larger integration time steps, which can be up to ten times greater than those required by the FOM.
\begin{table}[!htb]
    \centering
    \begin{tabular}{c|c|c|c|c}
        \toprule
        \multirow{2}*{Model} & Dimension & RHS Evaluation & BDF Scheme & Fastest Timescale \\
         & $d$ & \ordermagn{d^3} & \ordermagn{d^3} & $ 1/\lvert\lambda_{\min}\rvert$ \\
        \midrule
        FOM & 46 & $9.73\times 10^4$ & $9.73\times 10^4$ & $1.43\times 10^{-9}$ \\
        \arrayrulecolor{gray}\midrule
        \multirow{2}*{ROM-PG} & 9 & $7.29\times 10^2$ & $7.29\times 10^2$ & $5.39\times 10^{-9}$ \\
        & 5 & $1.25\times10^2$ & $1.25\times10^2$ & $2.27\times 10^{-8}$ \\
        \bottomrule
    \end{tabular}
    \caption{\textit{Comparison of computational cost: FOM vs. ROM-PG for the VC Model}. Total FLOPs for essential numerical operations and system stiffness estimates are provided for the FOM and various dimensions of the ROM-PG model.}
    \label{table:vc.rom.costs}
\end{table}

%% file: sections/5_conclusions.tex
\section{Conclusions}\label{sec:conclusions}
This work presents a data-driven Petrov-Galerkin reduced-order model (ROM) for detailed state-to-state kinetics in thermochemical nonequilibrium, avoiding the conventional assumptions often required in traditional models. 
In contrast to standard ROMs such as multi-temperature (MT) or coarse-grained (CG) models, which rely on empirical or physical assumptions, we sought a more flexible, assumption-free, and computationally efficient approach.
\par
To leverage the low-rank dynamics often observed in kinetic systems under thermal nonequilibrium, we used the recently-developed ``Covariance Balancing Reduction using Adjoint Snapshots'' (CoBRAS)~\cite{Otto2023ModelCovariance} technique as the foundation of our model reduction pipeline. 
This formulation allows for the computation of a linear projection operator for Petrov-Galerkin modeling by balancing the state and gradient covariances associated with the full-order system (the state-to-state kinetic equations in our specific case).
Our analysis revealed that these systems generally do not exhibit significant nonlinear behavior, allowing us to apply linearized equations around thermochemical equilibrium during the data-collection phase of CoBRAS. 
Experimental validation of our test cases confirmed that the linearized equations provided sufficiently accurate description of the subspace on which the nonlinear system evolves. 
While the linearization facilitated rapid data generation with minimal information loss, we acknowledge that we could also utilize the fully nonlinear system. 
However, this would lead to a much more computationally expensive training phase, with the computational cost seeing increases of two to four orders of magnitude.
The approach was applied to two distinct thermochemical systems: a pure oxygen rovibrational collisional (RVC) model and a vibrational collisional (VC) model. 
In both cases CoBRAS computed models with excellent predictive accuracy. 
In particular, these models exhibited relative errors of less than 1\% for macroscopic quantities (i.e., moments) and under 10\% for microscopic quantities (i.e., energy level population) along with excellent compression rates.
Additionally, our models significantly outperformed existing MT and CG models, as well as Galerkin model based on Proper Orthogonal Decomposition (POD).
\par
The proposed data-driven approach offers several advantages over traditional models, such as MT or CG. By relying solely on the structure of the underlying state-to-state governing equations, the proposed method eliminates the need for empirical or physical assumptions, resulting in enhanced accuracy and flexibility across various nonequilibrium conditions. Furthermore, it bypasses the time-consuming system analysis or optimization typically required to determine the optimal reduced representation, enabling the generation of the reduced model within a few minutes. Additionally, the proposed projection-based model avoids the computational costs and potential errors associated with interpolation needed for nonlinear models such as CG. 
For these reasons, we believe that this work lays the foundation for more efficient computational tools that preserve the fidelity of detailed state-to-state kinetics, while making them suitable for real-world applications in fields such as hypersonic flight, plasma dynamics, and combustion.
\par
While the proposed data-driven ROM shows great potential, further research is required to improve its performances. A significant limitation is that the current model does not preserve the positivity of the distribution function by design, which could result in negative values during the initial stages of the dynamics (e.g., $t < 10^{-8}$ s). 
Addressing this issue and testing the model in adiabatic systems will be priorities in future research to ensure the model’s applicability in real-world computational fluid dynamics applications.

%% file: sections/6_appendix.tex
\section{Proof of Proposition \ref{prop:adjoint}}\label{app:proof_of_prop}

\begin{proof}
    Given the exact solution 
\begin{equation}
    \vw(t) = \vw(t_0) + \int_{t_0}^t \vf(\vw(\tau);\vtheta)\,d\tau,
\end{equation}
to the ordinary differential equation \eqref{eq:fom.param}, 
it follows that a small perturbation $\delta \vw(t_0)$ induces a corresponding change $\delta \vw(t)$ according to
\begin{equation}
    \delta \vw(t) = \delta \vw(t_0) + \int_{t_0}^t D\vf(\vw(\tau);\vtheta)\delta \vw(\tau)\,d\tau \coloneqq \mathcal{L}(t,t_0) \delta\vw(t_0).
\end{equation}
Here, $\mathcal{L}(t,t_0):\mathbb{R}^N \to \mathbb{R}^N$  is the state transition matrix, which satisfies the linear time-varying dynamics
\begin{equation}
    \frac{d \mathcal{L}(t,t_0)}{d t} = D\vf(\vw(t);\vtheta)\mathcal{L}(t,t_0).
\end{equation}
Assuming an Euclidean inner product on $\mathbb{R}^N$, the adjoint $\mathcal{L}^\intercal(t_0,t)$ satisfies the adjoint equation
\begin{equation}
    \label{eq:adj_state_transition}
    -\frac{d \mathcal{L}^\intercal(t_0,t)}{d t} = D\vf(\vw(t);\vtheta)^\intercal \mathcal{L}^\intercal(t_0,t).
\end{equation}
Now, recalling that $y_j(t) =  \ve_j^\intercal\mC \vw(t)$, where $\ve_j$ is the $j$-th unit vector in the standard basis of $\mathbb{R}^{\mathrm{dim}(\vy)}$, we have 
\begin{equation}
    \delta y_j(t_k) = \ve_j^\intercal\mC \delta \vw(t_k) = \ve_j^\intercal \mC \mathcal{L}(t_k,t_0)\delta \vw(t_0) \coloneqq \langle \vg^{(k)}_j,\delta \vw(t_0)\rangle
\end{equation}
where $\vg^{(k)}_j$ is the gradient of $y_j(t_k)$ with respect to $\vw(t_0)$, and the last equality follows from the definition of the gradient.
Then, it follows immediately that
\begin{equation}
    \vg^{(k)}_j = \mathcal{L}^\intercal(t_0,t_k)\mC^\intercal \ve_j.
\end{equation}
Using \eqref{eq:adj_state_transition} we recover \eqref{eq:adjoint} and this concludes the proof.
\end{proof}

\section{Covariance matrices}\label{app:covmat}
The probability density functions used to compute the covariance matrices in equations \eqref{eq:state_cov} and \eqref{eq:grad_cov} are defined as follows
\begin{align}
    f_{\mathrm{T}_0}(t_0) & = \frac{1}{b_{t_0}-a_{t_0}}
    \eqspace , \\
    f_{\mTheta}(\vtheta) = f_{\mTheta}(T,\rho) & = \frac{1}{b_{T}-a_{T}}\frac{1}{b_{\rho}-a_{\rho}}
    \eqspace , \\
    f_{\mMu}(\vmu) = f_{\mMu}(w_{\atom_0}, T_0) 
        & = \frac{1}{b_{w_{\atom_0}}-a_{w_{\atom_0}}} \frac{1}{T_0\left[\ln\left(b_{T_0}\right)-\ln\left(a_{T_0}\right)\right]}
    \eqspace .
\end{align}
Here, $a_{(\cdot)}$ and $b_{(\cdot)}$ represent the distribution bounds provided in tables \ref{table:rvc.mu_space} and \ref{table:vc.mu_space} for $f_{\mTheta}$ and $f_{\mMu}$, while for $f_{\mathrm{T}_0}$, the bounds are $a_{t_0}=0$ s and $b_{t_0}=10^{-2}$ s.

%% file: sections/7_supplementary.tex
\section{O$_2$-O rovibrational collisional model}\label{suppl:rvc}
\subsection{Linearized FOM equations}\label{suppl:rvc.linfom}
Equations \eqref{eq:rvc.fom.ni} and \eqref{eq:rvc.fom.no} can be written in a compact matrix-vector form as follows
\begin{align}
    \frac{d\mathbf{n}}{dt}
        = & \;\left(
            {}^{\mathrm{a}}\mathbf{K}^{\mathrm{e}} - {}^{\mathrm{a}}\mathbf{K}^{\mathrm{d}}
        \right)\mathbf{n}n_\atom + {}^{\mathrm{a}}\mathbf{k}^\text{r}n^3_\atom \\
        = & \;\mathbf{A}_1\mathbf{n}n_\atom + \mathbf{b}_1n^3_\atom \eqspace, \label{eq:rvc.fom.vec.ni} \\
    \frac{dn_\atom}{dt} = & -\mathbf{2}^\intercal\frac{d\mathbf{n}}{dt}\label{eq:rvc.fom.vec.no}
    \eqspace,
\end{align}
where $\mathbf{A}_1\in\mathbb{R}^{(N-1)\times (N-1)}$ and $\mathbf{b}_1\in\mathbb{R}^{N-1}$. After substituting the following identities
\begin{align}
    \mathbf{n} & = \bvn + \tvn \eqspace , \label{eq:lin_ni} \\
    n_\atom & = \overline{n}_\atom + n^\prime_\atom \eqspace , \label{eq:lin_no}
\end{align}
into equations \eqref{eq:rvc.fom.vec.ni} and \eqref{eq:rvc.fom.vec.no} and removing terms that give zero rate of change at equilibrium, we get the following equations
\begin{align}
    \frac{d\tvn}{dt}
        = & \;\mathbf{A}_1\overline{n}_\atom\tvn
        + \left(
            \mathbf{A}_1\bbgam + 3\mathbf{b}_1
        \right)\overline{n}_\atom^2n^\prime_\atom + \text{h.o.t.} \\
        = & \;\mathbf{A} \tvn + \mathbf{b}n^\prime_\atom + \text{h.o.t.} \eqspace , \label{eq:ma.lin_fom.vec.ni} \\
    \frac{dn^\prime_\atom}{dt} =
        & -\mathbf{2}^\intercal \frac{d\tvn}{dt} \eqspace , \label{eq:ma.lin_fom.vec.no}
\end{align}
where
\begin{equation}
    \mathbf{A} = \mathbf{A}_1\overline{n}_\atom \eqspace,
    \quad
    \mathbf{b} = \left(
        \mathbf{A}_1\bbgam + 3\mathbf{b}_1
    \right)\overline{n}_\atom^2\eqspace.
\end{equation}
Here $\bbgam=\overline{\mathbf{n}}/\overline{n}_\atom^2= \vq/q_\atom^2$ is constant for a given temperature $T$, with $\vq\in\mathbb{R}^{N-1}$ and $q_\atom\in\mathbb{R}^1$ representing the total partition functions (including the zero-point energy) for the O$_2$ rovibrational energy levels and the O ground state, respectively. By combining equations \eqref{eq:ma.lin_fom.vec.ni} and \eqref{eq:ma.lin_fom.vec.no} and neglecting higher-order terms (h.o.t.), we obtain the fully linearized system for equations \eqref{eq:rvc.fom.vec.ni} and \eqref{eq:rvc.fom.vec.no}
\begin{equation}\label{eq:rvc.lin}
    \frac{d}{dt}
    \begin{bmatrix}
        \tvn \\
        n^\prime_\atom
    \end{bmatrix} \approx 
    \underbrace{\begin{bmatrix}
        \mA & \vb \\ 
        -\mathbf{2}^\intercal\mA & -\mathbf{2}^\intercal\vb
    \end{bmatrix}}_{\mA_{\vtheta}}
    \begin{bmatrix}
        \tvn \\
        n^\prime_\atom
    \end{bmatrix} \eqspace .
\end{equation}
We then apply the linear transformation \eqref{eq:n_to_w} to equation \eqref{eq:rvc.lin} to convert number densities to mass fractions.

\subsection{ROM equations}\label{suppl:rvc.rom}
After applying the method described in sections \ref{sec:rom:pg} to \ref{sec:rom:covmat}, we derive the following reduced system for equations \eqref{eq:rvc.fom.vec.ni} and \eqref{eq:rvc.fom.vec.no}
\begin{align}
    \frac{d\mathbf{z}}{dt}
        = & \;\mathbf{A}_{1r}\mathbf{z}n_\atom
        + \mathbf{b}_{1r}n^3_\atom \eqspace, \\
    \frac{dn_\atom}{dt} = & -\mathbf{m}_r^\intercal \frac{d\mathbf{z}}{dt} \eqspace,
\end{align}
where the compressed tensors can be precomputed as follows
\begin{align}
    \mathbf{A}_{1r} & = \psib^\intercal\mathbf{A}_1\phib \eqspace, \\
    \mathbf{b}_{1r} & = \psib^\intercal\mathbf{b}_1 \eqspace, \\
    \mathbf{m}_r^\intercal & = \mathbf{2}^\intercal\phib \eqspace.
\end{align}

\subsection{Maximum moment $m=1$}\label{suppl:rvc.mom}
\begin{figure}[H]
\centering
\begin{subfigure}[htb!]{0.23\textwidth}
    \caption*{\hspace{5mm}$T=6\,000$ K}
    \includegraphics[width=\textwidth]{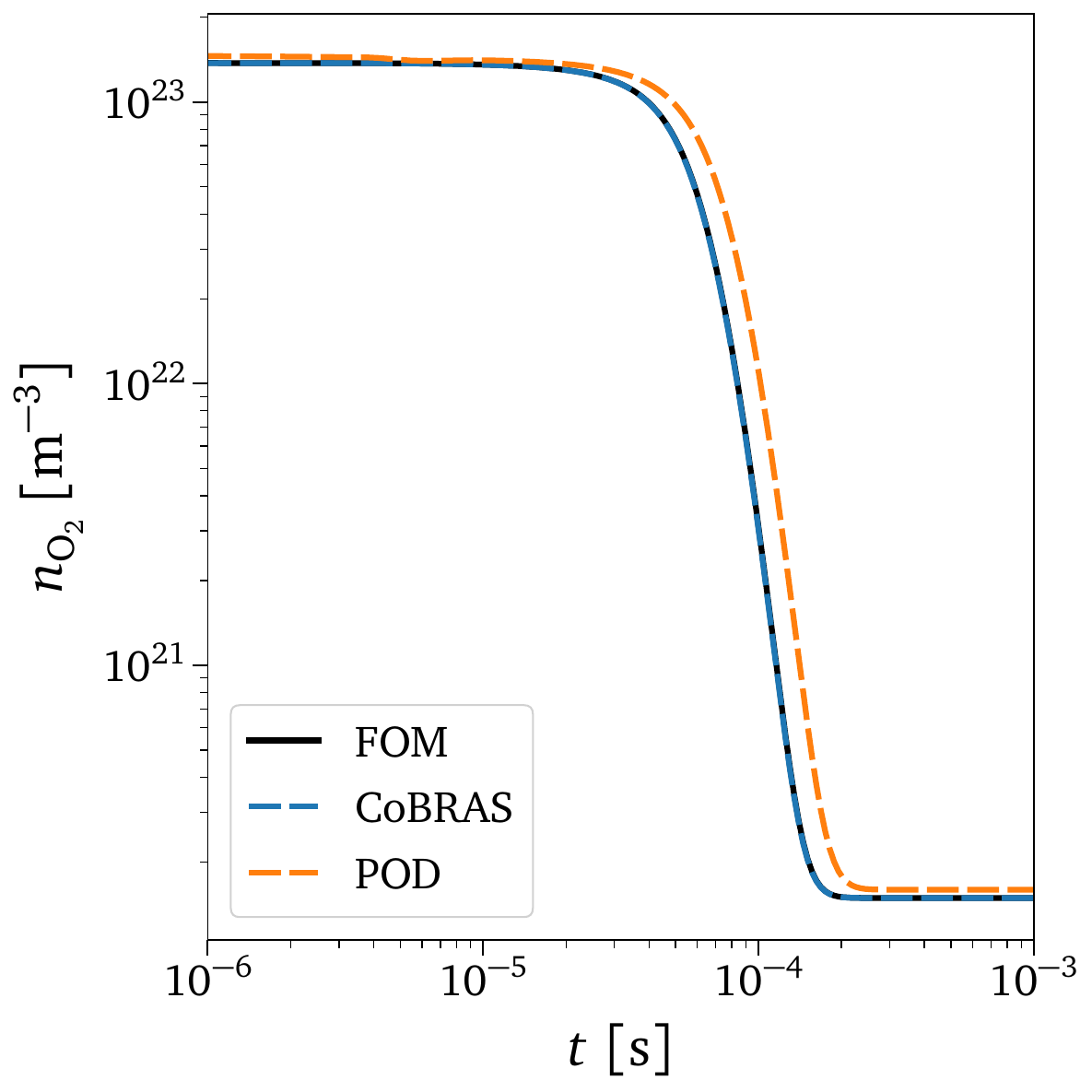}
\end{subfigure}
\begin{subfigure}[htb!]{0.23\textwidth}
    \caption*{\hspace{5mm}$T=8\,000$ K}
    \includegraphics[width=\textwidth]{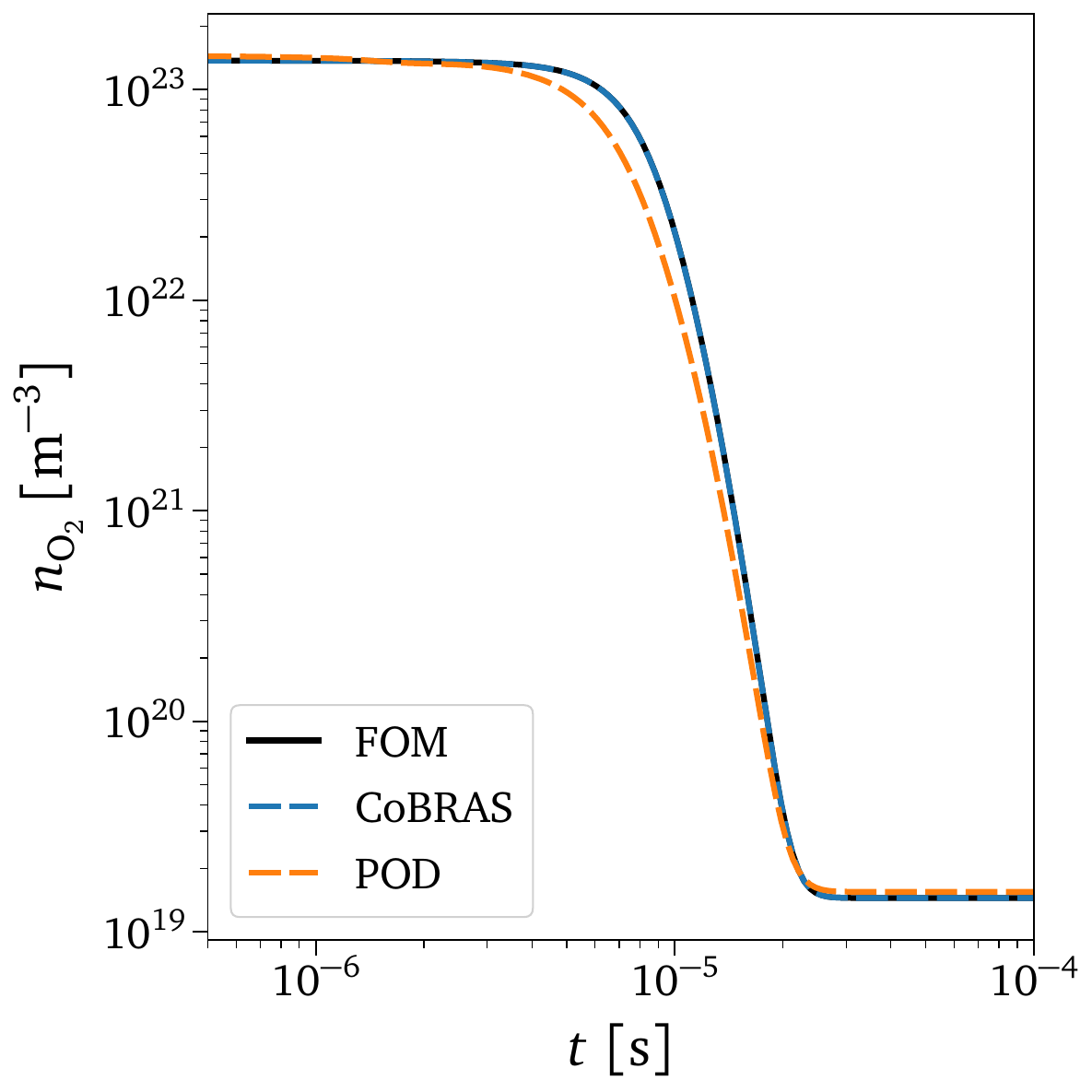}
\end{subfigure}
\begin{subfigure}[htb!]{0.23\textwidth}
    \caption*{\hspace{5mm}$T=12\,000$ K}
    \includegraphics[width=\textwidth]{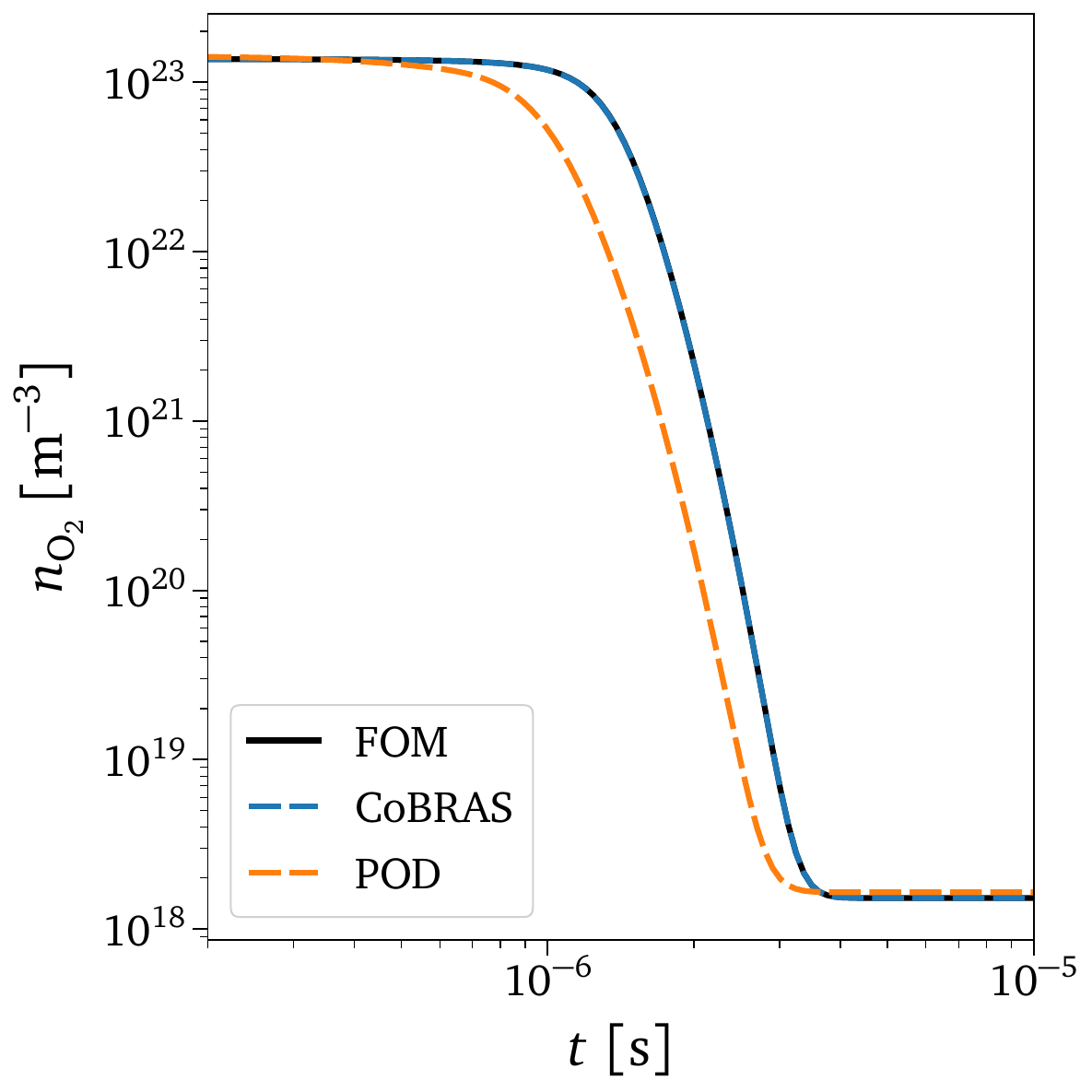}
\end{subfigure}
\begin{subfigure}[htb!]{0.23\textwidth}
    \caption*{\hspace{5mm}$T=14\,000$ K}
    \includegraphics[width=\textwidth]{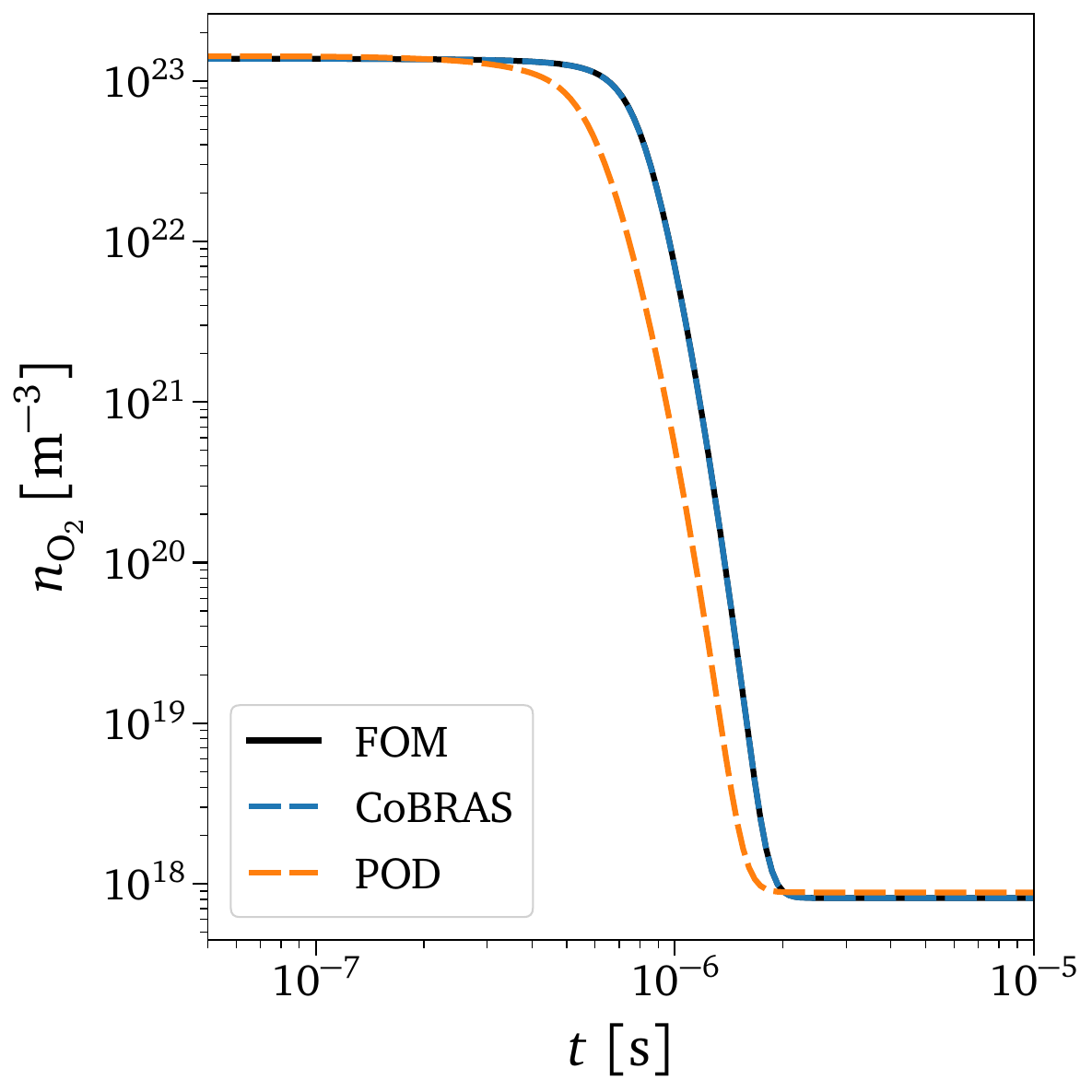}
\end{subfigure}
\\[5pt]
\begin{subfigure}[htb!]{0.23\textwidth}
    \includegraphics[width=\textwidth]{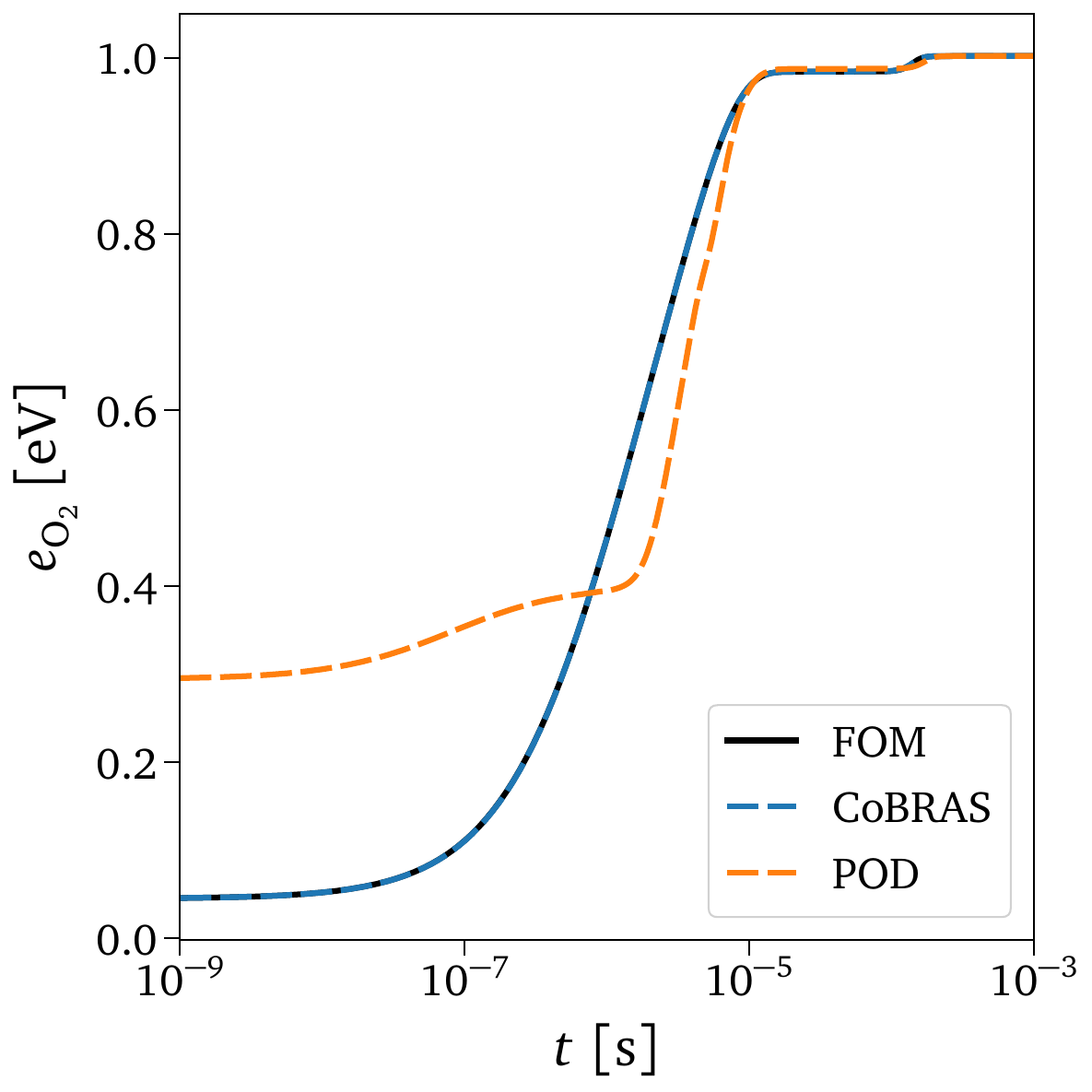}
\end{subfigure}
\begin{subfigure}[htb!]{0.23\textwidth}
    \includegraphics[width=\textwidth]{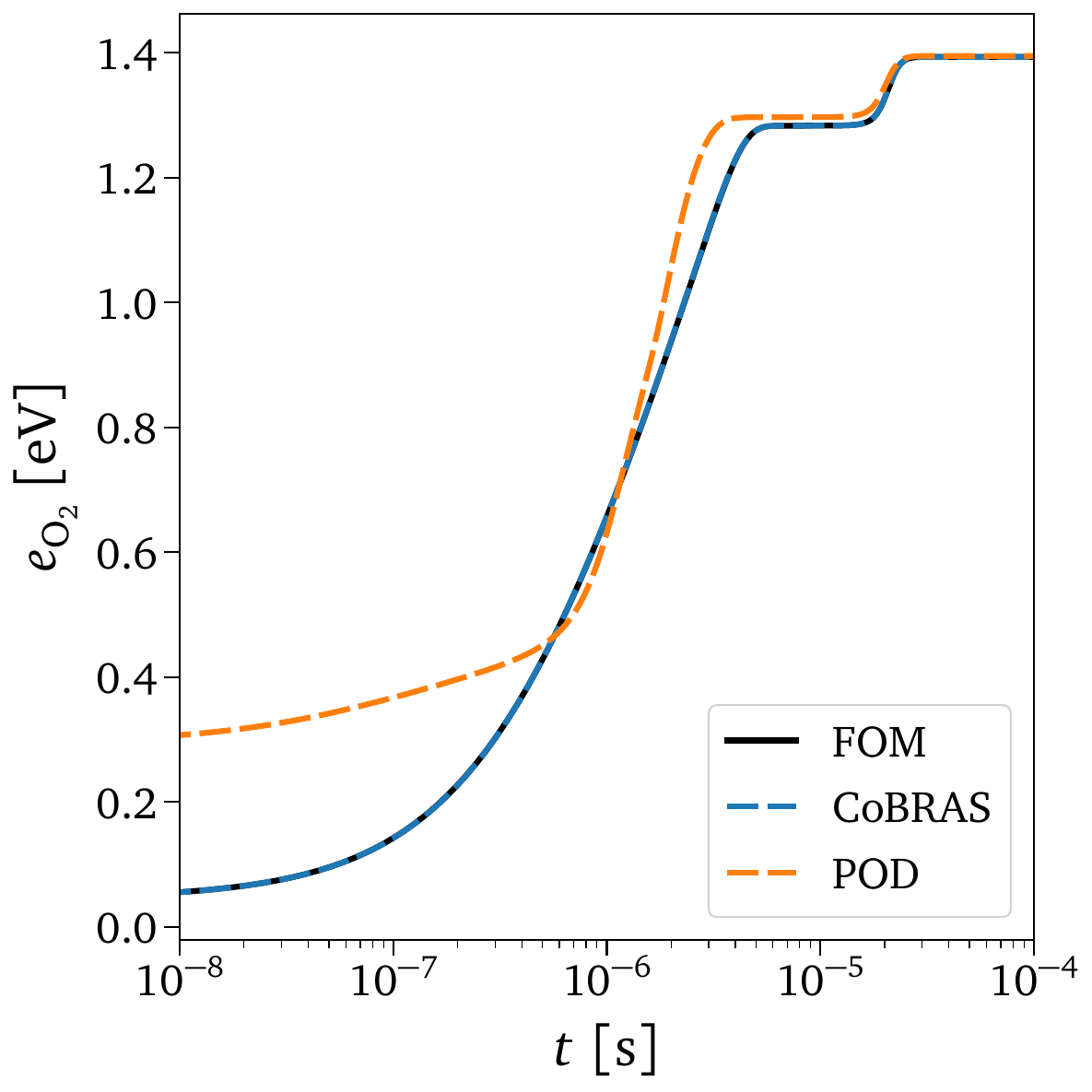}
\end{subfigure}
\begin{subfigure}[htb!]{0.23\textwidth}
    \includegraphics[width=\textwidth]{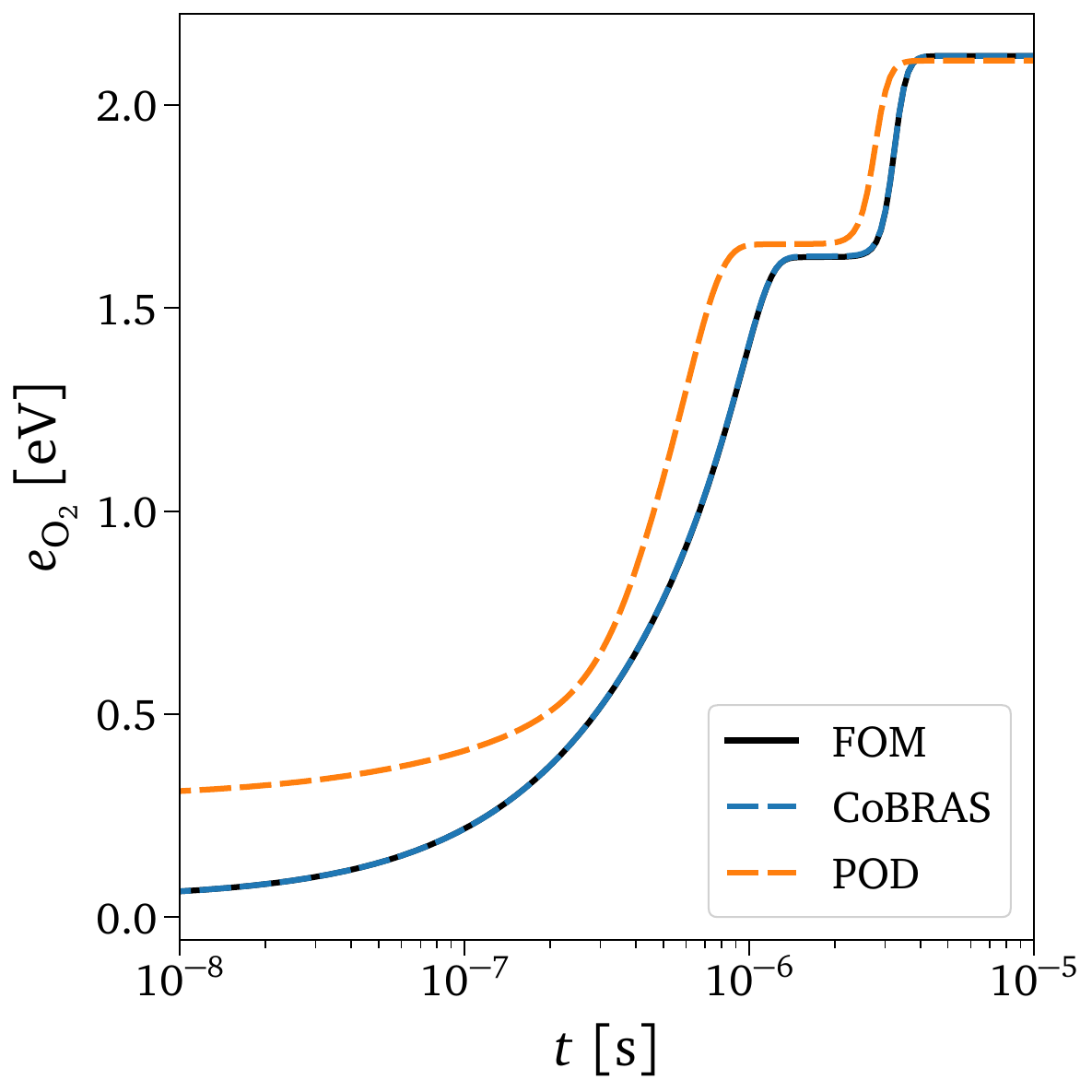}
\end{subfigure}
\begin{subfigure}[htb!]{0.23\textwidth}
    \includegraphics[width=\textwidth]{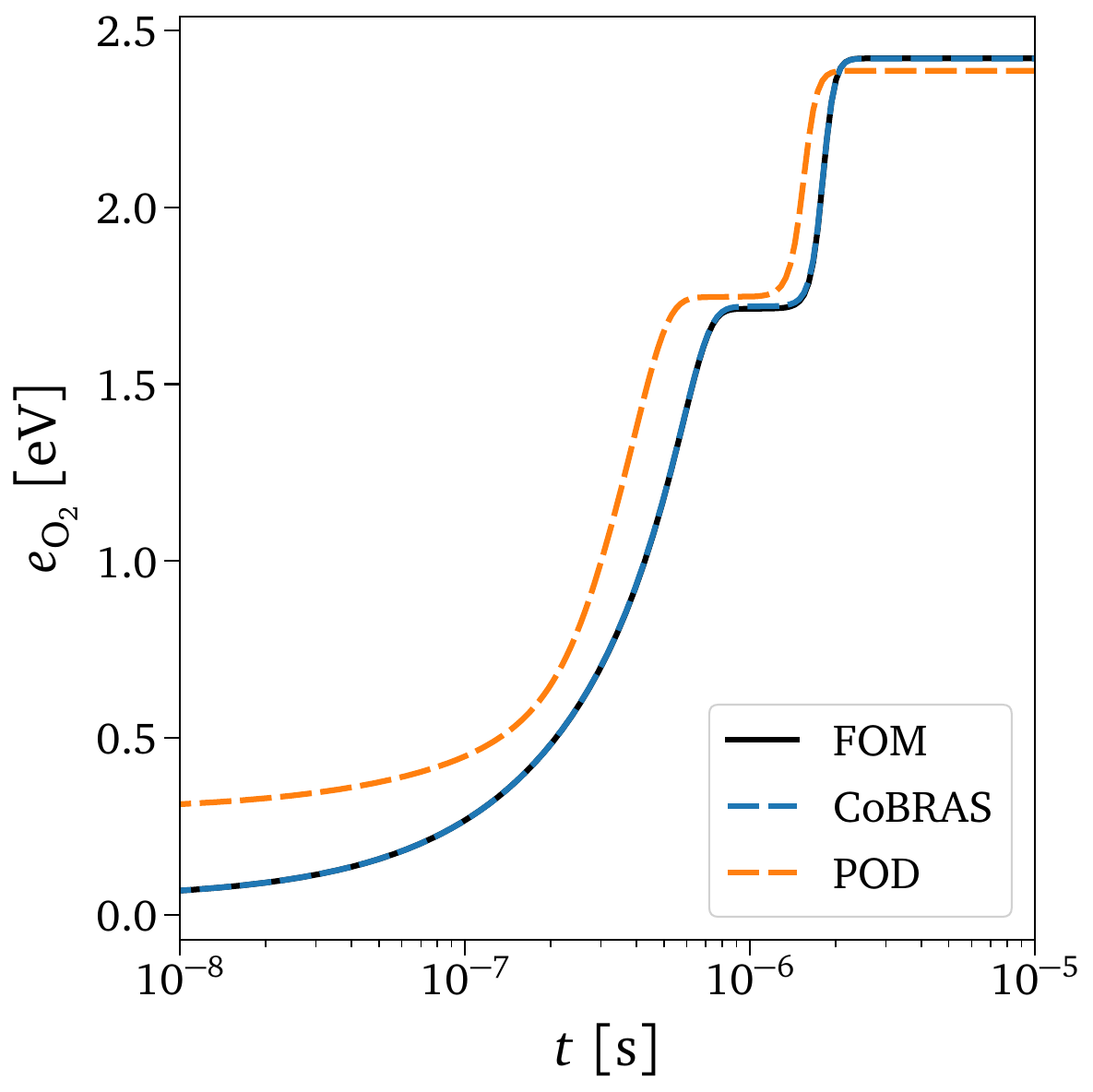}
\end{subfigure}
\caption{\textit{FOM vs. ROMs for RVC model: moments evolution}. Time evolution of the zeroth-order (top row) and first-order (bottom row) moments of the O$_2$ energy level population in an initially cold gas at different system temperatures, computed using the FOM and various ROMs with a reduced dimension of $r = 8$.}
\label{fig:rvc_mom_cold_temps}
\end{figure}

\subsection{Maximum moment $m=9$}\label{suppl:rvc.dist}
\begin{figure}[H]
\centering
\begin{subfigure}[htb!]{0.23\textwidth}
    \caption*{\hspace{5mm}$T=6\,000$ K}
    \includegraphics[width=\textwidth]{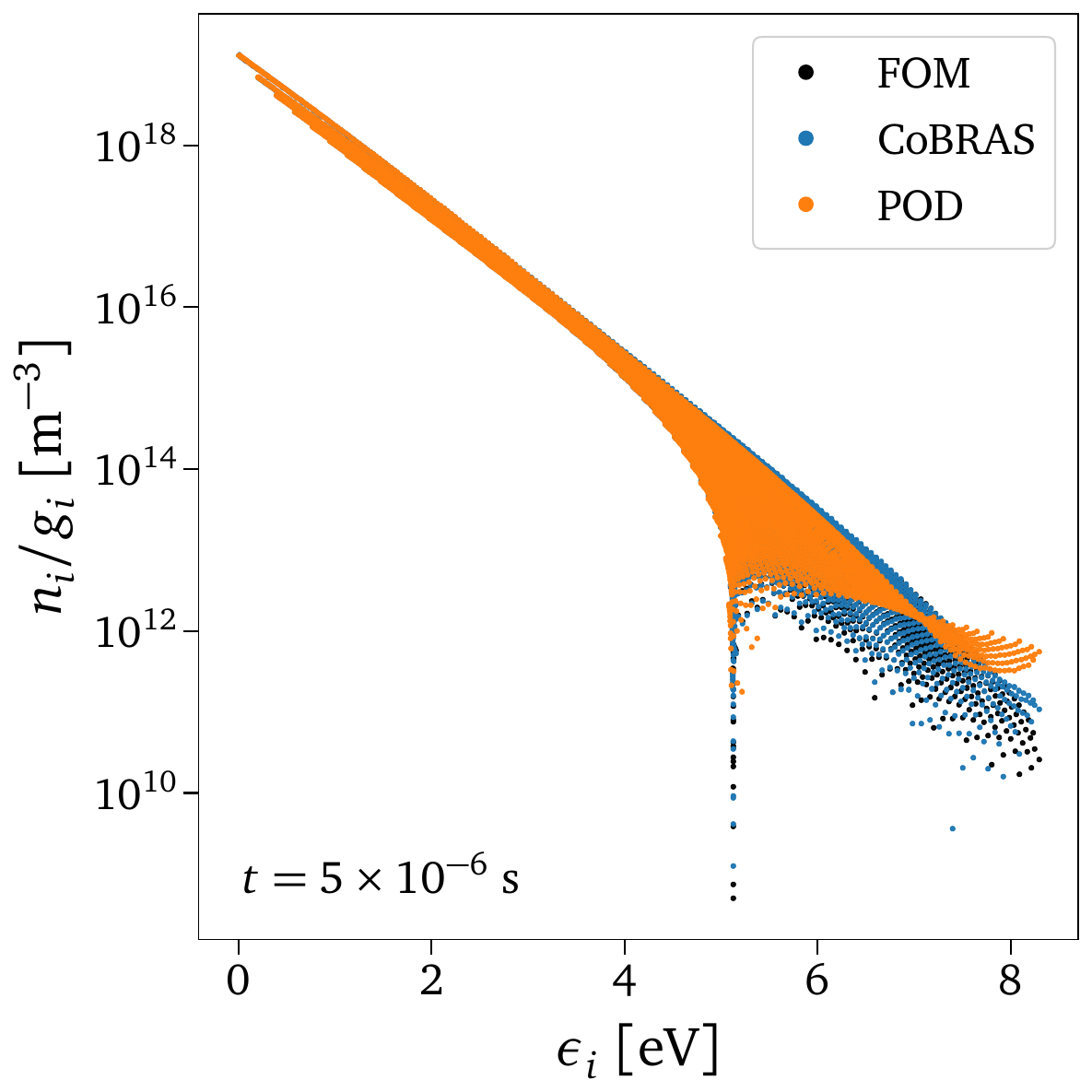}
\end{subfigure}
\begin{subfigure}[htb!]{0.23\textwidth}
    \caption*{\hspace{5mm}$T=8\,000$ K}
    \includegraphics[width=\textwidth]{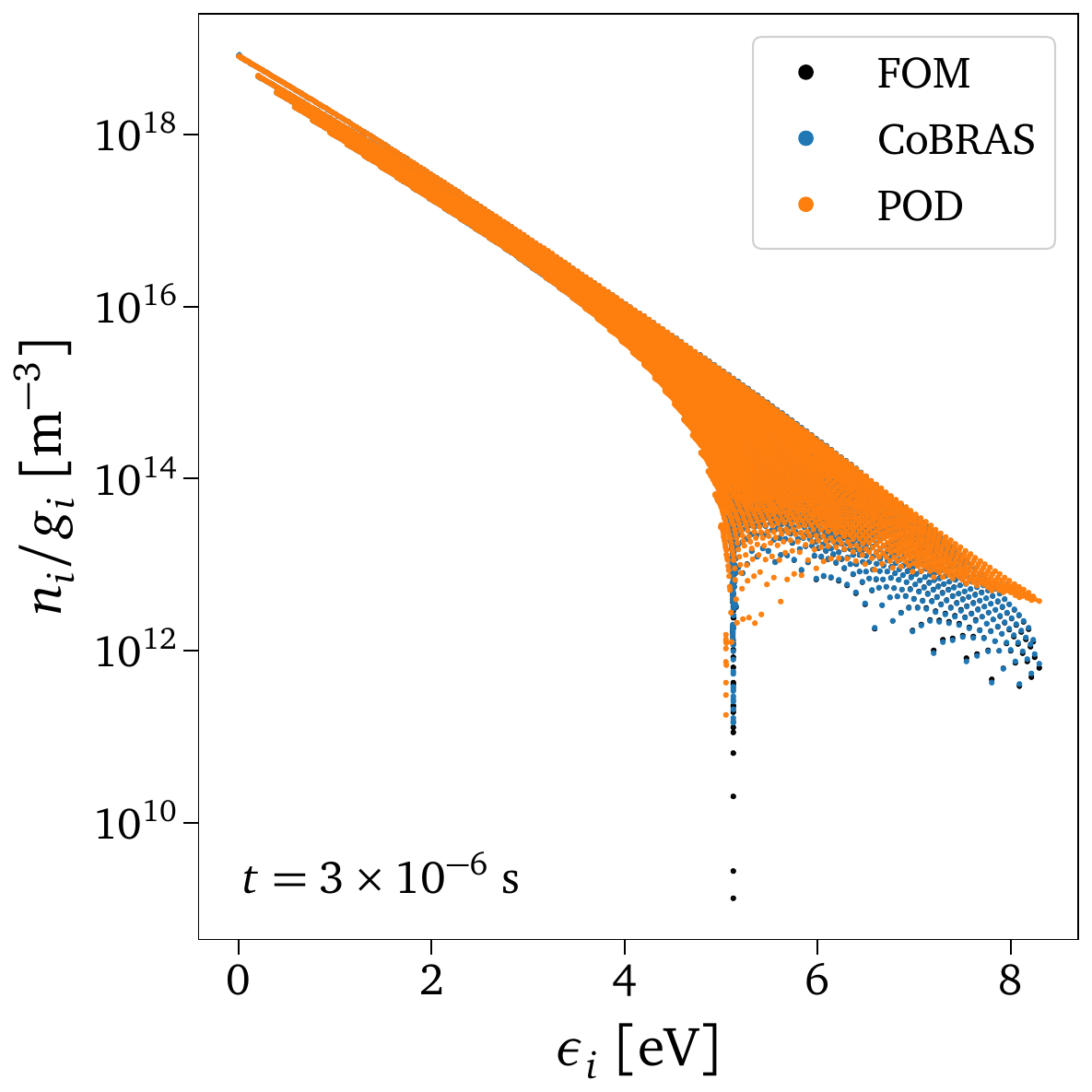}
\end{subfigure}
\begin{subfigure}[htb!]{0.23\textwidth}
    \caption*{\hspace{5mm}$T=12\,000$ K}
    \includegraphics[width=\textwidth]{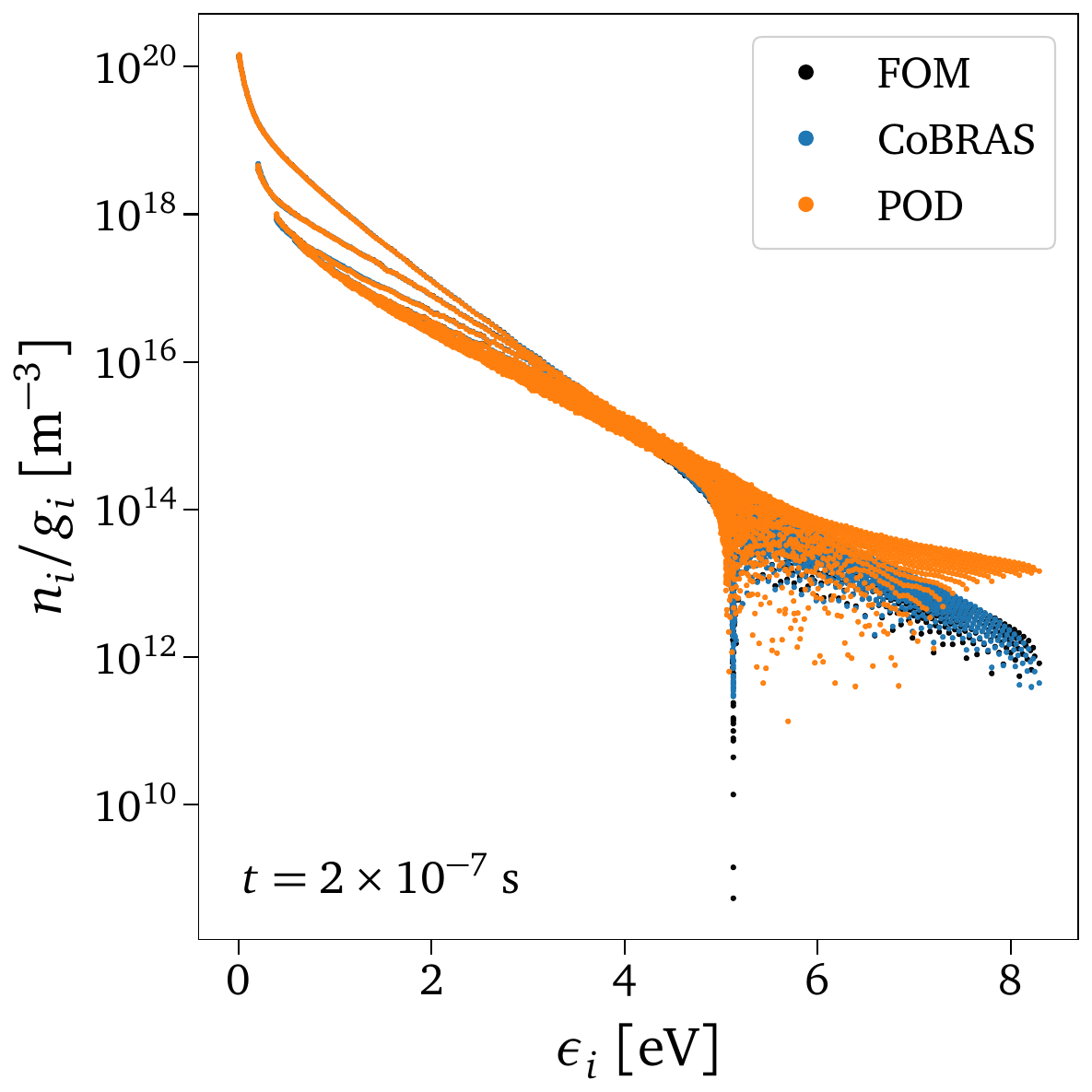}
\end{subfigure}
\begin{subfigure}[htb!]{0.23\textwidth}
    \caption*{\hspace{5mm}$T=14\,000$ K}
    \includegraphics[width=\textwidth]{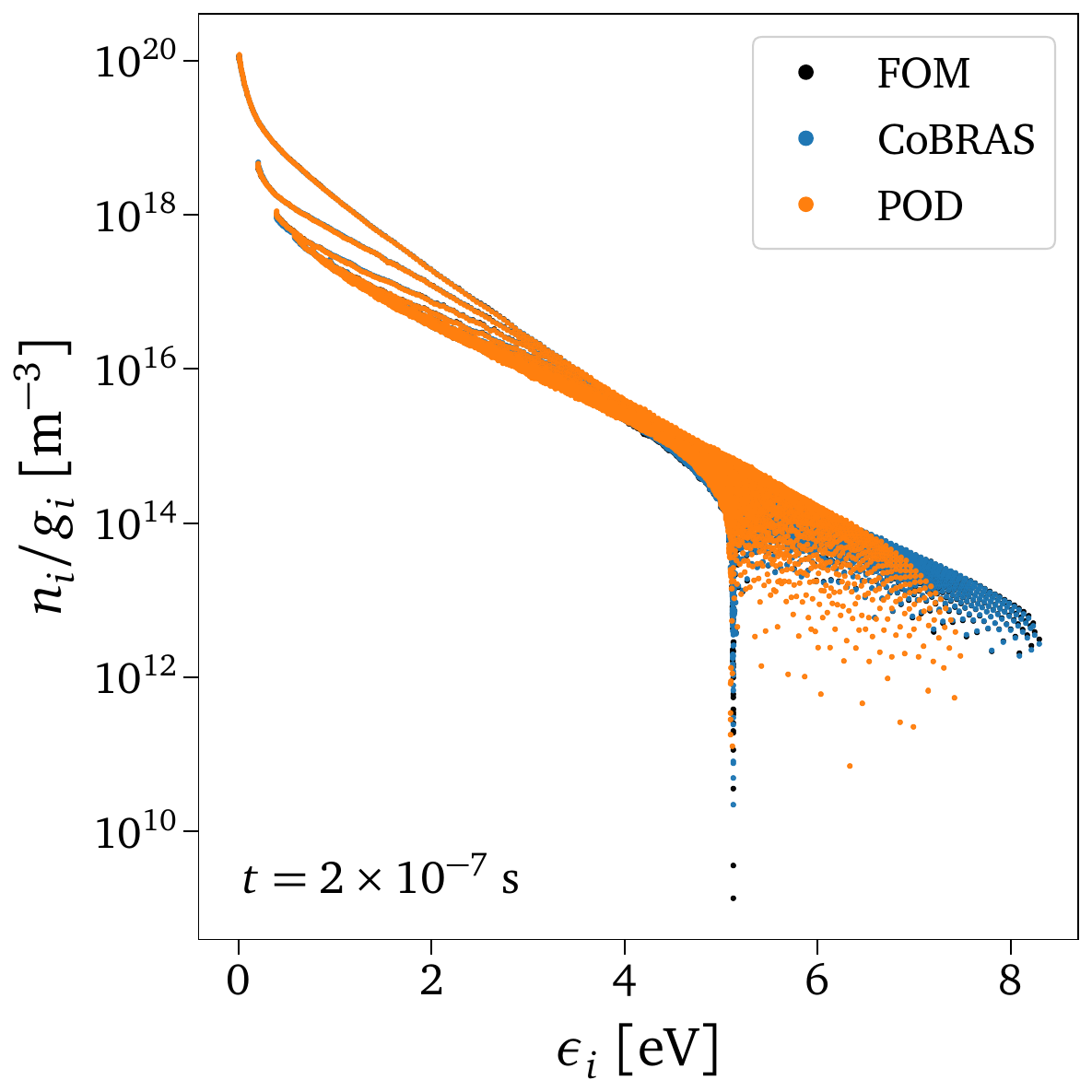}
\end{subfigure}
\\[5pt]
\begin{subfigure}[htb!]{0.23\textwidth}
    \includegraphics[width=\textwidth]{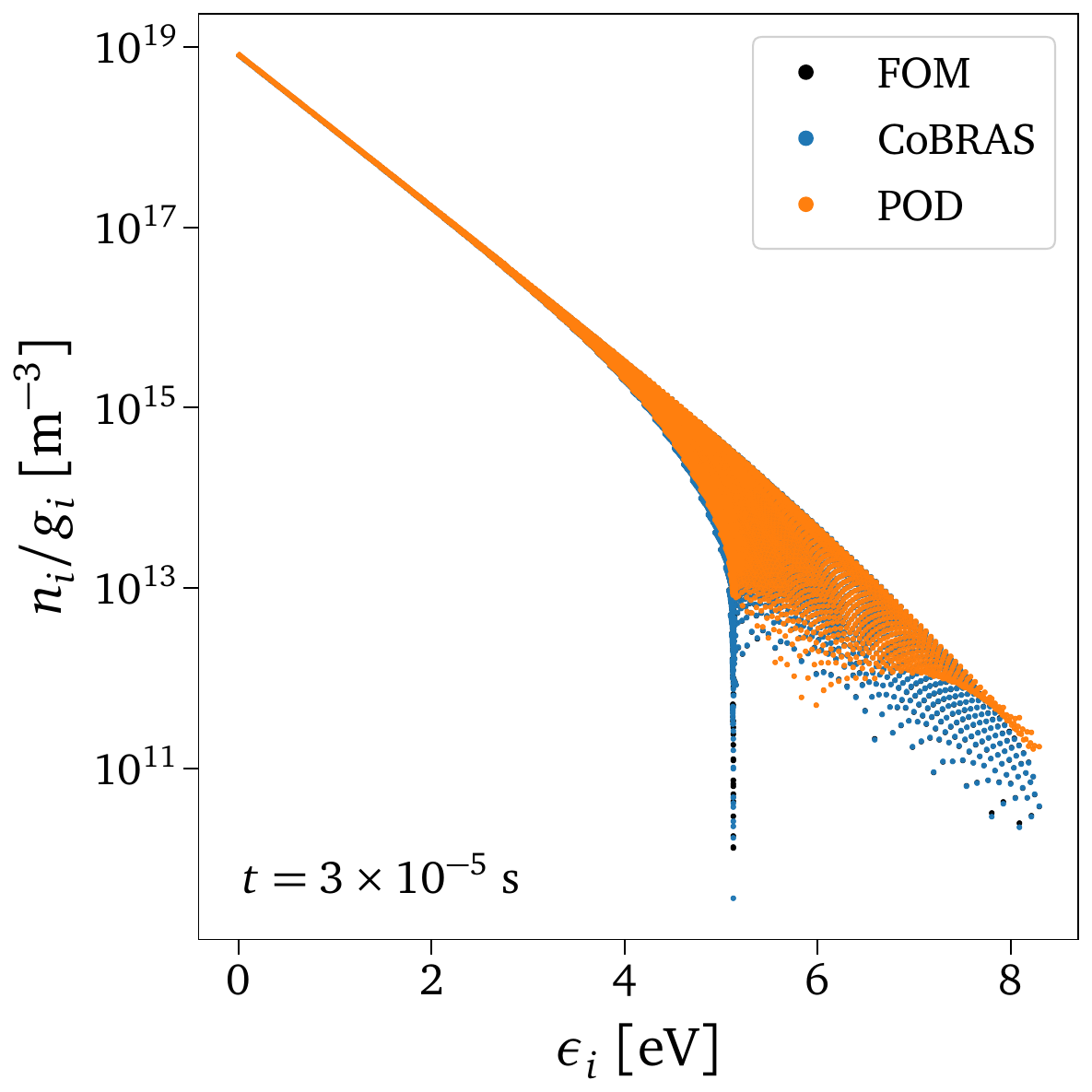}
\end{subfigure}
\begin{subfigure}[htb!]{0.23\textwidth}
    \includegraphics[width=\textwidth]{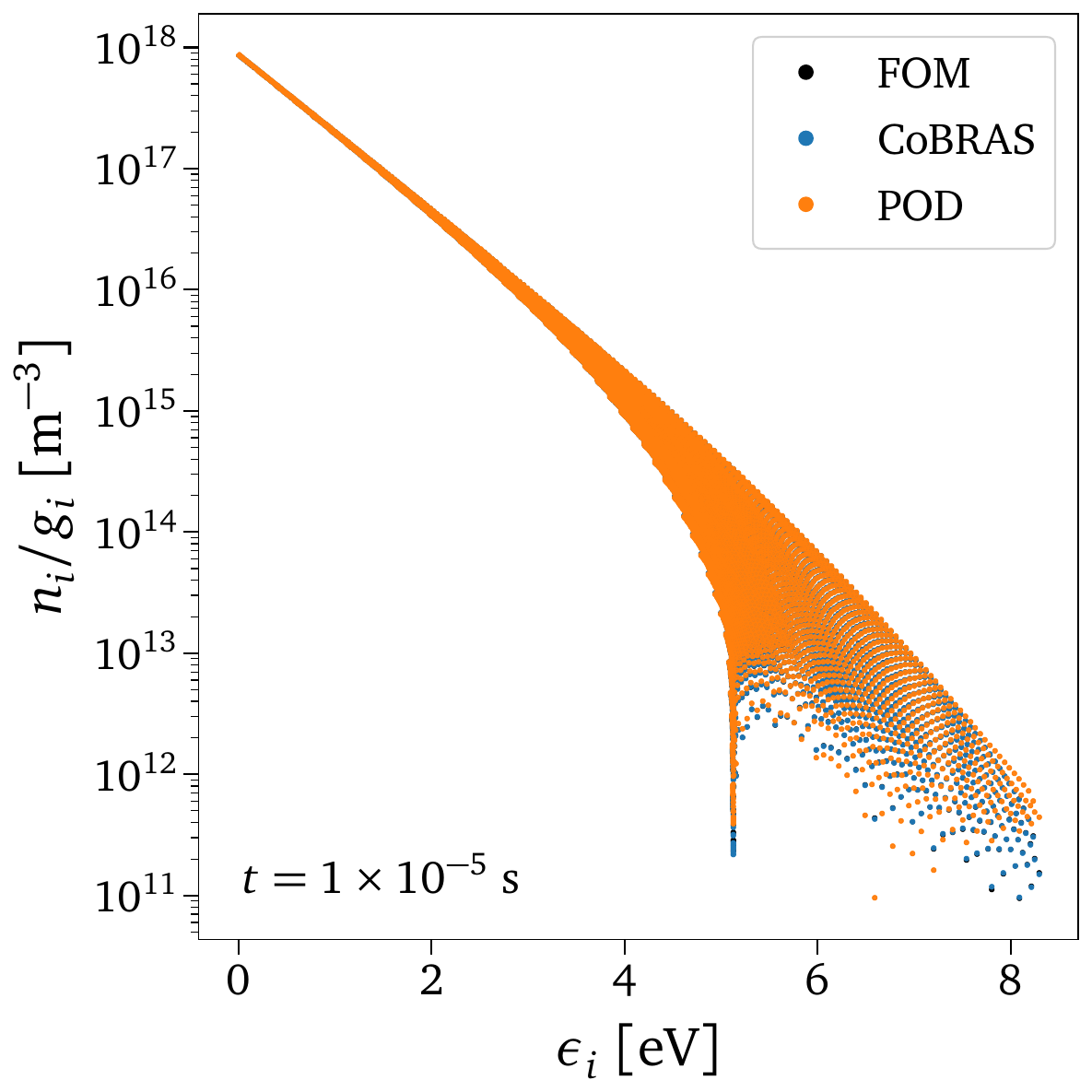}
\end{subfigure}
\begin{subfigure}[htb!]{0.23\textwidth}
    \includegraphics[width=\textwidth]{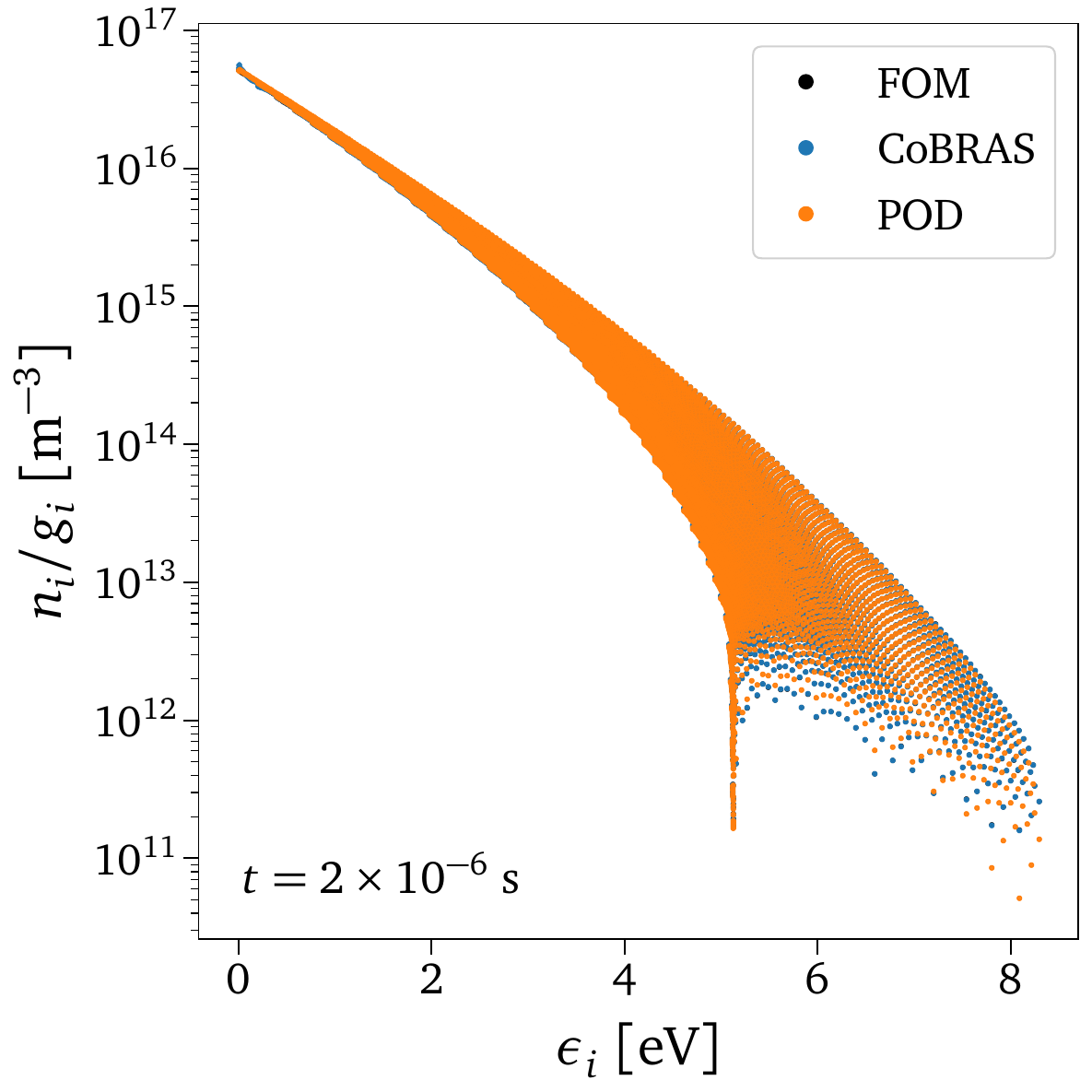}
\end{subfigure}
\begin{subfigure}[htb!]{0.23\textwidth}
    \includegraphics[width=\textwidth]{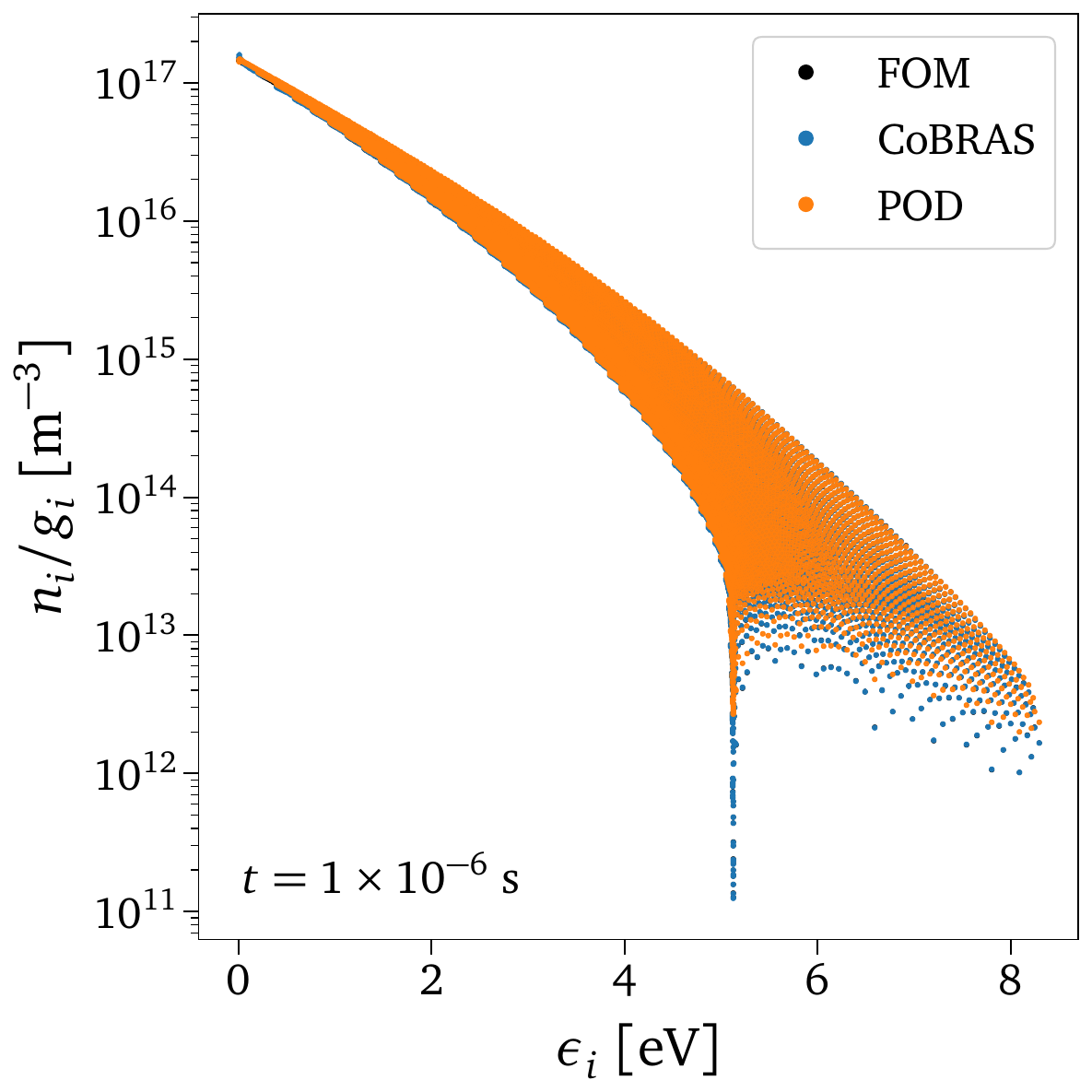}
\end{subfigure}
\caption{\textit{FOM vs. ROMs for RVC model: distribution snapshots}. Snapshots showing the time evolution of the O$_2$ energy level population distribution in an initially cold gas at various system temperatures. Results are computed using the FOM and various ROMs with a reduced dimensionality of $r = 24$. The second row illustrates the quasi-steady state (QSS) distribution.}
\label{fig:rvc_dist_cold_temps}
\end{figure}

\section{O$_2$-O$_2$ and O$_2$-O vibrational collisional model}\label{suppl:vc}
\subsection{Linearized FOM equations}\label{suppl:vc.linfom}
Equations \eqref{eq:vc.fom.ni} and \eqref{eq:vc.fom.no} can be written in a compact matrix-vector form as follows
\begin{align}
    \frac{d\mathbf{n}}{dt} =
        & \;\left(
            {}^{\mathrm{m}}\mathbf{K}^{\mathrm{e}}
            - {}^{\mathrm{m}}\mathbf{K}^{\mathrm{ed}}
            - {}^{\mathrm{m}}\mathbf{K}^{\mathrm{d}}
        \right)\colon\mathbf{n}\mathbf{n}^\intercal 
        + {}^{\mathrm{m}}\mathbf{K}^{\mathrm{er}}\mathbf{n}n^2_\atom \nonumber\\
        & + \left(
            {}^{\mathrm{a}}\mathbf{K}^{\mathrm{e}}
            - {}^{\mathrm{a}}\mathbf{K}^{\mathrm{d}}
        \right)\mathbf{n}n_\atom 
        + {}^{\mathrm{m}}\mathbf{k}^\text{r}n^4_\atom
        + {}^{\mathrm{a}}\mathbf{k}^\text{r}n^3_\atom \\
        = & \;\mathbf{A}_1\colon\mathbf{n}\mathbf{n}^\intercal 
        + \mathbf{A}_2\mathbf{n}n^2_\atom + \mathbf{A}_3\mathbf{n}n_\atom 
        + \mathbf{b}_1n^4_\atom + \mathbf{b}_2n^3_\atom
        \label{eq:vc.fom.vec.ni} \eqspace, \\
    \frac{dn_\atom}{dt}
        = & -\mathbf{2}^\intercal\frac{d\mathbf{n}}{dt}\label{eq:vc.fom.vec.no}
    \eqspace,
\end{align}
where $\mathbf{A}_1\in\mathbb{R}^{(N-1)\times (N-1)\times (N-1)}$, $\mathbf{A}_2\in\mathbb{R}^{(N-1)\times (N-1)}$, $\mathbf{A}_3\in\mathbb{R}^{(N-1)\times (N-1)}$, $\mathbf{b}_1\in\mathbb{R}^{(N-1)}$, $\mathbf{b}_2\in\mathbb{R}^{(N-1)}$. We linearize equations \eqref{eq:vc.fom.vec.ni} and \eqref{eq:vc.fom.vec.no} by following the same procedure used in \ref{suppl:rvc.linfom}, yielding the same equation as in \eqref{eq:rvc.lin}, with
\begin{align}
    \mathbf{A} & = \left(
        \mathbf{A}_1\bbgam
        + \mathbf{A}_1^\intercal\bbgam
        + \mathbf{A}_2
        + \frac{1}{\overline{n}_\atom}\mathbf{A}_3
    \right)\overline{n}_\atom^2 \eqspace, \label{eq:mat_a_vc_lin}\\
    \mathbf{b} & = \left(
        2\mathbf{A}_2\bbgam\overline{n}_\atom
        + \mathbf{A}_3\bbgam
        + 4\mathbf{b}_1\overline{n}_\atom
        + 3\mathbf{b}_2
    \right)\overline{n}_\atom^2\eqspace. \label{eq:mat_b_vc_lin}
\end{align}
The transpose operations on $\mathbf{A}_1$, as in equation \eqref{eq:mat_a_vc_lin}, are performed only on the last two dimensions of the 3D tensor.

\subsection{ROM equations}\label{suppl:vc.rom}
After applying the method described in sections \ref{sec:rom:pg} to \ref{sec:rom:covmat}, we derive the following reduced system for equations \eqref{eq:vc.fom.vec.ni} and \eqref{eq:vc.fom.vec.no}
\begin{align}
    \frac{d\mathbf{z}}{dt}
        = & \; \mathbf{A}_{1r}\colon\mathbf{z}\mathbf{z}^\intercal
        + \mathbf{A}_{2r}\mathbf{z} n^2_\atom
        + \mathbf{A}_{3r}\mathbf{z} n_\atom
        + \mathbf{b}_{1r}n^4_\atom + \mathbf{b}_{2r}n^3_\atom \eqspace, \\
    \frac{dn_\atom}{dt} = & -\mathbf{m}_r^\intercal \frac{d\mathbf{z}}{dt} \eqspace,
\end{align}
where the compressed tensors can be precomputed as follows
\begin{align}
    \mathbf{A}_{1r,pqs}
        & = \sum_{i,j,k}\psib_{ip}\mathbf{A}_{1,ijk}\phib_{jq}\phib_{ks}
        \eqspace, \\
    \mathbf{A}_{2r} & = \psib^\intercal\mathbf{A}_2\phib \eqspace, \\
    \mathbf{A}_{3r} & = \psib^\intercal\mathbf{A}_3\phib \eqspace, \\
    \mathbf{b}_{1r} & = \psib^\intercal\mathbf{b}_1 \eqspace, \\
    \mathbf{b}_{2r} & = \psib^\intercal\mathbf{b}_2 \eqspace, \\
    \mathbf{m}_r^\intercal & = \mathbf{2}^\intercal\phib \eqspace.
\end{align}